\documentclass[aps, prd, onecolumn, tightenlines, notitlepage, superscriptaddress, nofootinbib, preprintnumbers, floatfix,showkeys,11pt,altaffilletter]{revtex4-2}

\usepackage{upgreek}

\usepackage[official]{eurosym}
\usepackage[normalem]{ulem}
\usepackage{amstext}
\usepackage{graphicx}
\graphicspath{{figures/}}
\usepackage{url}
\usepackage{color}
\usepackage{ulem}
\usepackage[version=4]{mhchem}
\usepackage[utf8]{inputenc}
\usepackage{fontawesome}
\usepackage{yfonts}
\usepackage{amsmath}

\pdfoutput=1
\usepackage{textcomp}
\usepackage{comment}
\usepackage{yfonts}
\usepackage{epsfig,amsfonts,mathrsfs,graphicx,color,slashed,multirow}
\usepackage{latexsym,graphicx,slashed,color,enumerate,url,cancel,gensymb}
\usepackage{textcomp}

\usepackage{slashed}

\usepackage[x11names]{xcolor}
\usepackage[colorlinks,pdfstartview=FitV,breaklinks=true]{hyperref}
\usepackage{booktabs}
\usepackage{adjustbox}
\usepackage{textgreek} 
\usepackage{caption, subcaption} 
\usepackage{ragged2e} 
\DeclareCaptionJustification{justified}{\justifying}

\usepackage{booktabs} 
\usepackage{float}
\usepackage{orcidlink}
\usepackage{lmodern}
\usepackage{ae,aecompl}
\usepackage{appendix}
\usepackage{accents}


\newcommand*{\pbar}[1]{\accentset{(-)}{#1}}

\definecolor{persiangreen}{rgb}{0.0, 0.65, 0.58}
\definecolor{mediumpersianblue}{rgb}{0.0, 0.4, 0.65}

\makeatletter
    \newcommand{\colorboxed}[3][white]{\fcolorbox{#2}{#1}{\m@th$\displaystyle#3$}}
\makeatother

\AtBeginDocument{\hypersetup{citecolor=mediumpersianblue,linkcolor=mediumpersianblue,urlcolor=mediumpersianblue}}
\usepackage{appendix}

\begin{document}

\preprint{IPPP/24/23}

\title{{\Large Primordial black hole probes of heavy neutral leptons}}

\author{Valentina De Romeri~\orcidlink{0000-0003-3585-7437}}
\email{deromeri@ific.uv.es}

\affiliation{Instituto de F\'{i}sica Corpuscular (CSIC-Universitat de Val\`{e}ncia), Parc Cient\'ific UV C/ Catedr\'atico Jos\'e Beltr\'an, 2 E-46980 Paterna (Valencia) - Spain}

\author{Yuber F. Perez-Gonzalez~\orcidlink{0000-0002-2020-7223}}
\email{yuber.f.perez-gonzalez@durham.ac.uk
}
\affiliation{Institute for Particle Physics Phenomenology, Durham University, South Road, Durham DH1 3LE United Kingdom.}

\author{Agnese Tolino~\orcidlink{0009-0003-3278-0902}}
\email{atolino@ific.uv.es}
\affiliation{Instituto de F\'{i}sica Corpuscular (CSIC-Universitat de Val\`{e}ncia), Parc Cient\'ific UV C/ Catedr\'atico Jos\'e Beltr\'an, 2 E-46980 Paterna (Valencia) - Spain}

\keywords{primordial black holes, neutrinos, heavy neutral leptons, IceCube}

\begin{abstract}
Primordial black holes (PBH), while still constituting a viable dark matter component, are expected to evaporate through Hawking radiation. 
Assuming the semi-classical approximation holds up to near the Planck scale, PBHs are expected to evaporate by the present time, emitting a significant flux of particles in their final moments, if produced in the early Universe with an initial mass of $\sim 10^{15}$ g. 
These ``exploding'' black holes will release a burst of Standard Model particles alongside any additional degrees of freedom, should they exist. 
We explore the possibility that heavy neutral leptons (HNL), mixing with active neutrinos, are emitted in the final evaporation stages. 
We perform a multimessenger analysis. We calculate the expected number of active neutrinos from such an event, including contributions due to the HNL decay for different assumptions on the mixings, that could be visible in IceCube. We also estimate the number of gamma-ray events expected at HAWC. By combining the two signals,
we infer sensitivities on the active-sterile neutrino mixing and on the sterile neutrino mass.  We find that, for instance, for the scenario where $U_{\tau 4}\neq 0$, IceCube and HAWC could improve current constraints by a few orders of magnitude, for HNLs masses between 0.1 - 1 GeV, and a PBH explosion occurring at a distance of $\sim 10^{-4}$ pc from Earth.
\end{abstract}
\maketitle

\section{Introduction}

One major shortcoming of the Standard Model (SM) of particle physics is the lack of an explanation of the largest matter component of the Universe. Numerous observations at different cosmological scales indicate that most matter in the Universe is made up of dark matter (DM), whose nature and properties are still unknown~\cite{Aghanim:2018eyx}. 

While most observations seem to prefer a particle nature for DM, thus calling for the introduction of new particles beyond the Standard Model (BSM), null searches at typical DM direct detection experiments~\cite{Schumann:2019eaa,Billard:2021uyg} have recently revived the interest in alternative candidates. 
One such an option is that the DM is made not by BSM particles but by Primordial Black Holes (PBHs), see e.g. Refs.~\cite{Carr:2020xqk,Bird:2022wvk,Green:2024bam} for recent reviews on PBH-DM.

PBHs are hypothetical black holes that would have formed in the early Universe before the Big-Bang Nucleosynthesis (BBN) epoch~\cite{Zeldovich:1967lct,Hawking:1971ei,Carr:1974nx,Chapline:1975ojl}. 
Black holes have received renewed interest in recent years after the observation of gravitational waves in the LIGO-Virgo experiments~\cite{Abbott:2016blz,Abbott:2020gyp}, which have definitely proven their existence and opened the possibility of searching for black holes of primordial origin. 
There are different constraints on PBHs depending on their masses and whether they still exist at present.
Crucially, some of the strongest bounds rely on the particle production from the black hole gravitational field, i.~e. via Hawking radiation~\cite{Hawking:1974rv,Hawking:1974sw}.
As demonstrated by Hawking, black holes have a temperature
inversely proportional to their masses and they are expected to evaporate by emitting a flux of particles with thermal spectra.
Clearly, the observation of PBH evaporation would confirm this prediction and provide invaluable information about several aspects of physics. 
However, non-detection of PBH evaporation allows to impose a large number of constraints on the abundance of PBHs with masses in the range $10^{15}\sim 10^{17}$ g, using $\gamma$ rays~\cite{1976ApJ...206....8C,Lehoucq:2009ge,Wright:1995bi,Arbey:2019mbc,Ballesteros:2019exr,Laha:2020ivk}, electrons/positrons~\cite{Boudaud_2019,Dasgupta:2019cae,DeRocco:2019fjq,Laha:2019ssq} and neutrinos~\cite{Dasgupta:2019cae,Wang:2020uvi,DeRomeri:2021xgy,Bernal:2022swt}. Lighter PBHs, that would have evaporated at earlier times in the Universe's history, can be constrained through BBN~\cite{Carr:2009jm,Carr:2020gox,Keith:2020jww}, reionization~\cite{He:2002vz,Mack:2008nv} or CMB~\cite{Carr:2009jm,Carr:2020gox} measurements. See, e.g., Ref.~\cite{Auffinger:2022khh} for a comprehensive review. 

Non-rotating PBHs with masses $\lesssim 5 \times 10^{14}$ g would have completely evaporated by now~\cite{Page:1976df,Page:1976ki,MacGibbon:2007yq} and could not constitute the totality of DM. 
However, they could still give rise to interesting signals. Following an initial stage during which the evaporation proceeds slowly, as the PBH mass decreases, its temperature and flux increase, leading to a final stage that would resemble an \textit{explosion}~\cite{Hawking:1974rv}.
The luminosity of such exploding PBHs is established by the types of degrees of freedom that they can emit, together with their instantaneous properties (mass spectrum, charge and spin). 
Clearly, photons would be one smoking-gun signature of these explosions, having strong matches with a gamma-ray burst though with some intrinsic differences~\cite{Boluna:2023jlo}. Facilities such as H.E.S.S~\cite{Glicenstein:2013vha,Tavernier:2019exh}, Milagro~\cite{Abdo:2014apa}, VERITAS~\cite{Archambault:2017asc}, Fermi-LAT~\cite{Fermi-LAT:2018pfs}, and HAWC~\cite{HAWC:2013kzm,HAWC:2019wla} have already placed constraints on the rate of PBH explosions. Notice that, in order to be detectable by current (or even near-future) telescopes, they have to occur (at most) a few parsecs from Earth~\cite{Ukwatta:2010zn,Boluna:2023jlo,Perez-Gonzalez:2023uoi}. Currently,  the H.E.S.S array of imaging atmospheric Cherenkov telescopes~\cite{HESS:2023zzd} sets the strongest direct limit on the rate of exploding PBHs: $\dot n_\mathrm{PBH} < 2000~\mathrm{pc}^{-3} ~\mathrm{yr}^{-1}$ for a burst interval of 120 seconds, at the $95\%$ confidence level. 

Evaporating PBHs will radiate all elementary particles with a mass below their temperature. Among the SM states, neutrinos constitute one possible signature of exploding PBHs~\cite{Halzen:1995hu,Dave:2019epr,Capanema:2021hnm,Perez-Gonzalez:2023uoi,Calza:2023iqa}. On the other hand, new degrees of freedom that may appear in BSM scenarios could also be probed through the final stages of PBHs evaporation~\cite{Ukwatta:2015iba,Baker:2021btk,Calabrese:2021src,Calabrese:2022rfa,Baker:2022rkn,Calza:2021czr,Calza:2022ljw,Auffinger:2022khh,Calza:2023gws}. 
For instance, several BSM extensions call upon the introduction of SM-singlet degrees of freedom, dubbed sterile neutrinos or heavy neutral leptons (HNLs), to address neutrino masses and mixings, another shortcoming of the SM and actually the first laboratory probe of BSM physics. HNLs can have masses that vary over several orders of magnitude, depending on the actual mechanism that originates neutrino masses, and extended phenomenological implications. Despite intense experimental efforts in searching for these sterile fermions, no positive evidence has been found to date~\cite{Atre:2009rg,Drewes:2015iva,Abdullahi:2022jlv}. However, HNLs in the MeV-GeV mass range remain appealing BSM candidates, as they emerge rather naturally in several SM extensions (e.g.~\cite{Abdullahi:2022jlv}) besides being accessible also at terrestrial facilities.
Interestingly, the existence of HNLs and PBHs in the early Universe could affect leptogenesis scenarios, as noted in Refs.~\cite{Bugaev:2001xr, Fujita:2014hha, Morrison:2018xla, Ambrosone:2021lsx, Hooper:2020otu, Perez-Gonzalez:2020vnz, Bernal:2022pue,Calabrese:2023bxz,Calabrese:2023key,Ghoshal:2023fno}.

In this paper we consider the constraints that could be derived on HNLs after an exploding PBH in close proximity to Earth is observed in a neutrino telescope.
We estimate sensitivities at IceCube~\cite{IceCube:2018pgc,IceCube:2019cia} as a benchmark facility to observe high-energy neutrinos from a PBH burst. 
IceCube is an ice Cherenkov detector located in the South Pole with a size of about 1 ${\rm km^3}$. It has an impressive angular resolution of $\lesssim 1\degree$ for high-energy muon-tracks with energies $E\sim{\cal O}({\rm TeV})$. 
Thus, we expect IceCube to have the ability to determine the PBH position and discriminate neutrino events from possible backgrounds~\cite{Dave:2019epr}. 
While we expect the HNL contribution to be distinguishable in the neutrino signal, we further compute the photon events expected at HAWC, whose sensitivity falls in the range $[10^2, 10^5]$ GeV~\cite{HAWC:2019wla}. Photon events outnumber neutrinos by a few orders of magnitude: a multimessenger analysis allows to pinpoint the PBH position with more precision. 
At this scope, we will assess potential modifications to the neutrino spectra detected by IceCube if HNLs are generated during the final stages of PBH evaporation and subsequently decay into active neutrinos. Similarly, we will compute the photon spectra taking into account photon contributions arising in the HNL decay chain.
While unresolved issues persist regarding black hole evaporation physics, such as the information paradox~\cite{Hawking:1976ra,Almheiri:2020cfm,Buoninfante:2021ijy} or the thermal nature of the Hawking spectrum post-Page time~\cite{Page:1993wv,Page:2013dx}, and how alterations would impact the mentioned constraints --- see e.~g.~\cite{Alexandre:2024nuo,Thoss:2024hsr,Balaji:2024hpu,Haque:2024eyh} for an example of possible modifications in a specific prototype model --- we adopt a more neutral stance ahead, and assume the semi-classical approximation's validity until near the Planck scale.

The remainder of this work is organized as follows. We discuss the primary and secondary SM contributions to the muon neutrino spectra expected at IceCube in Section~\ref{sec:vemission}. Section~\ref{sec:HNLcontrib} is instead devoted to summarize the relevant secondary contributions from HNL decays to both neutrino and photon spectra. We present in Sec.~\ref{sec:analysis} the details of the statistical analysis performed to estimate the sensitivity at IceCube and HAWC. We show our results in Sec.~\ref{sec:results}, in terms of sensitivities on the HNL parameter space. Finally we conclude in Sec.~\ref{sec:conc}. 
Throughout this work, we consider natural units where $\hbar = c = k_{\rm B} = 1$, and define the Planck mass to be $M_p=1/\sqrt{G}$, with $G$ the gravitational constant.

\section{Standard Model contributions}
\label{sec:vemission}

Among different possibilities, large density fluctuations in the very early Universe may have led to the creation of a PBH population that could persist until the present day. 
Typically, the initial mass of PBHs is linked to the particle horizon mass when they formed. 
Given the lack of clear constraints on the Universe's evolution before BBN, PBHs could have masses ranging from around ${\cal O}(1)$ g to several orders of magnitude the Solar mass~\cite{Carr:2009jm,Carr:2020gox}. 
Within this broad range of masses, a PBH could have formed with precisely the right mass for its lifetime to match the age of the Universe, thus entering the final explosive stages of its evolution at present. 
In the standard evaporation scenario, considering only SM degrees of freedom emitted during evaporation, a PBH with a mass of approximately $\sim 10^{15}$ g would be evaporating today.

Therefore, by observing the absence of gamma-ray bursts in our Galaxy, we can potentially limit the quantity of evaporating PBHs at present. As mentioned in the Introduction, the current strongest direct limit on the rate of exploding PBHs is $\dot n_\mathrm{PBH} < 2000~\mathrm{pc}^{-3} ~\mathrm{yr}^{-1}$ for a burst interval of 120 seconds, at the $95\%$ confidence level~\cite{HESS:2023zzd}. 
In general, the rate of evaporating PBHs depends on the initial PBH mass distribution and current constraints on the maximum fraction of DM that could be composed of PBHs.  
Following Ref.~\cite{Boluna:2023jlo}, the evaporation rate is given by  
\begin{align}
    \dot{n}_\mathrm{PBH} = \rho_{\rm DM} \frac{\psi_i(M_\star)}{3 t_U},
\end{align}  
where $\rho_{\rm DM}$ denotes the local DM density, $\psi_i(M)$ represents the initial PBH mass distribution, and $M_\star \sim 5.3 \times 10^{14}~{\rm g}$ corresponds to the mass of a PBH with a lifetime equal to the age of the Universe, $t_U$, assuming the emission of only SM degrees of freedom.  
For a log-normal initial mass distribution~\cite{Green:2016xgy,Dolgov:2008wu,Cheek:2022mmy}, the evaporation rate is approximately~\cite{Boluna:2023jlo}  
\begin{align}
    \dot{n}_\mathrm{PBH} \sim \frac{1.2\times 10^{-3}~{\rm pc^{-3}~yr^{-1}}}{\sigma},
\end{align}  
where $\sigma$ denotes the width of the mass distribution.  
Thus, within a spherical volume of radius one parsec, current observational constraints allow the presence of approximately one evaporating PBH per year if $\sigma \lesssim 5\times 10^{-3}$.  
Naturally, this estimate depends on the assumed initial mass function and can vary significantly for alternative distributions, such as power-law or critical collapse models; see Ref.~\cite{Boluna:2023jlo} for further details.  Given these considerations,  it is conceivable that one of these PBHs may be in close proximity to Earth, near the end of its lifespan, presenting an opportunity for observation. 
This is the scenario we will presume henceforth.

Assuming a semi-classical picture, where matter fields, treated as quantum, propagate in a classical background describing the gravitational field of a black hole, Hawking determined the particle emission after the gravitational collapse~\cite{Hawking:1974rv,Hawking:1974sw}.
The particle emission rate depends on the black hole instantaneous properties, such as its mass $M$, angular momentum $J$ and charge $Q$. Nevertheless, the charge and angular momentum\footnote{However, note that if there exist a dark sector containing scalar degrees-of-freedom in large quantities $N> 100$, the angular momentum evaporates in a significantly different rate, see e.~g.~Refs.~\cite{Calza:2021czr,Calza:2022ljw, Calza:2023gws,Calza:2023iqa}.} are expected to be depleted in a faster rate than the mass.
Thus, we can safely assume that PBH reaching the final stages of their lifetime are Schwarzschild, described uniquely by their masses $M$.

If we accept the semi-classical approach as valid until close to the Planck scale, the PBH mass loss rate can be expressed as
\begin{align}
    \frac{d M}{d t} &= - \varepsilon(M)\frac{M_p^4}{M^2} \, ,
\end{align}
where $\varepsilon(M)$ is the evaporation function that describes the degrees-of-freedom (dof) that could be emitted for a given PBH mass~\cite{MacGibbon:1990zk,MacGibbon:1991tj,Cheek:2021odj,Cheek:2022mmy}
\begin{align}\label{eq:mass_rate}
    \varepsilon(M) &= \sum_{i = \text{dofs}}\frac{g_i}{128 \pi^3}\int_{0}^\infty \frac{x \,\Gamma_{s_i}(x)}{\exp(x) - (-1)^{2s_i}}\,d x \, .
\end{align}
Here $g_i$ are the internal dofs associated to the particle species $i$ having spin $s_i$, and $\Gamma_{s_i}$ indicates the absorption probability, also known as greybody factors, describing potential backscattering arising from centrifugal and/or gravitational potentials related to the curved spacetime around the black hole.
Such quantities are computed using a numerical approach as the one described in Refs.~\cite{Page:1976df,Page:1976ki,Page:1977um}.
For convenience, we rely on the absorption probabilities from the code {\tt BlackHawk}~\cite{Arbey:2019mbc,Arbey:2021mbl}, which are valid for massless particles. We have verified that using the correct probabilities for massive fermions from Refs.~\cite{Doran:2005vm,Dolan:2006vj} yields similar results to those obtained from the tables provided in {\tt BlackHawk}.
The variable $x$ is defined as the ratio $x=E/T$, where $E$ is the particle's energy and $T$ the Hawking temperature, related to the PBH mass via
\begin{align}
    T= \frac{1}{8\pi G M} \sim 1~{\rm TeV} \left(\frac{10^{10}~{\rm g}}{M}\right).
\end{align}
From this we can infer that dofs having masses $m\lesssim 1~{\rm TeV}$ will be emitted by a black hole with $M=10^{10}~{\rm g}$ without any Boltzmann suppression, even those that do not interact with the SM. If the number of dofs is constant during the whole PBH lifetime, the mass lose rate is exactly solvable,
\begin{align}
    M(t) = M_{\rm in} \left(1-\frac{t}{\tau_d}\right)^{1/3},
\end{align}
with $\tau_d$ representing the PBH lifetime.
The lifetime is given by
\begin{align}
    \tau_d= \frac{M_{\rm in}^3}{3 \varepsilon M_p^4} \sim 428~{\rm s} \left(\frac{4.07\times 10^{-3}}{\varepsilon}\right)\left(\frac{M_{\rm in}}{10^{10}~{\rm g}}\right)^3,
\end{align}
where we used $\varepsilon = 4.07\times 10^{-3} $ corresponding to the value of the evaporation function for the SM dofs. 

The instantaneous emission rate for a fermion (boson) $i$ from PBH evaporation is given by a Fermi-Dirac (Bose-Einstein) thermal spectrum corrected by the absorption cross section mentioned above,
\begin{align}
    \frac{d^2N_i}{dE dt}(M) = \frac{g_i}{2\pi} \frac{\Gamma_{s_i}(E)}{\exp(8\pi G M E) - (-1)^{2s_i}}.
\end{align}
In experimental terms, optical and neutrino telescopes are anticipated to detect the final burst from PBHs within a finite time frame, measuring a burst of photons or neutrinos over a specific duration. 
Therefore, integrating the instantaneous spectrum over the assumed time interval is essential to calculate the fluence for a given experiment. 
Since any particles emitted prior would evade detection, we set the initial mass of an exploding PBH to yield a lifetime matching the observation period.
For instance, if the observation time is 100 s, we adjust $M_{\rm in} = 6.2\times 10^9$ g.
The time-integrated spectrum in a time interval $\tau$ is therefore given by
\begin{align}\label{eq:gen_fluence}
    \frac{dN_i}{dE} = \int_0^\tau dt\, \frac{d^2N_i}{dE dt}(M(t)) \, .
\end{align}
Once such time-integrated spectrum is determined for a given particle physics model, it will be possible to estimate the observed events in a neutrino telescope. However, it is important to note that observable neutrinos, specifically active neutrinos, will be produced in two different forms: first, directly from evaporation, and secondly from the decay of unstable SM particles and HNL of interest. Next, we provide a detailed account of these emissions.

\subsection{Primary emission}
\label{subsec:primarySM}

Direct neutrino emission from evaporation has been considered since the discovery of such a phenomenon. 
Moreover, neutrinos were considered as a benchmark for analyzing the emission of massless fermions from a black hole, due to the belief, at the time, that neutrinos were massless. 
Nevertheless, following the discovery of neutrino oscillations and consequently neutrino masses, the primary emission model has been revised in Ref.~\cite{Lunardini:2019zob}.
The main consequence of being massive particles for the present case study is that neutrinos are assumed to be emitted as mass eigenstates $\nu_{1,2,3}$, instead of the flavour eigenstates $\nu_{e,\mu,\tau}$ considered in early studies.

To detect the primary neutrinos from evaporation, we would still need to utilize weak interactions.
The time-integrated spectrum of an observable state with flavour $\alpha$ is then obtained by projecting the spectra of the mass eigenstates into the flavour basis by employing the Pontecorvo-Maki-Nakagawa-Sakata (PMNS) mixing matrix $U_{\alpha i}$,
\begin{align}
    \left.\frac{dN_{\nu_\alpha}}{dE}\right|_{\rm pri} = \sum_{i=1}^3 |U_{\alpha i}|^2 \frac{dN_{\nu_i}}{dE},
\end{align}
with $dN_{\nu_i}/dE$ the time-integrated spectrum for the massive state $\nu_i$ given by Eq.~\eqref{eq:gen_fluence} for the fermionic case.
The primary emission reflects the thermal spectrum, corrected by the greybody factors, expected from the evaporation.
Therefore, such emission will be peaked close to a value of $\sim 6 T_{\rm PBH}^{\rm in}$, $T_{\rm PBH}^{\rm in}$ being the initial PBH temperature.
For the exploding PBHs that we are interested in, we find that such a primary spectrum is peaked at large energies, $E\sim {\cal O}(10~{\rm TeV})$, but, as we will see shortly, is subdominant to secondary emission.

\subsection{Secondary emission}
\label{subsec:secondarySM}

The largest part of the overall time-integrated spectrum of neutrinos does not come from the direct black hole emission but from the weak decays of unstable particles produced during the evaporation~\cite{MacGibbon:1990zk,MacGibbon:1991tj}. If we only consider the SM particle content for the moment, quark hadronization and hadron posterior decays, together with heavy leptons and massive gauge bosons decays, will produce a large amount of neutrinos since these dofs dominate the SM spectrum.
To estimate the time-integrated spectrum of these neutrinos, known as the secondary component, we need to use tools that compute the spectra from hadronization plus decay considering SM gauge interactions, that are convoluted with the Hawking spectrum of parent particles  $j$,
\begin{align}\label{eq:sec_PBH}
    \left.\frac{dN^{\nu_{\beta}}_{\rm PBH}}{dE}\right|_{\rm sec} = \sum_{j}\int_0^\infty dE_{j} \frac{dN_j}{dE_j} \frac{dN (j\longrightarrow \nu_{\beta})}{dE}(E,E_j),
\end{align}
where $dN(j \longrightarrow\nu_{\beta})/dE$ are the neutrino spectra from the hadronization/decay of the parent particle $j$ and $dN_j/dE_j$ is its primary time-integrated spectrum. Finally, the subscript ``PBH'' indicates that the time-integrated spectrum is computed close to the PBH.
As an example, we have that quark hadronization will produce large amounts of $\pi^{\pm}$ that subsequently decay generating neutrinos in their flavour eigenstates, $\pi^\pm\to \mu^{\pm} + \pbar{\nu}_\mu \to e^{\pm} + \pbar{\nu}_e + \pbar{\nu}_\mu$, thus producing three neutrinos per each pion produced.
Therefore, we expect the secondary neutrino spectrum to dominate over the primary, specially at low energies.
We use {\tt BlackHawk v2.3}~\cite{Arbey:2019mbc,Arbey:2021mbl} to evaluate both the primary and secondary spectra of neutrinos. In particular, we rely on \texttt{HDMSpectra}~\cite{Bauer:2020jay}, already integrated in {\tt BlackHawk}, to compute the secondary emission through the hadronization of primary particles, whose range of validity extends up to the Planck scale.

As such secondary neutrinos are emitted as flavour eigenstates, we need to consider neutrino oscillations to compute the correct spectrum.
Given the distances involved here, significantly larger than oscillation distances $\sim {\cal O}({\rm km})$, neutrino oscillations undergo decoherence in their route from the PBH to the Earth.
The secondary contribution far from the PBH, including decohered oscillations, is given by
\begin{equation}
    \left.\frac{dN_{\nu_{\alpha}}}{dE}\right|_{\rm sec} = \sum_{\beta = e,\mu,\tau}\sum_{i=1,2,3} |U_{\alpha i}|^2|U_{\beta i}|^2\left.\frac{dN^{\nu_{\beta}}_{\rm PBH}}{dE} \right|_{\rm sec}.
\end{equation}
Thus, events from a PBH burst would consist in the combined contribution of both primary and secondary contributions defined above. 
However, if additional BSM dofs exist which decay producing additional neutrinos, the prediction of the secondary contribution would be significantly modified, and therefore the observation of a PBH burst could provide additional information to constrain the presence of such BSM states.
We will consider next a simple, but well motivated, scenario for BSM physics, and the implications in a PBH burst. 

\section{Heavy Neutral Lepton contributions}
\label{sec:HNLcontrib}

\subsection{Heavy Neutral Lepton Generalities}\label{subsec:hnl_gens}

The experimental observation of neutrino oscillations has firmly established the existence of neutrino masses, indicating the existence of BSM physics. 
To accommodate neutrino masses, a minimal possibility consists in including $n$ right-handed fermions $N_i$, $i=\{1,2,\ldots, n\}$, singlets under the SM gauge group. A simple model that can realize this setup is the type-I \emph{seesaw}
mechanism~\cite{Minkowski:1977sc,Yanagida:1979as,
  GellMann:1980vs,Mohapatra:1979ia,Schechter:1980gr}.
Crucially, the SM symmetries allow for masses $M^{ij}_R$ for such right-handed neutrinos (RHN) of the Majorana type. The most general mass lagrangian is given by
\begin{align}\label{eq:RHN_lag}
   \mathscr{L}_{\rm RHN}^m = -Y_{\alpha i} \overline{L}_\alpha \widetilde{H}N_i - \frac{1}{2} M^{ij}_R \overline{N_i^c} N_j + {\rm h.c.},
\end{align}
with $\overline{L}_\alpha$ the lepton doublets, $\widetilde{H}$ the conjugate Higgs doublet, and $Y_{\alpha i}$ Yukawa parameters related to the couplings between the aforementioned doublets and the RHNs.
To address the observed neutrino oscillation pattern~\cite{deSalas:2020pgw,Capozzi:2021fjo,Esteban:2020cvm}, where two non-zero quadratic mass differences have been measured, the minimal number of RHNs is $n=2$. However, this does not forbid the existence of additional singlets.

After electroweak symmetry breaking, the lagrangian above can be rewritten as
\begin{align}\label{eq:RHN_lag_aEWSB}
    \mathscr{L}_{\rm RHN}^m = - \frac{1}{2} \overline{{\cal N}_L^c} M_\nu {\cal N}_L+ h.c.,
\end{align}
where
\begin{align}\label{eq:mass_matrix}
    {\cal N}_L = \begin{pmatrix}
        \nu_L\\
        (N_R)^c
    \end{pmatrix},\quad
    M_\nu = \begin{pmatrix}
       \mathbf{0}_3 & Y v/\sqrt{2}\\
        Y^T v/\sqrt{2} & M_R
    \end{pmatrix} \, ,
\end{align}
being $\nu_L = (\nu_e, \nu_\mu, \nu_\tau)^T$ and $N_R = \{N_1,\ldots,N_n\}$ active and sterile neutrino states, respectively, and $v$ the Higgs vacuum expectation value.
After diagonalizing the Lagrangian in Eq.~\eqref{eq:RHN_lag_aEWSB} to form $M_\nu = {\mathcal{U}} M_\nu^{\rm diag} {\mathcal{U}}^T$, utilizing a unitary rotation matrix ${\mathcal{U}}$ with $(3+n)\times(3+n)$ entries, the resulting spectrum would encompass $3+n$ eigenstates denoted as ${\cal N}_L^m=\{\nu_1,\ldots,\nu_{3+n}\}$. Among these, 3 are considered ``active'', representing weakly interacting states, while the remaining $n$ are predominantly ``sterile'', as their interactions with other SM dofs occur solely through mixing parametrized via ${\mathcal{U}}$. 
The modified PMNS matrix now corresponds to a $3\times 3$ block of ${\mathcal{U}}$ and can show deviations from unitarity.

If we assume a large hierarchy between the RHN mass and electroweak scales, i.e., $M_R \gg v Y$, Majorana masses for the active neutrinos $\{\nu_1,\nu_2,\nu_3\}$ can be explicitly written as\footnote{Note that while being physical, related to CP-properties of active neutrinos, the negative sign can be reabsorbed through a redefinition of fields.},
\begin{align}\label{eq:seesaw}
    m_\nu \approx -\frac{v^2}{2} Y M_R^{-1} Y^T.
\end{align}
In the scenario where the Yukawa matrix has entries of $|Y_{\alpha i}| \sim  {\cal O}(1)$, we would require $M_R\sim {\cal O}(10^{12})~{\rm GeV}$, that is, the RHN masses are close to the grand-unification scale, to have active neutrino masses $m_\nu \sim {\cal O}({\rm eV})$, consistent with oscillation experiments. 
This case, corresponding to the well-known seesaw mechanism, would then explain the observed smallness of neutrino masses.
If such large hierarchy does exist, then the mixing between active and RHN would be extremely suppressed, ${\mathcal{U}}\approx \frac{v}{\sqrt{2}} Y M_R^{-1}$.
Therefore, the RHN would have tiny mixings with the active sector, making a direct experimental observation of such dofs quite challenging.

However, since the SM does not constrain either the scale of Majorana masses or the values of the Yukawa parameters $Y_{\alpha i}$ in Eq.~\eqref{eq:RHN_lag}, it is possible for $M_R$ to be at the electroweak scale and still explain neutrino masses. For instance, this occurs in low-energy realizations of the seesaw scenario, like the inverse~\cite{Mohapatra:1986bd} and the linear seesaw~\cite{Akhmedov:1995ip,Akhmedov:1995vm,Malinsky:2005bi}. In such a case, the seesaw formulae for the masses and mixing are not valid and need to be extended. 
Thus, in general, the diagonalization of Eq.~\eqref{eq:RHN_lag_aEWSB} would imply the presence of light sterile neutrinos, which could exhibit substantial mixing with the active states. 
This opens up the possibility of experimental detection of these light sterile states, i.~e., HNLs.

Various experimental searches of these states have been conducted and are planned for future facilities, as comprehensively reviewed in Ref.~\cite{Abdullahi:2022jlv}. In this context, our focus lies on understanding the impact of such additional states on the observed neutrino events arising from a PBH burst. To simplify matters, we assume the presence of a single HNL state $\nu_4$ with a mass $m_4$, which interacts with active neutrinos through the mixing $U_{\alpha 4}$, where $\alpha = {e, \mu, \tau}$. In such a simplified phenomenological scenario
\begin{align}\label{eq:mix_matrix4}
    {\cal U} = \begin{pmatrix}
       U_{\alpha i} & U_{\alpha 4}\\
        U_{4 i} & U_{4 4}
    \end{pmatrix} \, ,
\end{align}
while the flavor states $\nu_\alpha$ can be written as a superposition of the mass eigenstates as 
$\nu_\alpha = \sum_{i=1}^3 U_{\alpha i} \nu_{i} + U_{\alpha 4} \nu_4$.

Before we explore the effects of an additional HNL on a PBH burst, it is important to address the unitarity of the neutrino mixing matrix used to calculate the active neutrino events in Sec.~\ref{sec:vemission}.
In a standard scenario, where no sterile states are included, the mixing angles are fixed according to \cite{deSalas:2020pgw}, and the PMNS is a priori unitary. 
However, in a non-standard scenario, where sterile neutrinos are introduced in the picture and mix with at least one neutrino species, the unitarity of the $3\times 3$ block of ${\mathcal{U}}$ is not guaranteed, and hence 
\begin{align}
    \sum_{i=1}^3|U_{\alpha i}|^2  \neq  1 \, ,
\end{align}
in general.
Nevertheless, we have verified that taking the $3\times 3$ ``active" block of ${\mathcal{U}}$ as unitary is a good approximation as long as the mixing between active neutrinos and the additional state is smaller than $|U_{\alpha 4}|^2 \lesssim 10^{-3}$. 

Should HNLs exist and be produced during PBH evaporation, they will impact its mass evolution and give rise to additional secondary contributions. 
The rate at which PBHs lose mass is primarily influenced by the number of emitted degrees of freedom~\cite{Baker:2021btk}, with only minor dependence on the properties of the emitted species, whether fermionic or bosonic, and their actual spin~\cite{Page:1976df}.
As a consequence, if the SM particle content is extended only by one type of HNL, the PBH mass evolution will not be significantly modified. On the other hand, the possible emission of HNLs in multiple generations would increase the evaporation rate accelerating the PBH explosion. We have checked that a large ($\gtrsim 10$) number of generations is required in order to observe sizeable effects in the primary emission, in agreement with~\cite{Baker:2021btk}. 

Since we will work under the assumption that only one generation of HNLs is present, the primary neutrino spectra emitted in the PBH explosion are the same as in the SM. However, the HNLs produced during the PBH evaporation will generate an additional secondary flux of active neutrinos from their decays. The morphology of such a secondary contribution, expected to induce visible signals at IceCube, will depend on the HNL mass regime, on the available decay channels, and which active neutrino flavour the sterile neutrino mixes with.

At the scope of our analysis we assume an observation time of 100 s in a neutrino telescope like IceCube, corresponding to the explosion of a PBH with mass $M \sim 6\times 10^9$ g. This will constrain the maximum HNL mass that can be produced in the evaporation process to  $m_4 \lesssim 2$ TeV, a value that is dictated by the temperature of the PBH, as above that threshold the HNL emission is heavily suppressed. HNLs lighter than this value can be produced and give rise to an observable flux of daughter active neutrinos.
For the sake of illustration, in the following we will consider two separate HNL regimes: a light-mass ([0.1-1] GeV) and a heavy-mass ([0.5-2] TeV) regime. 

\subsection{Light-mass regime: [0.1-1] GeV}
\label{sub:light_mass_regime}

HNL decay modes have been studied extensively in the literature, see e.g. Refs.~\cite{Shrock:1980ct,Atre:2009rg,Helo:2010cw,Bondarenko:2018ptm,Bryman:2019ssi,Bryman:2019bjg,Mastrototaro:2019vug,Coloma:2020lgy,Abdullahi:2022jlv,Akita:2023iwq}. 
HNLs with masses $\mathcal{O} (100)$ MeV mainly decay into final states that include one lepton and one meson or three leptons. Below $1$ GeV, the following two channels with active neutrinos in the final state dominate over the others
\begin{eqnarray}
\label{eq:lightHNLdecaymodes}
    \nu_4 &\to& \nu_\alpha \pi^0 \, , \\
    \nu_4 &\to& \nu_\alpha \nu_\ell \bar{\nu}_\ell \,~~ (\ell = e, \mu, \tau) \, , \nonumber
\end{eqnarray}
where $\alpha$ indicates the neutrino flavor accessible depending on which HNL-active neutrino mixing $U_{\alpha 4}$ is activated.
In what follows, we assume neutrinos to be Majorana, so that the charge-conjugated process will be also taken into account for each HNL decay process that includes neutrinos in the final states. 
Moreover, we will consider three different scenarios for the mixings: a first case where only the mixing with the electron neutrino is activated, $|U_{e4}|^2 \neq 0, |U_{\mu 4}|^2 = 0, |U_{ \tau 4}|^2 = 0$ (dubbed \textit{1:0:0}); a second case where only mixing with muon neutrinos is allowed, $|U_{e4}|^2 = 0, |U_{\mu 4}|^2 \neq 0, |U_{ \tau 4}|^2 = 0$ (case \textit{0:1:0}); and case \textit{0:0:1}, $|U_{e4}|^2 = 0, |U_{\mu 4}|^2 = 0, |U_{ \tau 4}|^2 \neq 0$. While in the case \textit{0:1:0} the muon neutrino contribution is directly produced, in the other two cases $\nu_\mu$ appears due to oscillations of the other flavor components during their propagation from the PBH to the Earth.

Given these considerations, we estimate the time-integrated secondary spectrum of neutrinos with flavor $\alpha$, produced through the decay of HNLs, as~\cite{Oberauer:1993yr,Mastrototaro:2019vug,Syvolap:2023trc,Akita:2023iwq}
\begin{equation}
    \frac{dN_{\alpha}}{dE} = \mathcal{B}_a \int d\cos{\theta} \int_{E_{s, \rm min}}^{E_{s, \rm max}}dE_s \frac{1}{\gamma_s \left( 1 + \beta_s \cos{\theta} \right)} \frac{dN_s}{dE_s} \mathcal{F}_{\alpha} \left[\frac{E}{\gamma_s \left( 1 + \beta_s \cos{\theta} \right)}, \cos{\theta}\right] \, ,
    \label{Eq:dNdE_from_HNL_decay}
\end{equation}
where $E$, $E_s$ are the energies of $\nu_\alpha$ and the HNL in the laboratory frame, respectively,
and $\theta$ is the angle formed between the direction of $\nu_\alpha$ in the HNL rest frame and its velocity in the laboratory frame~\cite{Mastrototaro:2019vug}.
Moreover, in Eq.~\eqref{Eq:dNdE_from_HNL_decay}, $\gamma_s = E_s/m_4$ and $\beta_s = p_s/E_s$, $m_4$ are the HNL mass and $p_s$ its momentum, while $\mathcal{B}_{\alpha}$ = $\Gamma_{\alpha}/\Gamma_\mathrm{tot}$ indicates the branching ratio of the decay process under consideration, with $\Gamma_\mathrm{tot}$ and $\Gamma_{\alpha}$ being respectively the total and partial decay widths. $\frac{dN_s}{dE_s}$, instead, denotes the total primary spectrum of HNLs directly emitted by the PBH through Hawking evaporation.

The integration limits $E_{s, \rm min}$ and $E_{s, \rm max}$  in Eq.~\eqref{Eq:dNdE_from_HNL_decay} refer to the minimal and maximal energy that the HNL can have in the laboratory frame, respectively. 
In principle, these values can vary in the range $[10^{-5} - 10^{5}]$ $ T^{\rm in}_{\mathrm{PBH}}$ ~\cite{Arbey:2019mbc,Arbey:2021mbl}, as for a Schwarzschild black hole the Hawking radiation is mainly peaked around its temperature. However, they will eventually depend on the specific features of the decay channel under consideration. In particular, $E_{s, \rm min}$ is the minimal energy the HNL should reach in order to produce an active neutrino $\nu_\alpha$ with energy $E$, whereas $E_{s, \rm max}$ is the maximal energy value above which $\frac{dN_s}{dE_s}$ becomes negligible.
Finally, $\mathcal{F}_{\alpha}$ in Eq.~\eqref{Eq:dNdE_from_HNL_decay} represents the angular and energetic distribution $\mathcal{F}_{\alpha} \equiv dN_{\alpha}/dE'd\rm cos\theta$ that the active neutrino has in the HNL rest frame, and depends on the specific decay channel under consideration~\cite{Mastrototaro:2019vug,Akita:2023iwq,Syvolap:2023trc}. 

Note that the energies of the active neutrino in the laboratory frame $E$ and in the HNL rest frame $E'$, the sterile neutrino energy $E_s$ and $\theta$ are not independent quantities, but they are all related through the following boost relation
\begin{equation}
    E = \gamma_s E'\left(1 + \beta_s \cos{\theta} \right) .
    \label{eq:boost}
\end{equation}
Therefore, Eq.~\eqref{Eq:dNdE_from_HNL_decay} can be conveniently rewritten in terms of the energy $E'$ that the active neutrino $\nu_{\alpha}$ has in the HNL rest frame by simply performing the following change of variables~\cite{Mastrototaro:2019vug}
\begin{equation}
\label{eq:costhchange}
    d\cos{\theta} = - \frac{1}{\beta_s \gamma_s}\frac{E}{E^{'2}}dE' \, .
\end{equation}
By substituting Eq.~\eqref{eq:costhchange} into Eq.~\eqref{Eq:dNdE_from_HNL_decay} we obtain
\begin{equation}
    \frac{dN_{\alpha}}{dE} = \mathcal{B}_{\alpha} \int_{E'_\mathrm{min}}^{E'_\mathrm{max}} dE' \int_{E_{s, \rm min}}^{E_{s, \rm max}}dE_s \frac{1}{\beta_s \gamma_s} \frac{1}{E'} \frac{dN_s}{dE_s} \mathcal{F}_{\alpha}\left( E', \cos\theta \right)  \, .
\end{equation}
Taking into account that $\gamma_s \beta_s = p_s/m_4$, Eq.~\eqref{Eq:dNdE_from_HNL_decay} can finally be written as
\begin{equation}
     \frac{dN_{\alpha}}{dE} = \mathcal{B}_{\alpha} m_4 \int_{E'_\mathrm{min}}^{E'_\mathrm{max}} dE' \int_{E_{s, \rm min}}^{E_{s, \rm max}}dE_s \frac{1}{p_s} \frac{1}{E'} \frac{dN_s}{dE_s} \mathcal{F}_{\alpha}\left( E',\cos\theta  \right) \, .
     \label{Eq:dNdE_from_HNL_torefer}
\end{equation}
The minimal and maximal energy $E'_\mathrm{min/max}$ that an active neutrino can obtain in the HNL rest frame depend on the kinematic features of the decay under consideration, and more precisely on the shape of $\mathcal{F}_{\alpha}$ and on the energetic limits imposed by the boost in Eq.~\eqref{eq:boost}. Indeed, regardless of which distribution $\mathcal{F}_{\alpha}$ the active neutrino may follow, the energy of the latter in the HNL frame can never exceed 
$E/\gamma_s \left(1 - \beta_s \right)$ nor be smaller than $E/\gamma_s \left(1 + \beta_s \right)$.

As previously mentioned, in the mass range $[0.1-1]$ GeV the two-body decay $\nu_4 \rightarrow \nu_{\alpha} \pi^0$ and the three-body decay $\nu_4 \rightarrow \nu_{\alpha} \nu_{\ell} \bar{\nu}_{\ell} $ in Eq.~\eqref{eq:lightHNLdecaymodes} dominate over all the other decay modes. Given the different number of final particles and, consequently, the different kinematic features, the two channels will be separately discussed in Subsections \ref{subsubsec:2bdecays} and \ref{subsubsection:3body}.

\subsubsection{Two-body decays}
\label{subsubsec:2bdecays}

In the two-body decay process $\nu_4 \rightarrow \nu_{\alpha} \pi^0$ the daughter active neutrino $\nu_{\alpha}$ has in the HNL rest frame a Dirac delta-like distribution in energy and angle~\cite{Akita:2023iwq,Mastrototaro:2019vug,Syvolap:2023trc,Mastrototaro:2021wzl} 

\begin{equation}
    \mathcal{F}_{\alpha}^\mathrm{2b} = \left. \frac{dN_{\alpha}}{dE'd \rm cos\theta}\right|^\mathrm{2b}_{\alpha} = \frac{1}{2}\delta \left( E' - \frac{m_4^2 - m_{\pi^0}^2}{2 m_4} \right) \, .
    \label{Eq:fa_2body}
\end{equation}
Therefore, the active neutrino $\nu_{\alpha}$ produced by the HNL together in a two-body decay with a $\pi_0$ will have the following energy spectrum~\cite{Mastrototaro:2019vug,Mastrototaro:2021wzl,Abdullahi:2022jlv,Syvolap:2023trc}
\begin{equation}
     \left. \frac{dN_{\alpha}}{dE} \right|^{\rm 2b} = \frac{\mathcal{B}_{\alpha} m_4^2}{m_4^2 - m_{\pi^0}^2} \int_{E_{s, \rm min}}^{E_{s, \rm max}} dE_s \frac{1}{p_s}  \frac{dN_s}{dE_s} \, .
     \label{Eq:dNdE_from_HNL_twobody}
\end{equation}
The decay rate for this two-body channel is~\cite{Coloma:2020lgy}
\begin{equation}
\label{eq:2bodydecaywidth}
    \Gamma_{\alpha} \left(\nu_4 \rightarrow \nu_{\alpha} \pi^0 \right) = 2 \frac{G_F^2 m_4^3}{32 \pi}f_\pi^2 |U_{\alpha 4}|^2 \left[1 - \left(\frac{m_{\pi_0}}{m_4}\right)^2 \right]^2 \, ,
\end{equation}
where the factor 2 takes into account that for Majorana neutrinos the charged-conjugated channel $\nu_4 \rightarrow \bar{\nu}_{\alpha} \pi^0$ is open as well \cite{Coloma:2020lgy,Mastrototaro:2019vug,Fischer:2022zwu}, and yields the same value as the main channel: $ \Gamma_{\alpha} \left( \nu_4 \rightarrow \nu_{\alpha} \pi^0 \right) =  \Gamma_{\alpha} \left(\nu_4 \rightarrow \bar{\nu}_{\alpha}  \pi^0 \right)$. In Eq.~\eqref{eq:2bodydecaywidth}, we assume the meson decay constant $f_\pi = 130$ MeV~\cite{ParticleDataGroup:2022pth,Coloma:2020lgy}.
The branching ratio $\mathcal{B}_{\alpha}$ in Eq.~\eqref{Eq:dNdE_from_HNL_twobody} has been computed following \cite{Coloma:2020lgy} considering all the relevant decay processes between 0.1 and 1.0 GeV. 

Finally,  in a two-body decay the minimal energy of the HNL in the laboratory frame is obtained by maximizing the boost in Eq.~\eqref{eq:boost} ($\cos{\theta}$ = 1), giving~\cite{Mastrototaro:2019vug}
\begin{align}
    E_{s,\rm min} = \frac{m_4^2 - m_{\pi_0}^2}{4E} + \frac{E m_4^2}{m_4^2 - m_{\pi_0}^2} \, .
\end{align}
The maximal energy of the HNL, instead, has been set to $E_{s,\rm max} = 10^{5}~T^{\mathrm in}_{\mathrm{PBH}}$.

\subsubsection{Three-body decays}
\label{subsubsection:3body}

In the case of the three body decay $\nu_4 \rightarrow \nu_{\alpha} \nu_{\ell} \bar{\nu}_{\ell} $, the produced active neutrino and antineutrino follow two different angular and energetic distributions in the HNL rest frame~\cite{Mastrototaro:2019vug,Syvolap:2023trc,Akita:2023iwq}
\begin{align}
    \mathcal{F}_{\mathrm{I}, \rm \alpha}^{\mathrm{3b}}\left( E'  \right) =  \left. \frac{dN_{\alpha}}{dE'd \rm cos\theta}\right|^\mathrm{3b}_{\mathrm{I}} = \frac{1}{2}16 \frac{E'^2}{m_4^3}\left(3 - 4 \frac{E^{'}}{m_4}\right) \label{eq:f1} \, ,\\
    \mathcal{F}^{\mathrm{3b}}_{\mathrm{II}, \rm \alpha}\left( E'  \right) = \left. \frac{dN_{\alpha}}{dE'd \rm cos\theta}\right|^\mathrm{3b}_{\mathrm{II}} = \frac{1}{2}96 \frac{E'^2}{m_4^3}\left(1 - 2 \frac{E'}{m_4}\right) \, .
    \label{eq:f2}
\end{align}
Notice that $\mathcal{F}_{\mathrm{I}, \rm \alpha}^{\mathrm{3b}}$ and $\mathcal{F}^{\mathrm{3b}}_{\mathrm{II}, \rm \alpha}$ are double differential distributions in energy $E'$ and angle $\cos{\theta}$ [or equivalently, in $E$ and $\cos{\theta}$, see Eq.~\eqref{eq:boost}]. For this reason, they require to be normalized to 1 when integrated over both energy and angle. 
However, we assume that the decay products are isotropically distributed~\cite{Oberauer:1993yr,Mastrototaro:2019vug}, given that these daughter particles 
inherit the isotropicity of the initial HNL distribution. 
Therefore, when integrated only over the energy $E'$, $\mathcal{F}_{\mathrm{I}, \rm \alpha}^{\mathrm{3b}}$ and $\mathcal{F}^{\mathrm{3b}}_{\mathrm{II}, \rm \alpha}$ in Eqs.~\eqref{eq:f1} and~\eqref{eq:f2} must be normalized to 1/2 in the integration range $[0,m_4/2]$ \cite{Akita:2023iwq}, which is the only kinematically allowed energy range for the active neutrino in the HNL rest frame.  

The active neutrinos $\nu_{\alpha}$ and $\nu_{\ell}$ produced in the decay process $\nu_4 \rightarrow \nu_{\alpha} \nu_{\ell} \bar{\nu}_{\ell}$ follow $\mathcal{F}_{\mathrm{I}, \rm \alpha}^{\mathrm{3b}}$, while $\mathcal{F}^{\mathrm{3b}}_{\mathrm{II}, \rm \alpha}$ describes the spectrum of the antineutrino $\bar{\nu}_{\ell}$. The opposite applies to the charge-conjugated process $\nu_4 \rightarrow \bar{\nu}_{\alpha} \bar{\nu}_{\ell} \nu_{\ell}$, where $\mathcal{F}_{\mathrm{I}, \rm \alpha}^{\mathrm{3b}}$ traces the spectrum of the antineutrinos $\bar{\nu}_{\alpha}, \bar{\nu}_{\ell}$ whereas the neutrino $\nu_{\ell}$ follows the distribution $\mathcal{F}^{\mathrm{3b}}_{\mathrm{II}, \rm \alpha}$~ \cite{Mastrototaro:2019vug,Akita:2023iwq}.

As previously considered for the two-body decay of Subsec.~\ref{subsubsec:2bdecays}, we will focus on scenarios in which only one neutrino species $\alpha$ mixes with the HNL. Under this assumption and given the above considerations, the total angular and energetic distribution for the two neutrinos with flavor $\ell$ --- that do not mix with the HNL --- is given by~\cite{Mastrototaro:2019vug}
\begin{equation}
   \mathcal{F}_{\ell}^\mathrm{3b}\left(E'\right) = \frac{1}{8}\left( \mathcal{F}^\mathrm{3b}_{\mathrm{I}} +  \mathcal{F}^\mathrm{3b}_{\mathrm{II}} \right) \, .
   \label{Eq:3body_fa_not_mixing}
\end{equation}
The active neutrino with flavor $\alpha$ that mixes with the HNL will instead have the following distribution function, in the HNL rest frame
\begin{equation}
    \mathcal{F}_{\alpha}^\mathrm{3b}\left(E'\right) = \frac{1}{4}\left(3 \mathcal{F}^\mathrm{3b}_{\mathrm{I}} +  \mathcal{F}^\mathrm{3b}_{\mathrm{II}} \right) \, .
    \label{Eq:3body_fa_mixing}
\end{equation}
Thus, the active neutrino time-integrated spectrum for the three-body decay is given by
\begin{equation}
    \left. \frac{dN_{\alpha (\ell)}}{dE} \right|^{\rm 3b}\ = \mathcal{B}_{\alpha}m_4 \int_{E_s,\rm min}^{E_s,\rm max} dE_s \frac{1}{p_s} \frac{dN_s}{dE_s} \int_{E'_{\rm min}}^{E'_{\rm max}}dE' \frac{1}{E'} \mathcal{F}_{\alpha (\ell)}^\mathrm{3b} \left(E'\right) \, ,
    \label{Eq.:dNdE_3b_torefe}
\end{equation}
as it can be derived from Eq.~\eqref{Eq:dNdE_from_HNL_torefer}, for an active neutrino produced in the three-body decay that mixes with the HNL ($\nu_\alpha$) or
not ($\nu_\ell$).

The minimal and maximal HNL energies in the laboratory frame are set to the standard values of $E_{s, \rm min}$ = $10^{-5}$ $T^{\rm in}_{\mathrm{PBH}}$ and $E_{s, \rm max}$ =  $10^{5}$ $T^{\rm in}_{\mathrm{PBH}}$, differently from the two-body case.

The minimal energy of the active neutrino $E'_{\rm min}$ in the HNL rest frame is obtained by inserting the maximal cosine of the angle $\theta$, i.e. $\cos{\theta}$ = 1, into Eq.~\eqref{eq:boost}
\begin{equation}
    E'_{\rm min} = \frac{E}{\gamma_s \left(1 + \beta_s\right)} \, .
\end{equation}
On the other hand, the maximal energy of the active neutrino $E'_{\rm max}$ is set to $m_4/2$ as it constitutes the upper limit of the kinematically allowed energy range for $\nu_{\alpha}$ in the HNL rest frame~\cite{Akita:2023iwq}.
Similarly to the two-body decay, the branching ratio of the process $\mathcal{B}_{\alpha}$ has been computed following Ref.~\cite{Coloma:2020lgy}.

The partial decay width for the three-body channel reads
\begin{equation}
    \Gamma_{\alpha} \left(\nu_4 \rightarrow \nu_{\alpha} \sum_{\ell} \nu_{\ell} \bar{\nu}_{\ell}\right)  = \sum_{\ell} \left[\Gamma\left(\nu_4 \rightarrow \nu_{\alpha}  \nu_{\ell} \bar{\nu}_{\ell} \right) + \left(\nu_4 \rightarrow \bar{\nu}_{\alpha}  \nu_{\ell} \bar{\nu}_{\ell} \right)\right] =  2\frac{G_F^2 m_4^5}{64 \pi^3} |U_{\alpha 4}|^2 \, .
    \label{Eq:3nuGamma}
\end{equation}

\subsection{Heavy-mass regime: [0.5-2] TeV}
\label{sub:heavy_mass_regime}

Above the electroweak scale, HNLs mainly decay into bosonic states~\cite{Atre:2009rg}. In this scenario, we assume for simplicity that only the muonic neutrino mixes with the HNL, i.e., we consider only the case 0:1:0. Under this assumption,
the three dominant channels that contribute to the muonic neutrino production in the [0.5-2] TeV mass range are 
\begin{eqnarray}
\label{eq:heavyHNLdecaymodes}
    \nu_4 &\to& W^\pm \mu^\mp \, , \\
    \nu_4 &\to& Z^0 \nu_\mu \, , \nonumber\\
    \nu_4 &\to& H^0 \nu_\mu \, . \nonumber
\end{eqnarray}

The produced bosons and muons will subsequently hadronize and/or decay producing additional fluxes of visible $\nu_\mu$, similarly to the secondary contributions already present in the SM and previously described in Sec.~\ref{subsec:secondarySM}. In the following, we will dub the intermediate states of the previous processes, $W^\pm$, $\mu^\pm$, $Z^0$ and $H^0$, as ``\textit{i.s.}". 

Similarly to the light-mass regime [see Eq.~\eqref{Eq:dNdE_from_HNL_torefer}], the secondary spectrum of $\nu_\mu$ from heavy-mass HNLs can be obtained by a convolution of the HNL primary spectrum, $\frac{dN_s}{dE_s}$, with the spectrum of $\nu_\mu$ produced at a fixed parent energy: $\frac{dN}{dE'}(\nu_4 \to i.s.\to \nu_\mu)$. The latter corresponds to the energetic and angular distribution that the muon neutrino can have in the HNL rest frame, and will be evaluated using numerical tools.

To compute such energetic and angular distributions, we use  \texttt{PPPC4DMID}~\cite{Cirelli:2010xx}, a code that relies on high-statistics simulations and the \texttt{PYTHIA}~\cite{Sjostrand:2014zea} event generator. As mentioned before, given the observation time frame under consideration, the production of HNLs with masses above 1 TeV is heavily suppressed. 
Although \texttt{PPPC4DMID} aims at computing the secondary emission from DM annihilation/decay, it can be used for PBH evaporation following few steps.  

We compute the muonic neutrino spectra from each bosonic or leptonic intermediate state separately, i.e. $\frac{dN}{dE'} (\nu_4 \to W/Z/H\to \nu_\mu)$ and $\frac{dN}{dE'}(\nu_4 \to \mu/\nu_\mu \to \nu_\mu)$ from DM decay with a centre-of mass energy equal to the energy of the intermediate states in the HNL rest frame. Afterwards, we combine the resulting neutrino spectra for each HNL decay channel to obtain the overall neutrino spectrum.

For the sake of comparison and to test our method, we computed the total muon neutrino spectrum $\frac{dN}{dE'} (\nu_4 \to i.s.\to \nu_\mu)$ for $m_4 = 1$ TeV using another tool, \texttt{HDMSpectra}~\cite{Bauer:2020jay} and checked that the two methods provide same results\footnote{Contrarily to \texttt{PPPC4DMID}, \texttt{HDMSpectra}'s range of validity is above the scale of electroweak symmetry breaking, and in particular $m_\mathrm{DM} \gtrsim$ TeV. Differences in the spectra among the two codes due to the implementation of electroweak corrections turn out not to affect the overall spectra of our interest.}.

The resulting time-integrated spectrum in the laboratory frame of active muonic neutrinos from the decay of HNLs with masses above the electroweak scale is
\begin{equation}
    \frac{dN_{\nu_\mu}}{dE} = \sum_{i.s.} \mathcal{B}(\nu_4 \rightarrow i.s. )\; m_4 \int_{E_s,\rm min}^{E_s,\rm max} dE_s \frac{1}{p_s} \frac{dN_s}{dE_s} \int_{E'_{\rm min}}^{E'_{\rm max}}dE' \frac{1}{E'} \frac{dN}{dE'}\left(\nu_4 \rightarrow i.s. \rightarrow \nu_{\mu} \right) \, ,
    \label{Eq:activedNdE_fromHeavy}
\end{equation}
where $\mathcal{B}(\nu_4 \rightarrow i.s. )$ is the branching ratio of the process $\nu_4 \to i.s.$ 
computed taking into account all the heavy HNL decay modes of Eq.~\eqref{eq:heavyHNLdecaymodes} in the total decay width.

The partial decay widths for each of these processes have been taken from Ref.~\cite{Atre:2009rg} and read 
\begin{align}
    \Gamma \left( \nu_4 \to \mu W_L\right)  &=  2\frac{
    g^{2}}{64\pi M^{2}_W}|U_{\mu 4}|^{2}m_4^{3} \left[ 1 - \left(\frac{M_W}{m_4} \right)^2  \right]^{2} \, \nonumber,\\
    \Gamma \left( \nu_4 \to \mu W_T\right)  &=  2\frac{
    g^{2}}{32\pi}|U_{\mu 4}|^{2}m_4 \left[ 1 - \left(\frac{M_W}{m_4}\right)^2 \right]^{2} \, \nonumber,\\
    \Gamma \left( \nu_4 \to \nu_\mu Z^0_L\right) &= \frac{
    g^{2}}{64\pi M^{2}_Z}|U_{\mu 4}|^{2}m_4^{3} \left[ 1 - \left(\frac{M_Z}{m_4} \right)^2 \right]^{2} \, \nonumber,\\
    \Gamma \left( \nu_4 \to \nu_\mu Z^0_T\right)  &= \frac{
    g^{2}}{32\pi\cos^{2}\theta_W}|U_{\mu 4}|^{2}m_4 \left[ 1 - \left(\frac{M_Z}{m_4}\right)^2 \right]^{2}\, \nonumber,\\
    \Gamma \left( \nu_4 \to \nu_\mu H^0\right)  &= \frac{
    g^{2}}{64\pi M^{2}_H}|U_{\mu 4}|^{2}m_4^{3} \left[1 - \left(\frac{M_H}{m_4}\right)^2 \right]^{2} \, , 
    \label{eq:heavydecaywidths}
\end{align}
where $M_{W,Z}$ are respectively the mass of W and Z bosons, $\cos \theta_W$ is the cosine of the Weinberg angle, and $g$ is the weak coupling. 
The Z and W unpolarized decay widths are obtained by summing the decay widths into the transverse $X_T$ and longitudinal $X_L$ components, $X=\{W,Z\}$, of Eq.~\eqref{eq:heavydecaywidths}~\cite{Cirelli:2010xx}. In the decay widths into a W boson, the factor 2 accounts for the two possible charged channels $\nu_4 \to \mu^\pm W^\pm$.

\subsection{Summary of neutrino spectra from HNL decays}
\label{subsec:summary_neutrino_spectra}
The overall muon neutrino time-integrated spectrum expected from an evaporating PBH will be given by the sum of all SM contributions $\left. \frac{dN_{\nu_\mu}}{dE} \right|_\mathrm{SM}$ and the secondary contributions from HNL decay modes $\left. \frac{dN_{\nu_\mu}}{dE} \right|_\mathrm{HNL}$ 
described above
\begin{equation}
    \left. \frac{dN_{\nu_\mu}}{dE} \right|_\mathrm{tot} = \left. \frac{dN_{\nu_\mu}}{dE} \right|_\mathrm{SM} + \left. \frac{dN_{\nu_\mu}}{dE} \right|_\mathrm{HNL} \, .
    \label{eq:dNdE_contributions}
\end{equation}
More specifically, in the case of HNLs with masses in the range $[0.1-1]$ GeV the muon-neutrino time-integrated spectrum from HNL decays will receive contributions from both the two- and three-body decays
\begin{equation}
    \left. \frac{dN_{\nu_\mu}}{dE} \right|_\mathrm{HNL} = \left. \frac{dN_{\nu_\mu}}{dE} \right|^\mathrm{2b} + \left. \frac{dN_{\nu_\mu}}{dE} \right|^\mathrm{3b} \, ,
\end{equation}
whereas the spectrum of muon neutrinos from heavy HNLs with masses in the $[0.5 - 2]$ TeV range is directly given in Eq.~\eqref{Eq:activedNdE_fromHeavy}.

Note that under the assumption that only one active neutrino species $\nu_\alpha$ mixes with the HNL, the branching ratios of Eq.~\eqref{Eq:dNdE_from_HNL_twobody}, Eq.~\eqref{Eq.:dNdE_3b_torefe} and Eq.~\eqref{Eq:activedNdE_fromHeavy} do not depend on $|U_{\alpha 4}|^2$, as the latter simplifies in the ratio of the partial over the total decay widths of the decay under consideration. As a consequence, the overall neutrino spectrum from both HNL and SM contributions do not depend on the neutrino mixing $|U_{\alpha 4}|^2$ in both light and heavy HNL mass regimes.
This is related to the fact that HNL production is purely gravitational in our case, so it does not depend on the mixing angle, unlike usual experiments.

\begin{figure}[!htb]
    \centering
    \begin{subfigure}{0.49\textwidth}
        \includegraphics[width=\textwidth]{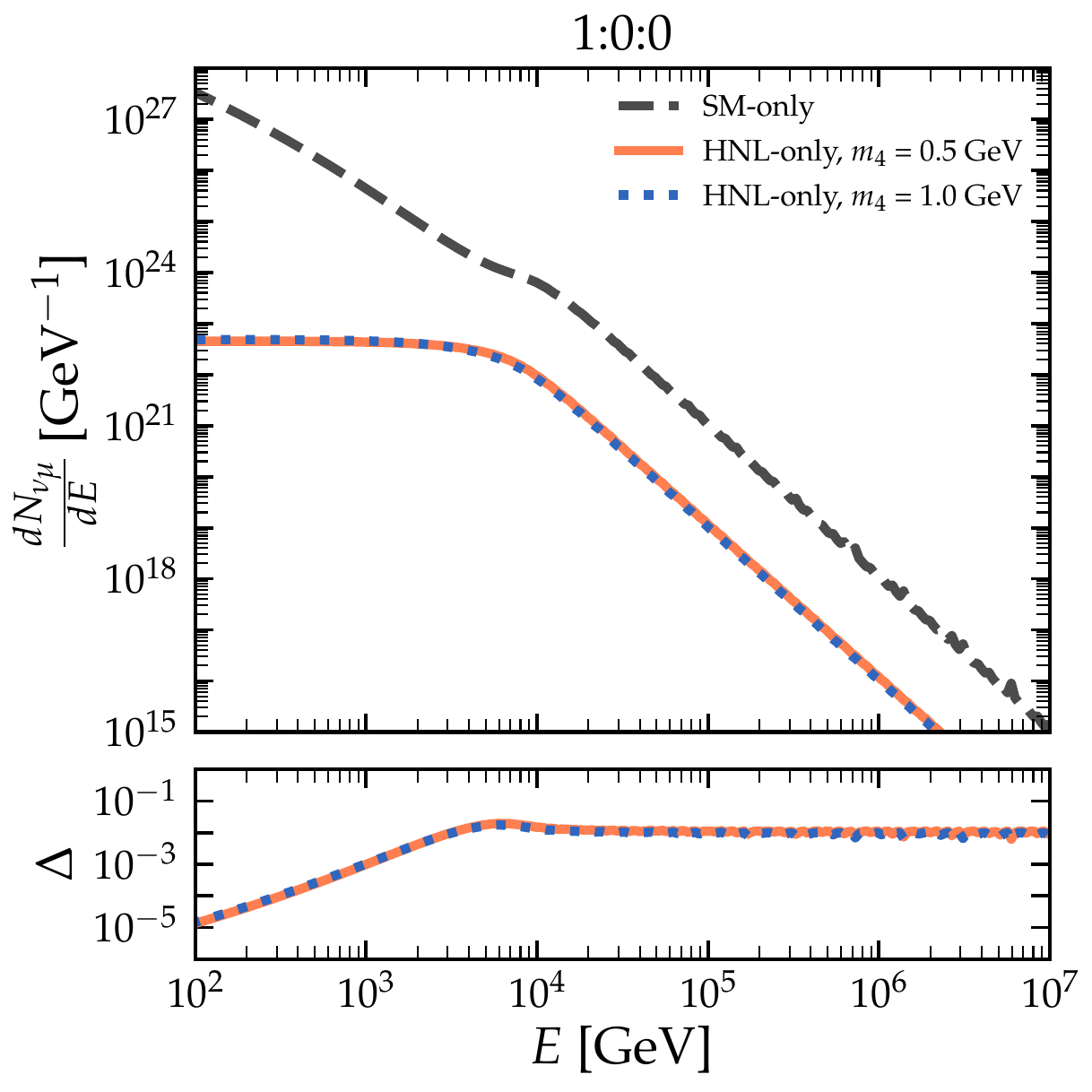}
    \end{subfigure}
    \hfill
     \begin{subfigure}{0.49\textwidth}
        \includegraphics[width=\textwidth]{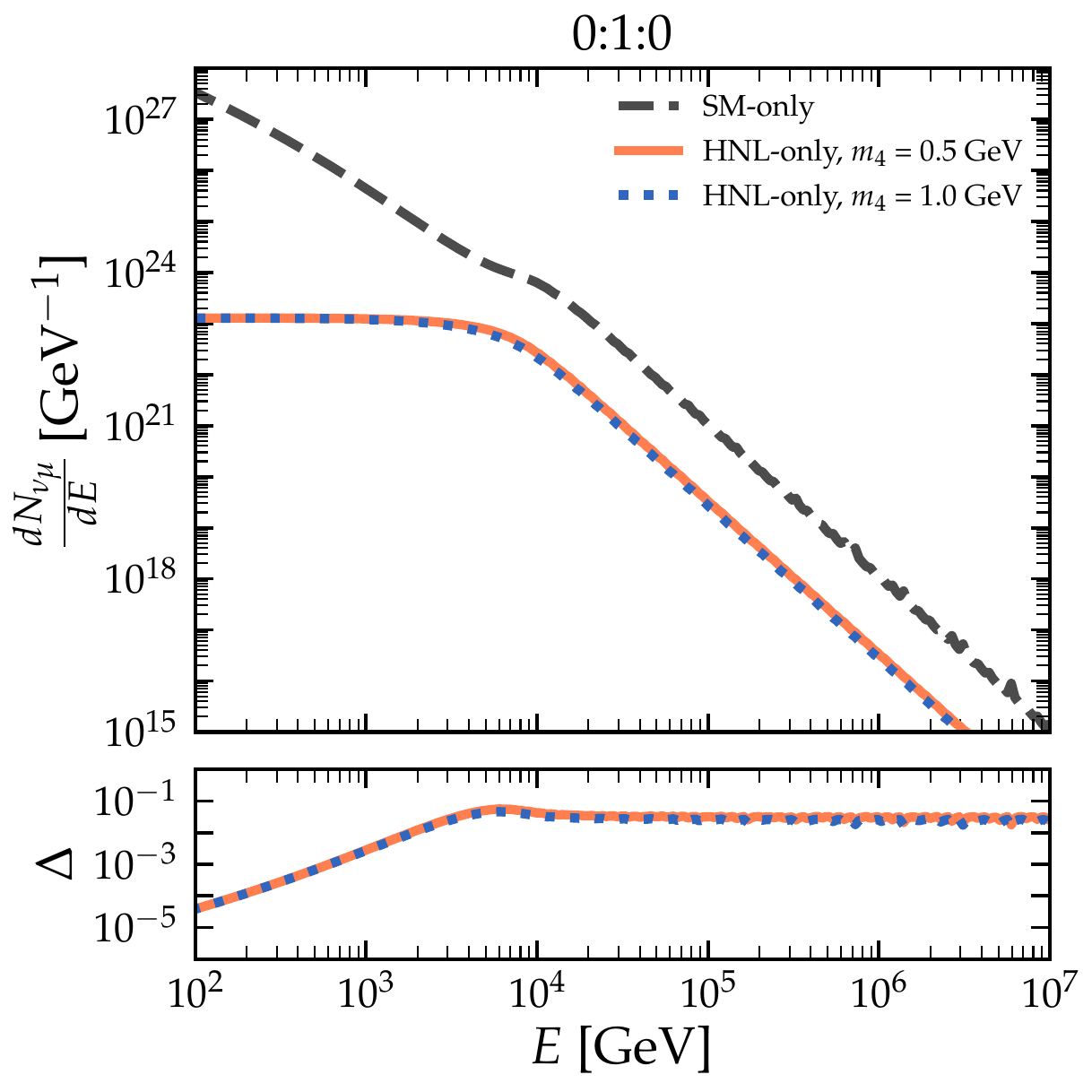}
    \end{subfigure}
     \begin{subfigure}{0.49\textwidth}
        \includegraphics[width=\textwidth]{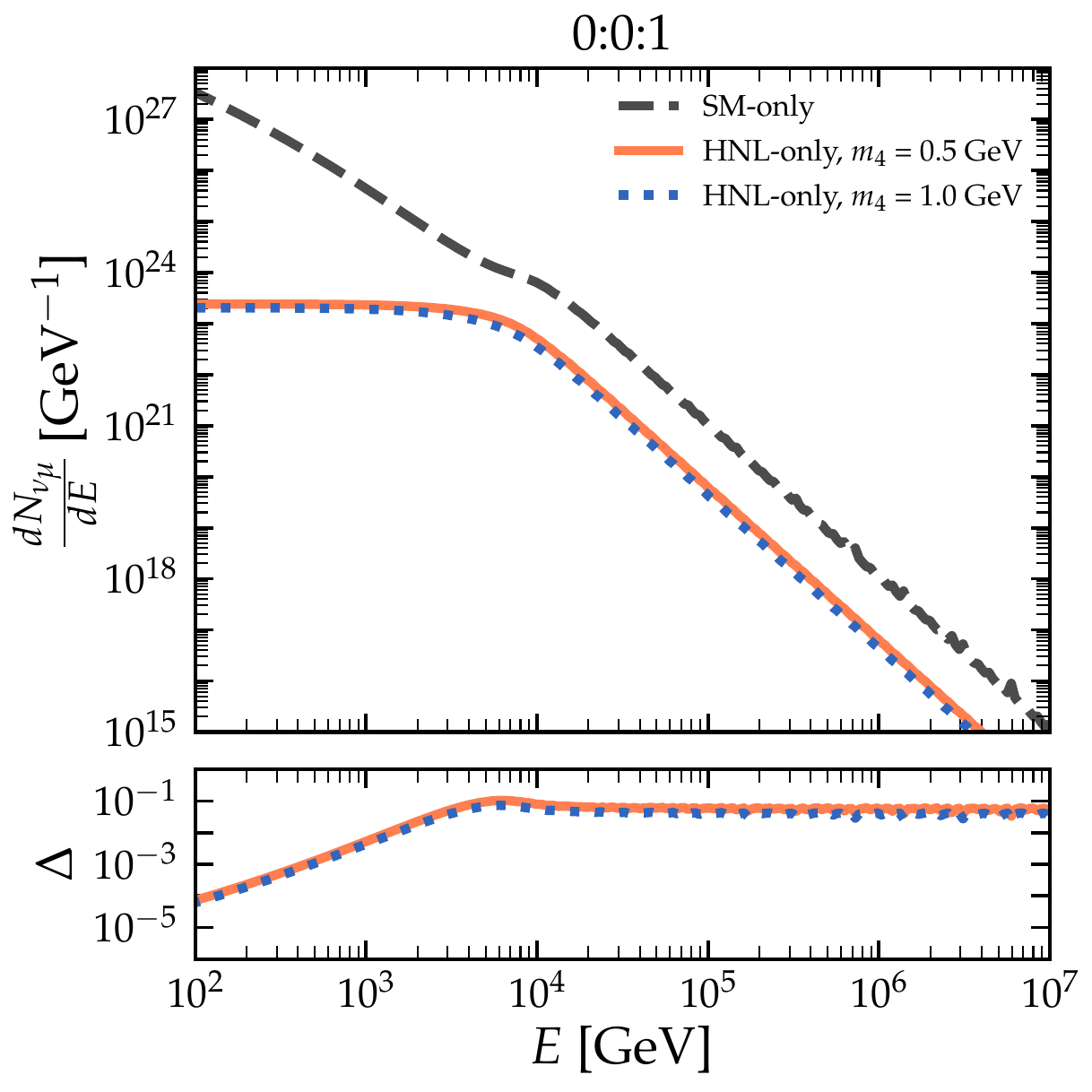}
        \end{subfigure}
    \hfill
    \begin{subfigure}{0.49\textwidth}
        \includegraphics[width=\textwidth]{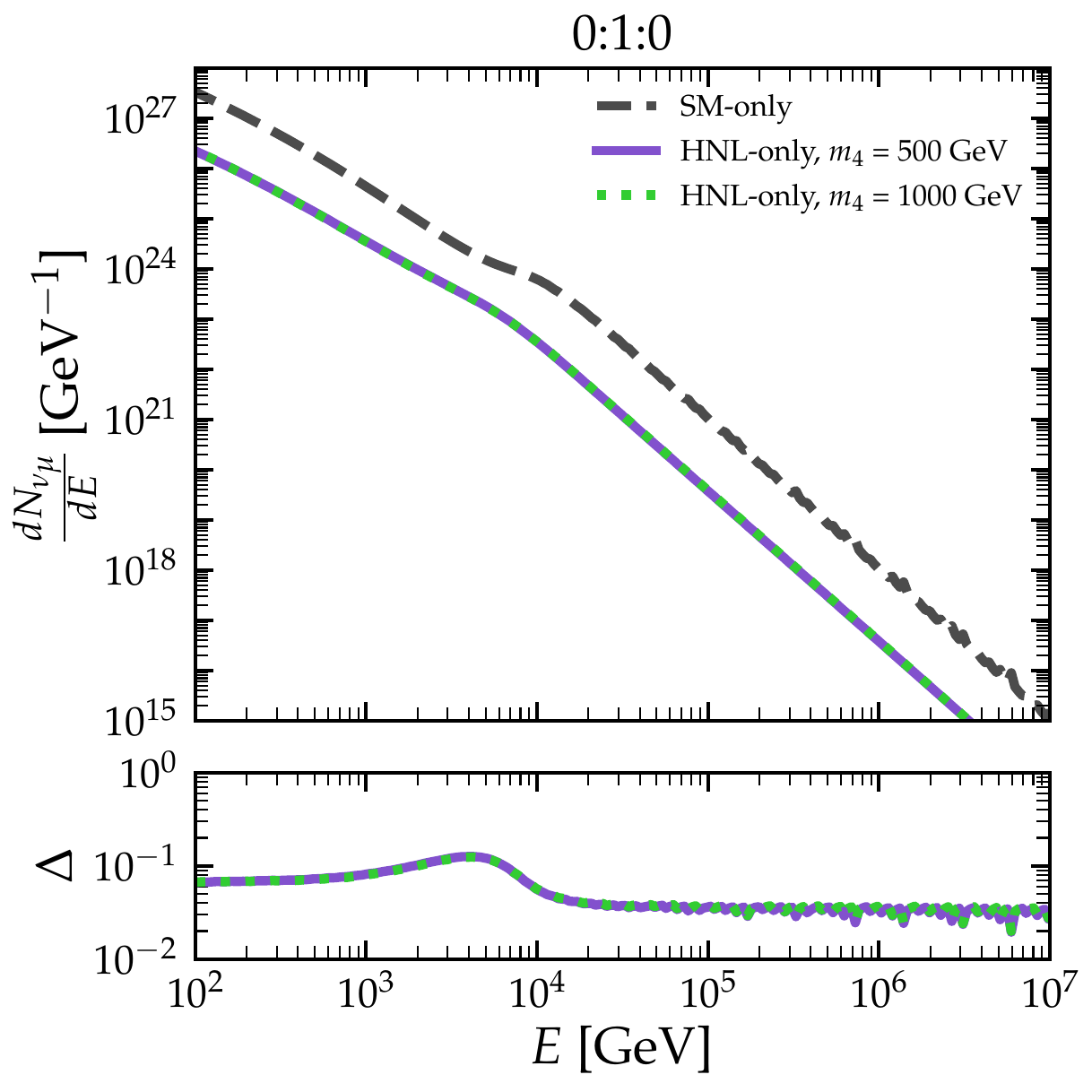}
    \end{subfigure}
    \caption{Total time-integrated spectrum of muon neutrinos expected from the evaporation of a PBH, for an observation time of $\tau = 100$ s. In each panel, we show the SM-only contribution (black, dashed) and the HNL-only contribution to the muon neutrino spectrum for different HNL benchmark masses (solid/dotted, color) as indicated in the caption. The smaller panels below each figure depict the ratio between the HNL-only and SM-only contributions, defined as $\Delta$ (see main text for more details). In the small panel on the left (right), the cyan (dark green) dotted curve corresponds to $m_4 = 1~(1000)$ GeV, while the dark red (pink) solid one refers to  $m_4 = 0.2~(500)$ GeV. }
    \label{fig:dNdE}
\end{figure}

We show in Fig.~\ref{fig:dNdE} the total muon neutrino time-integrated spectrum for both SM-only and HNL-only contributions as function of the muon neutrino energy.  The upper left (1:0:0), upper right (0:1:0) and lower left (0:0:1) panels refer to the expected time-integrated spectrum in the presence of low-mass HNLs, for three different mixing scenarios. Namely, we fix two HNL benchmark masses $m_4 = 0.2$ (coral solid) and $m_4 = 1$ (blue dotted) GeV. The black, dashed curve indicates the SM-only contribution, while the colored dotted and solid lines represent the HNL-only contributions, i.e. the muonic neutrinos generated from the HNL decay through the different modes we already discussed. Let us recall also that we are considering an observation time of $\tau = 100$ s.
In all panels, it is possible to notice a bump in the SM-only contribution around energies of $E\sim {10^4~{\rm GeV}}$. Such increase in the spectrum is caused by the peak of primary neutrinos, that is around the temperature of the PBH ($\sim$ TeV).
In addition, on the bottom of each panel in Fig.~\ref{fig:dNdE}, we present the ratio between the HNL-only neutrino spectrum and the SM-only neutrino spectrum. This ratio is defined as
\begin{equation}
\label{Eq:Delta}
    \Delta =\frac{\left.\frac{dN_{\nu_\mu}}{dE} \right|_\mathrm{HNL} }{\left.\frac{dN_{\nu_\mu}}{dE}\right|_\mathrm{SM}}\, .
\end{equation}
We can see that, regardless of the mixing mode, the light HNLs induce a time-integrated spectrum that resembles the SM-only one, however being subdominant in magnitude. The difference among the two spectra is even more pronounced below the primary bump, where the SM secondary contribution dominates. 
For $m_4 =  0.2$ GeV, the HNLs mainly contribute to the muon-neutrino time-integrated spectrum through the three-body decay given in Eq.~\eqref{Eq.:dNdE_3b_torefe}.

Higher HNLs masses instead give rise to additional two-body decay contributions that more sizeably affect the expected time-integrated spectrum. However, such increase is non-linear, as at higher HNLs masses other decay channels without neutrinos as final states open and the branching ratios of the two and three-body decay are dumped. Eventually, the expected spectrum for $m_4 = 1$ GeV shows mild differences with respect to the $m_4 = 0.2$ GeV case.

Finally, the lower-right panel of Fig.~\ref{fig:dNdE} shows the muon-neutrino time-integrated spectrum with contributions from HNLs in the mass range $[0.5 -2]$ TeV. In particular, we fix two HNL benchmark masses $m_4 = 500$ (purple solid) and $m_4 = 2000$ (green dotted) GeV and we consider  
the mixing scenario 0:1:0. Similarly to the light-mass regime, the HNL decays induce a secondary muon neutrino spectrum that is similar in shape to the SM one, especially at higher energy, although a factor of a few smaller.
With respect to the light mass regime, it can be noticed that even below the primary bump the HNL-only contribution behaves similarly to the SM-only spectrum. The reason behind this difference lies in the different angular distributions $\mathcal{F}_{\alpha}$ adopted in the light and in the heavy mass regimes. In the light regime, the angular distributions only depend on the energy of the neutrino in the HNL rest frame. In the heavy regime, the angular distribution is obtained by considering all the secondary and subsequent processes that give rise to muon neutrino from a initial bosonic state. This procedure relies on a numerical approach as discussed in section~\ref{sub:heavy_mass_regime} and it is similar to how the secondary component of the SM-only spectrum is evaluated.

\subsection{Gamma-ray spectra from HNL decays}
\label{subsec:photon_spectra}

Neutrinos are not the only expected signature of exploding PBHs, as in the final instants of the evaporation stage abundant fluxes of high-energy photons will also be emitted. Telescopes dedicated to the search of gamma rays are particularly suited to observe a PBH explosion, such as, for instance, the High Altitude Water Cherenkov (HAWC) observatory~\cite{HAWC:2013kzm,HAWC:2019wla}. HAWC is a ground-based air shower array located on the Sierra Negra volcano in
Mexico, with a large field-of-view of about 2 sr. It is sensitive to gamma rays from the Northern hemisphere sky with energies $\sim 100 - 10^5$ GeV, corresponding to the high end of the PBH burst spectrum. It also features an excellent angular resolution, ranging from $0.2\degree$ to $1\degree$, thus making it an ideal tool to search for the high-energy emission from gamma-ray transients such as exploding PBHs.\\

We estimate the expected photon spectra similarly to the neutrino case. We evaluate the SM primary and secondary contributions using {\tt BlackHawk v2.3}~\cite{Arbey:2019mbc,Arbey:2021mbl}. For the contributions from HNL decays, we proceed as follows.
In the HNL light-mass regime, the main channel contributing to the gamma-ray spectrum is $\nu_4 \to \pi^0 \nu_\alpha \to \gamma \gamma \nu_\alpha$.
We evaluate the photon spectrum from this decay channel as~\cite{Ukwatta:2015iba}
\begin{equation}
   \left. \frac{dN_\gamma}{dE}  \right|_\mathrm{HNL} = 2 \int_{E_{\rm min}}^{\infty} dE_{\pi^0} \frac{dN_{\pi^0}}{dE_{\pi^0}}\frac{1}{p_{\pi^0}} \, ,
\end{equation}
where $E_{\rm min} = E_{\gamma} + m_{{\pi^0}}^2/(4E_{\gamma} )$ \cite{Ukwatta:2015iba} and the factor 2 accounts for both photons reaching the detector due to the high boost. The pion spectrum is evaluated as 
\begin{equation}
    \frac{dN_{\pi^0}}{dE} =  \frac{1}{2}\mathcal{B}\left(\nu_4 \rightarrow {\pi^0} \nu_\alpha  \right)    \int dE_s \left (\frac{E \overline{E} - \gamma_s m_{\pi^0}^2}{\gamma_s \beta_s \overline{p}^3} \right)  \frac{\overline{p}}{\gamma_s(\overline{p}-\overline{E}) + E} \frac{dN_s}{dE_s} \,, 
\end{equation}
with
\begin{equation}
    \overline{E} = \frac{m_4^2 + m_{\pi^0}^2}{2 m_4}\, .
\end{equation}

In the HNL high-mass regime, we compute the photon energetic and angular distributions using \texttt{PPPC4DMID}~\cite{Cirelli:2010xx} for all HNL decay channels listed in Eq.~\eqref{eq:heavyHNLdecaymodes}, following the same procedure adopted for neutrinos. In this case, high-energy photons will be produced in further steps of the decay chain. In both regimes, the overall gamma-ray time-integrated spectrum expected from an evaporating PBH is finally given by the sum of all SM and HNL contributions, in full similarity to the neutrino signal:
\begin{equation}
    \left. \frac{dN_{\gamma}}{dE} \right|_\mathrm{tot} = \left. \frac{dN_{\gamma}}{dE} \right|_\mathrm{SM} + \left. \frac{dN_{\gamma}}{dE} \right|_\mathrm{HNL} \, .
    \label{eq:dNdE_contributions_photon}
\end{equation}

\begin{figure}[!htb]
    \centering
    \begin{subfigure}{0.49\textwidth}
        \includegraphics[width=\textwidth]{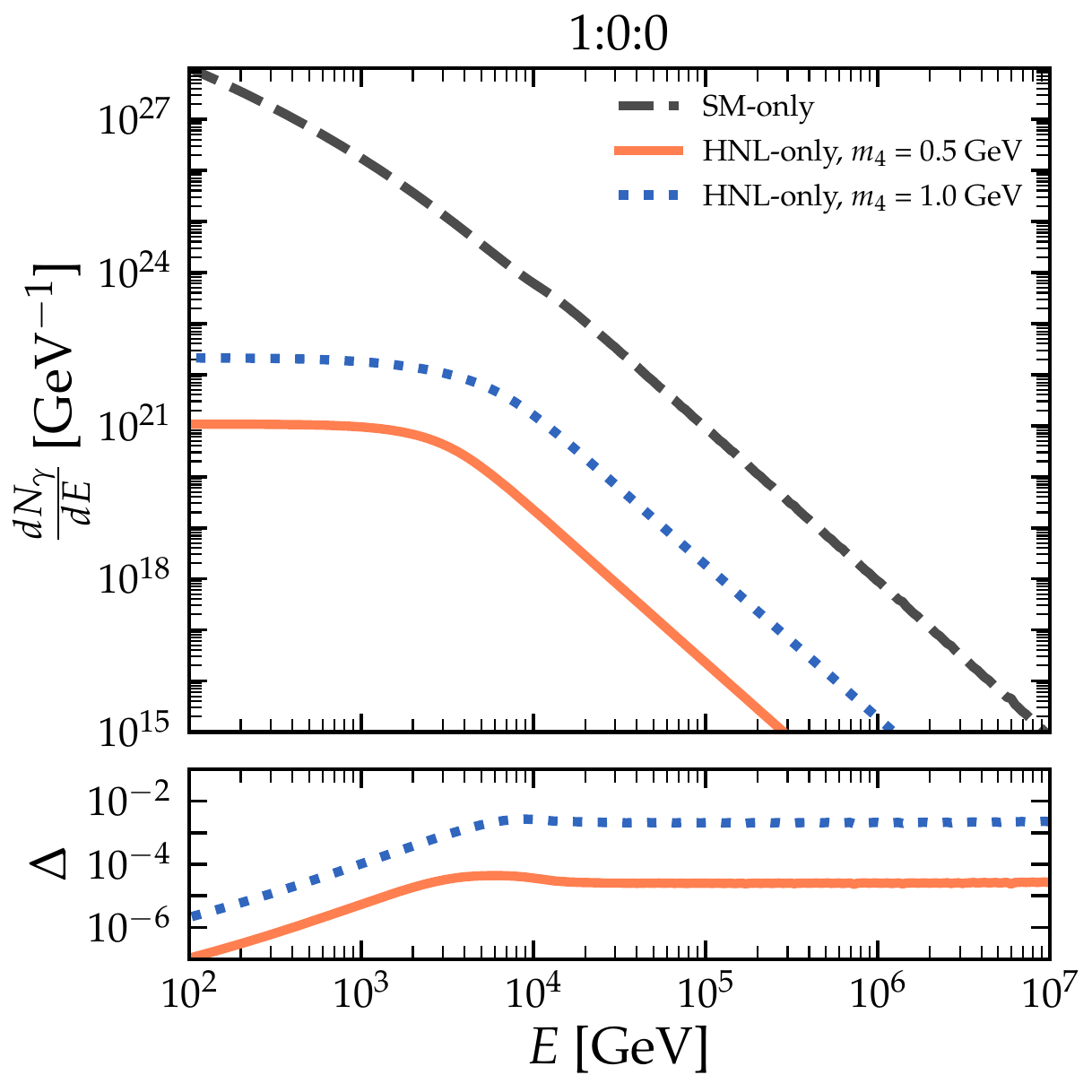}
    \end{subfigure}
    \hfill
     \begin{subfigure}{0.49\textwidth}
        \includegraphics[width=\textwidth]{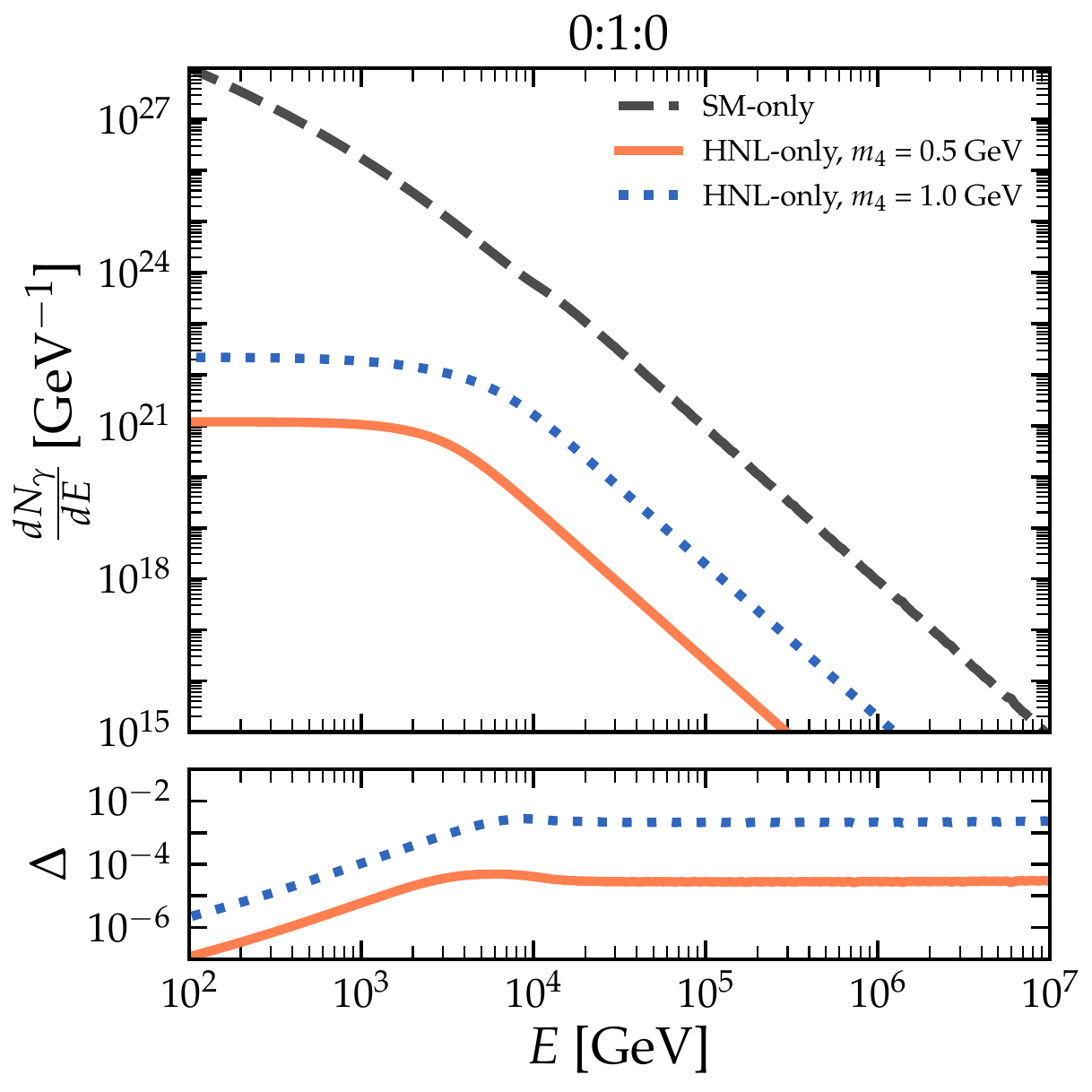}
    \end{subfigure}
     \begin{subfigure}{0.49\textwidth}
        \includegraphics[width=\textwidth]{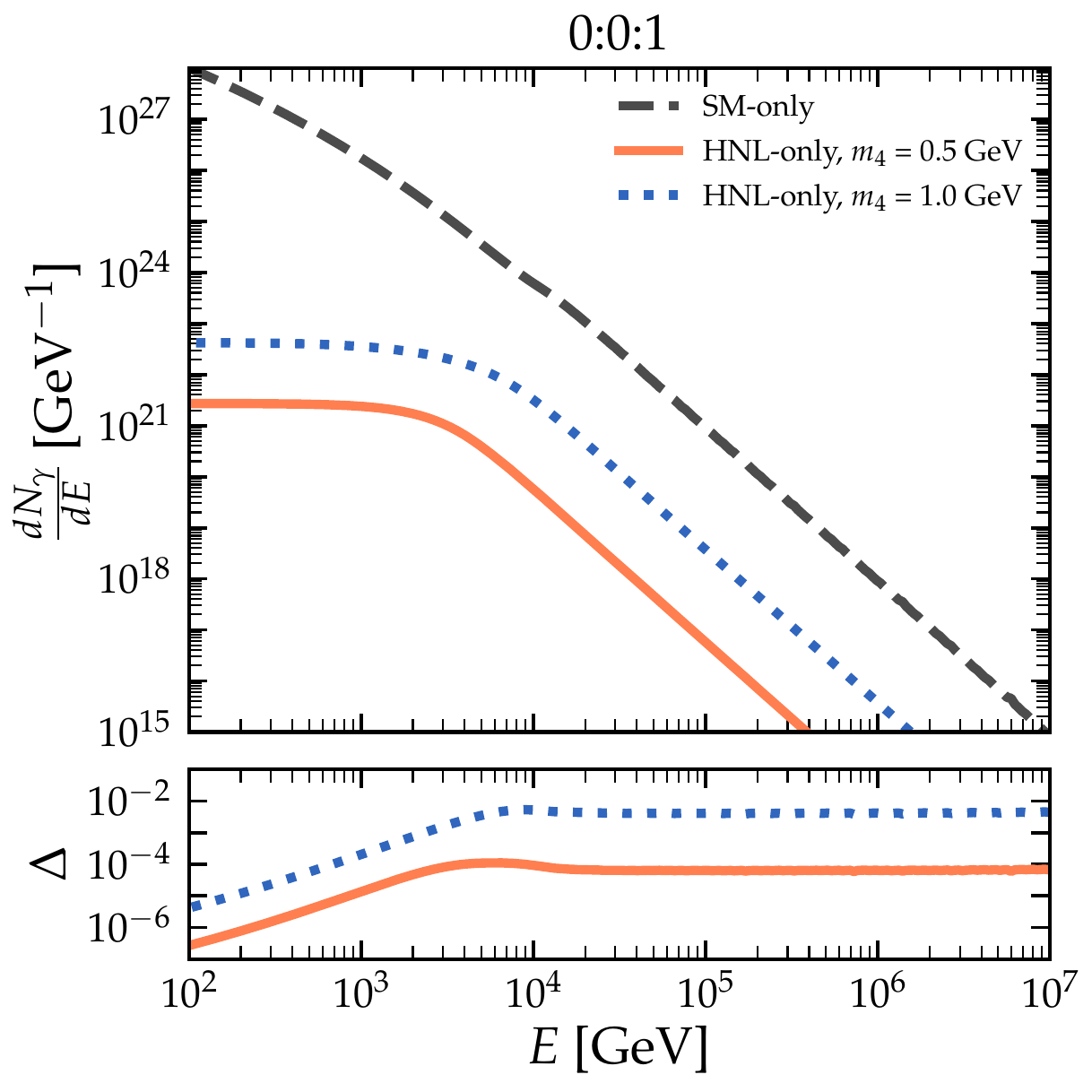}
        \end{subfigure}
    \hfill
    \begin{subfigure}{0.49\textwidth}
        \includegraphics[width=\textwidth]{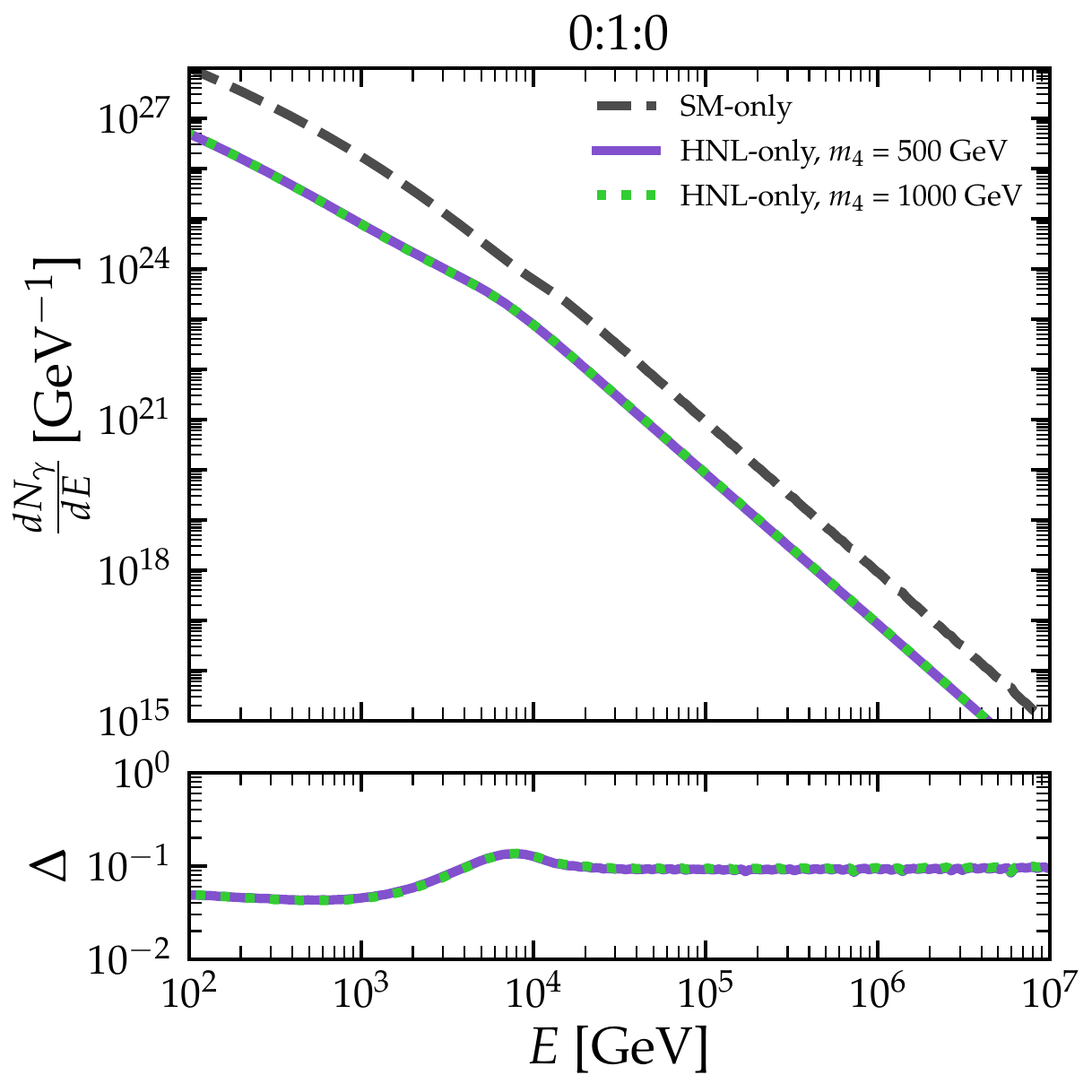}
    \end{subfigure}
    \caption{ Same as Fig.~\ref{fig:dNdE}, but for the time-integrated spectrum of photons.}    \label{fig:dNdE_photon}
\end{figure}

We show in Fig.~\ref{fig:dNdE_photon} the total time-integrated photon spectra for both SM-only and HNL-only contributions, as function of the photon energy.  Similar to Fig.~\ref{fig:dNdE}, we consider different mixing scenarios and both HNL mass regimes. We also display at the bottom of each panel the ratio $\Delta$ as defined in Eq.~\eqref{Eq:Delta}, this time for the photon emission. In the light-mass regime, we see that secondary gamma-ray time-integrated spectrum highly depends on the value of the HNL mass, in contrast with the neutrino case; indeed, we find more than one order of magnitude difference between the $m_4 = 0.5$ GeV curves (orange plain) and $m_4 = 1$ GeV curves (blue dotted), in the first three panels. The pion, indeed, starts to be produced when $m_4 \gtrsim 200$ MeV, but it will be boosted to high energies only above $m_4 \gtrsim 0.6$ GeV, leading to a reduced photon spectrum below this HNL mass.
Regarding the high-mass regime instead, the photon time-integrated spectrum from HNL decays is similar in shape to the SM contribution, for the same reasons as in the neutrino case, and contributes substantially to the overall number of gamma-ray events.

\section{Statistical Analysis}
\label{sec:analysis}
Neutrino telescopes like IceCube~\cite{IceCube:2018pgc,IceCube:2019cia} 
may be able to observe exploding PBHs as transient point sources. We are interested in evaluating the signatures at IceCube of muon neutrinos produced from the last stages of a PBH evaporation, more specifically during an observation time of 100 s. Muon-tracks events at IceCube benefit from a high-quality angular resolution~\cite{Capanema:2021hnm} that allows to pinpoint with precision the declination angle, i.e. the angle with respect to the ground, of the incoming astrophysical muon neutrinos.
We evaluate the fluence of $\nu_\mu$ from the last instants of an evaporating PBH located at a distance $d_\mathrm{PBH}$ from Earth as
\begin{equation}
F_{\nu_\mu} (E) =\frac{1}{4 \pi d_\mathrm{PBH}^2} \left. \frac{dN_{\nu_\mu}}{dE} \right|_\mathrm{tot}  \, ,
\label{eq:def_fluence}
\end{equation}
where $\left. \frac{dN_{\nu_\mu}}{dE} \right|_\mathrm{tot}$ takes into account all HNL and SM contributions as defined in Eq.~\eqref{eq:dNdE_contributions}.

The total number of muon-neutrino events expected at IceCube, for a given declination angle $\delta$, is obtained as
\begin{equation}
\label{eq:Nnumu}
    N_{\nu_\mu}(\delta) = \int_{E_\mathrm{min}}^{E_\mathrm{max}} dE\;F_{\nu_\mu} (E) \mathcal{A}_\mathrm{eff} (E, \delta) \, ,
\end{equation}\\
where the integration limits are set to ${E_\mathrm{min}} = 100$ GeV, that is the energy threshold of IceCube, and ${E_\mathrm{max}} = 10^8$ GeV, corresponding to the energy above which the total muon-neutrino fluence becomes negligible.
IceCube’s effective area, $\mathcal{A}_\mathrm{eff} (E, \delta)$, indicates the efficiency to observe
an astrophysical neutrino flux as a function of energy and declination\footnote{The declination angle $\delta$ is related to the zenith angle $\zeta$ through $\cos \zeta = \cos \left(\delta  -\phi\right)$, where $\phi$ is the observer's latitude on Earth.}. We use the publicly available effective area given in Fig. 1 in the supplemental material of~\cite{IceCube:2019cia}\footnote{Notice that the effective area strongly depends on the analysis cuts and detector effects. We rely on the effective area for the IC86 2012-2018 event selection, corresponding to an analysis sensitivity for a point-like neutrino source that most resembles our case.} and at the scope of example, we fix the declination angle to be in the range $[30 \degree < \delta < 90 \degree]$. This choice corresponds to muon neutrinos from a portion of the Northern sky and allows us to neglect the atmospheric muon background with more reliability, since almost all atmospheric muons from the Northern hemisphere are filtered out by the Earth.
Certainly, this choice is a simplification, as a detailed analysis should instead consider all possible declination-angle windows, so to maximize the sensitivity of the telescope to a PBH explosion. Note, however, that the inferred sensitivities would remain almost unchanged for a PBH explosion occurring anywhere in the Northern hemisphere ($\delta \gtrsim -5\degree$), while they would decrease if the PBH explodes in the other half of the hemisphere, in which case the effective area would be smaller.

\begin{figure}[!tb]
    \centering
    \begin{subfigure}{0.48\textwidth}
        \includegraphics[width=\textwidth]{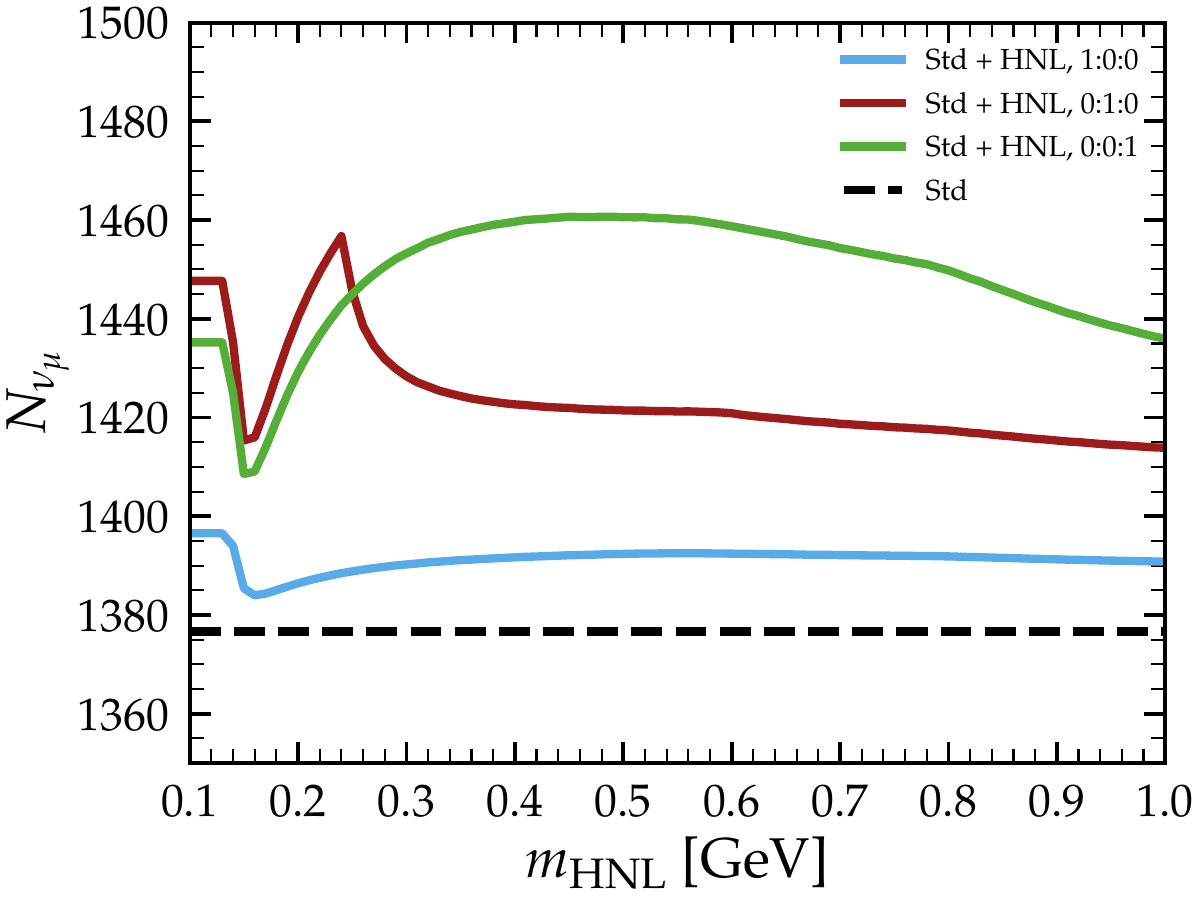}
    \end{subfigure}
    \hfill
    \begin{subfigure}{0.48\textwidth}
         \includegraphics[width=\textwidth]{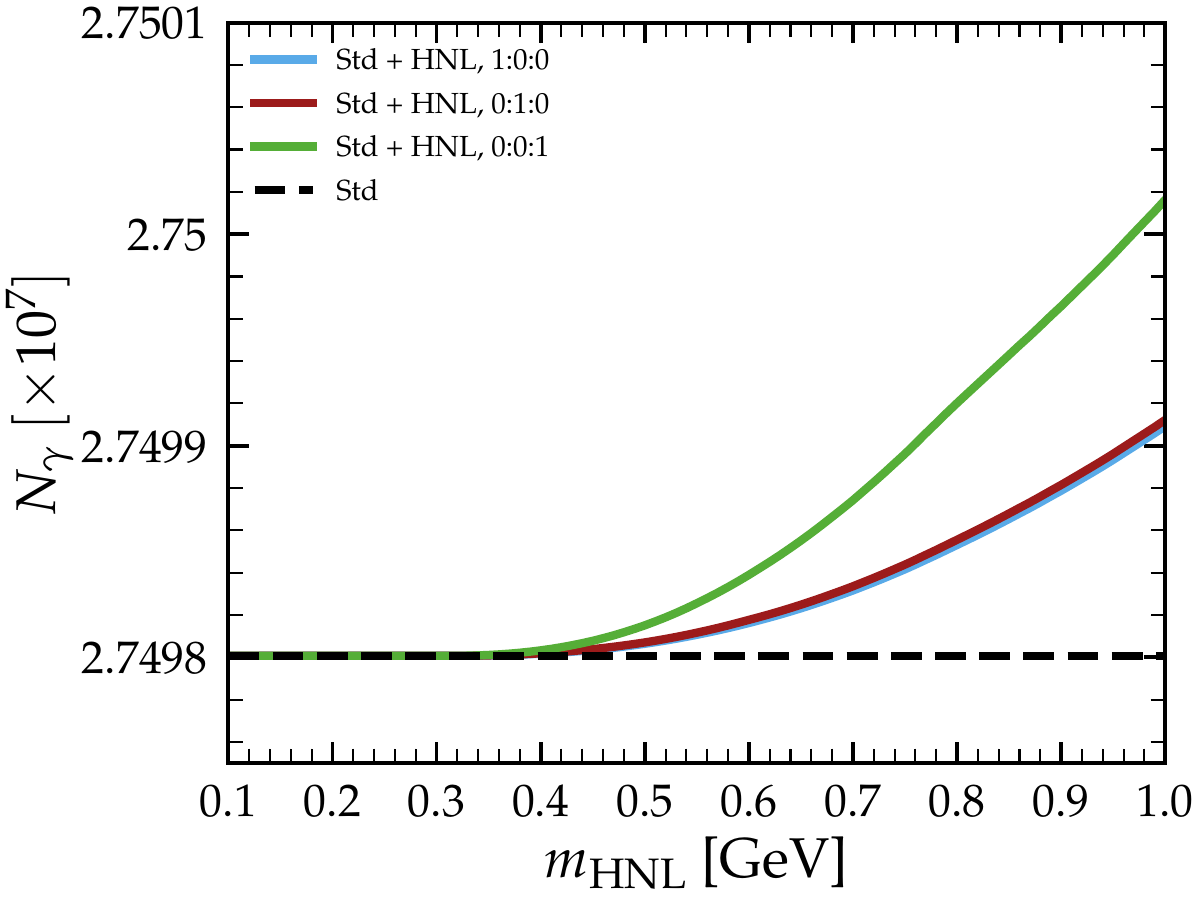}
    \end{subfigure}
    \caption{Expected number of muon-neutrino events at IceCube (left panel) and  photon events at HAWC (right panel), as a function of the HNL mass, from the last stages (100 s) of an evaporating PBH located at a distance $d_\mathrm{PBH} = 10^{-4}$ pc from Earth. Neutrinos and photons are expected to arrive from a declination angle $[30 \degree < \delta < 90 \degree]$. The black dashed curve corresponds to the SM-only case. }
    \label{fig:Numberofnumu_100s_0.0001pc}
\end{figure}

We present in Fig.~\ref{fig:Numberofnumu_100s_0.0001pc} (left panel) the number of muon neutrino events expected at IceCube from an evaporating PBH located at $d_\mathrm{PBH} = 10^{-4}$ pc from Earth, as a function of the HNL mass. As before, the PBH burst is assumed to last 100 s. The different colored curves represent the number of muon-tracks events from HNL in the $[0.1,1]$ GeV range for three mixing scenarios, 1:0:0 (cyan), 0:1:0 (red), and 0:0:1 (green).  The black dashed curve indicates the SM-only contribution. 
The behavior of the curves with HNLs is mainly determined by the oscillations among active states and the interplay between the two- and three-body decay contributions. The dip visible around $m_4 \sim 0.15$ GeV marks the transition from the regime where the three-body decay $\nu_4 \to \nu_\alpha \nu_\ell \bar{\nu}_\ell$ dominates to the regime where  $\nu_4 \to \nu_\alpha \pi^0$ is instead more important. A similar dependence of $N_{\nu_\mu}$ on the HNL mass can be seen in muon-tracks events where the muon neutrino does not directly mix with the HNL, i.e. in the 1:0:0 and 0:0:1 mixing modes. 
However, since $\theta_{23}$ is close to maximal, the number of events for the 0:0:1 case is larger than for 1:0:0.

Concerning HNLs with masses above the electroweak scale, we consider only the mixing 0:1:0 for illustration and restrict our analysis to HNLs with masses above 500 GeV, to ensure the dominance of the bosonic channels in Eq.~\eqref{eq:heavyHNLdecaymodes}. In this scenario, the number of expected muon-tracks events in the presence of HNLs hardly depends on the HNL mass, since the mass dependence in the different channels of Eq.~\eqref{eq:heavydecaywidths} is approximately the same and $m_4$ eventually simplifies in $\mathcal{B}_\alpha$. We predict $N_{\nu_\mu}^\mathrm{SM + HNL} \sim 1470$, independently of $m_4$, in the mass range $[0.5,2]$ TeV.

Note that, as previously discussed for the neutrino spectrum of Fig.~\ref{fig:dNdE}, the number of events $N_{\nu_\mu}^\mathrm{SM + HNL}$ does not depend on the mixing angle, for HNL masses both above and below the electroweak scale. Indeed, under the assumption that only one active neutrino species mixes with the HNL, $|U_{\alpha 4}|^2$ cancels out in the branching ratio that enters in the computation of $N_{\nu_\mu}^\mathrm{HNL}$.
We have also quantified how much using different ranges of declination angles affect the number of muon neutrinos expected at IceCube. In particular, declination angles between $-5 \degree$ and $30 \degree$, i.e. covering the remaining portion of the Northern sky, would imply $8\%$ more events. On the other hand, neutrino fluxes incoming from the Southern sky, i.e.,  $[-90 \degree < \delta < -5 \degree]$, are almost completely shadowed by the background of atmospheric muons leading to $N_{\nu_\mu} < 100$.

Regarding photons, we estimate the total number of gamma-ray events expected at HAWC through
\begin{equation}
\label{eq:Ngamam}
    N_{\gamma}(\delta) = \int_{E_\mathrm{min}^\gamma}^{E_\mathrm{max}^\gamma}dE\; F_{\gamma} (E) \mathcal{A}_\mathrm{eff} (E, \delta) \, ,
\end{equation}
where $\mathcal{A}_\mathrm{eff} (E, \delta)$ is taken from Fig. 2 in~\cite{HAWC:2011gts}. We choose HAWC as an example of gamma-ray telescope in the Northern hemisphere, that could therefore observe the same portion of sky as IceCube (or at least part of it). The chosen effective area for HAWC corresponds to zenith angles in the range $[0 \degree < \zeta < 26 \degree]$. Also in this case, the choice of the effective area is very relevant at the scope of the analysis. We choose the largest effective area, thus resulting in the largest photon flux expected at HAWC. We have checked that using a smaller effective area (namely, the smallest one from Fig. 2 in~\cite{HAWC:2011gts}, corresponding to larger declination angles) would imply up to a factor 100 less photon events. However, photons would still outnumber neutrinos by orders of magnitude and, as a consequence, the inferred sensitivities would remain substantially unchanged. Returning to Eq.~\eqref{eq:Ngamam}, in this case, $E_\mathrm{min}^{\gamma} = 100$ GeV and $E_\mathrm{max}^{\gamma} = 10^5$ GeV. 

We show in Fig.~\ref{fig:Numberofnumu_100s_0.0001pc} (right panel) the number of gamma-ray events expected at HAWC from an evaporating PBH located at $d_\mathrm{PBH} = 10^{-4}$ pc from Earth, as a function of the HNL mass in the light-mass regime. As we can note, the impact of HNL decays start to be visible above $m_4 \gtrsim 0.5$ GeV. Similar to the neutrino case, in the heavy-mass regime the number of overall photons does not strongly depend on the mass, and is around $N_{\gamma}^\mathrm{SM + HNL} \sim 2.87 \times10^7$. In both cases, the PBH burst is assumed to last 100 s.\\

To estimate the combined sensitivity at IceCube and HAWC we rely on the following $\chi^2$ test-statistics
\begin{align}
\label{eq:chi2}
\begin{split}
 \chi^2 \left (d_{\rm PBH}, m_4 \right) =  &   
 {\rm min}_\alpha\left. \left\{\frac{\left[N_{\gamma}^{\mathrm{SM + HNL}}(m_4, d_{\rm PBH}(1 + \alpha) )- N_{\gamma}^{\rm SM}(d_{\rm PBH})\right]^2}{\sigma_{\gamma}^2} \right. \right. \\ 
 & \qquad\quad + \left. \frac{\left[N^{\mathrm{SM + HNL}}_{\nu_\mu}(m_4, d_{\rm PBH}(1 + \alpha))- N_{\nu_\mu}^{\rm SM}(d_{\rm PBH})\right]^2}{\sigma^2_{\nu_\mu}} \right\}  \, ,
 \end{split}
\end{align}
that we minimize over the nuisance parameter $\alpha$. We find it to vary around $\alpha \sim 10^{-7}-10^{-5}$ in the light-mass regime, whereas larger values of $\alpha \sim 10^{-2}$ are required in the high-mass regime. In the previous expression, $N^{\mathrm{SM + HNL}}_{\nu_\mu}$ is the total number of $\nu_\mu$ events expected from the SM (primary + secondary) and HNL contributions, whereas $N^{\mathrm{SM}}_{\nu_\mu}$ indicates the SM-only contribution. In the case of neutrinos, the main source of background could be high-energy atmospheric neutrino events, however, the short duration of the PBH evaporation and the choice of declination angle make this observation essentially background-free~\cite{Capanema:2021hnm,Perez-Gonzalez:2023uoi}. Indeed, atmospheric neutrino events are estimated to be $\mathcal{O} (10^{-4})$ for an observation time of 100 s~\cite{Capanema:2021hnm,Esmaili:2012us,Honda:2015fha}. 
Given these premises, we fix $\sigma_{\nu_\mu}$ to be dominated by statistical errors, $\sigma_{\nu_\mu} = \sqrt{N^{\mathrm{SM}}_{\nu_\mu}(d_{\rm PBH})}$. Similarly, we expect backgrounds to be negligible also for gamma rays, given the short duration of the burst, and we fix  $\sigma_{\gamma} = \sqrt{N^{\mathrm{SM}}_{\gamma}(d_{\rm PBH})}$.

A multimessenger analysis based on two separate datasets (gamma rays and neutrinos) is essential to erase the possible degeneracy between the PBH position and the HNL contribution, and thus to identify $d_\mathrm{PBH}$ with more precision. 

\section{Results}
\label{sec:results}

We compute the combined sensitivity at IceCube and HAWC using the $\chi^2$ function defined in Eq.~\eqref{eq:chi2} and varying the following free parameters: $m_4$ and $d_\mathrm{PBH}$. While relevant to obtain the partial decay widths of the HNL, $|U_{\alpha 4}|^2$ does not affect the estimation of the muon-track events, as already explained in Sections ~\ref{subsec:summary_neutrino_spectra} and ~\ref{sec:analysis}. 
However, $|U_{\alpha 4}|^2$ enters as a relevant ingredient in the computation of the HNL lifetime. Therefore, if this mixing is small, the decay length may exceed $d_{\rm PBH}$, causing the HNL to decay after traversing the Earth. This dependence on the mixing will put a lower bound to our sensitivity as shown below.

\begin{figure}[!htb]
    \centering
    \begin{subfigure}{0.49\textwidth}
        \includegraphics[width=\textwidth]{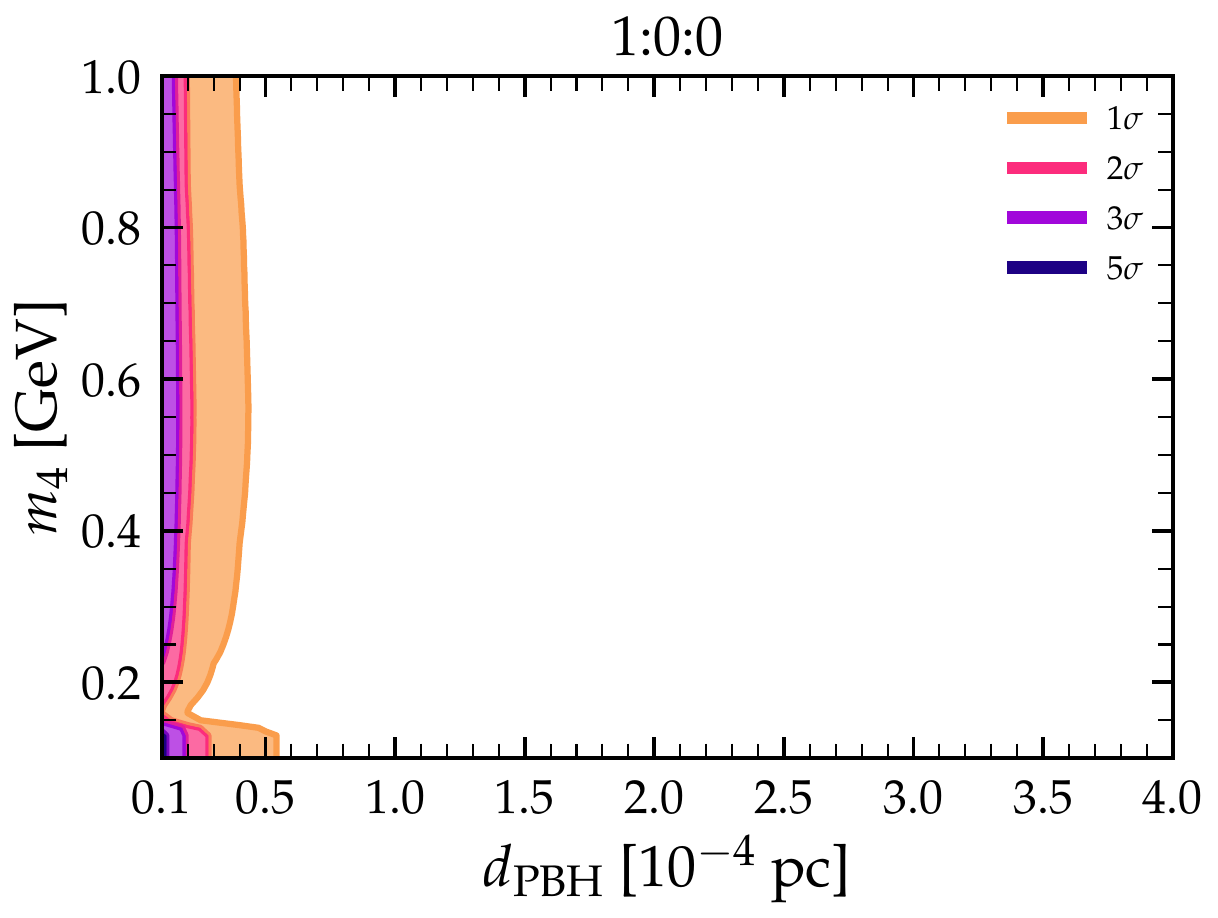}
    \end{subfigure}
    \hfill
    \begin{subfigure}{0.49\textwidth}
        \includegraphics[width=\textwidth]{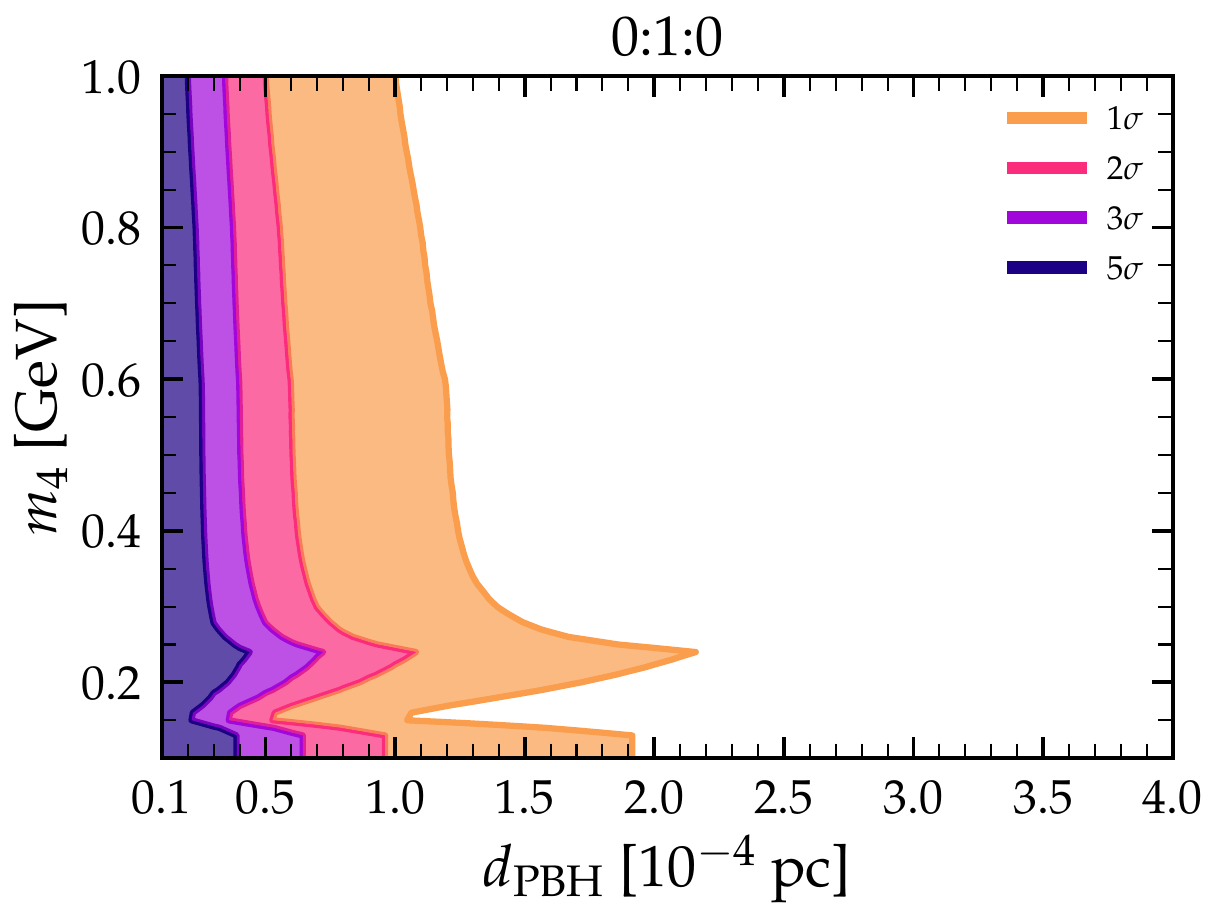}
    \end{subfigure}
    \begin{subfigure}{0.49\textwidth} 
        \includegraphics[width=\textwidth]{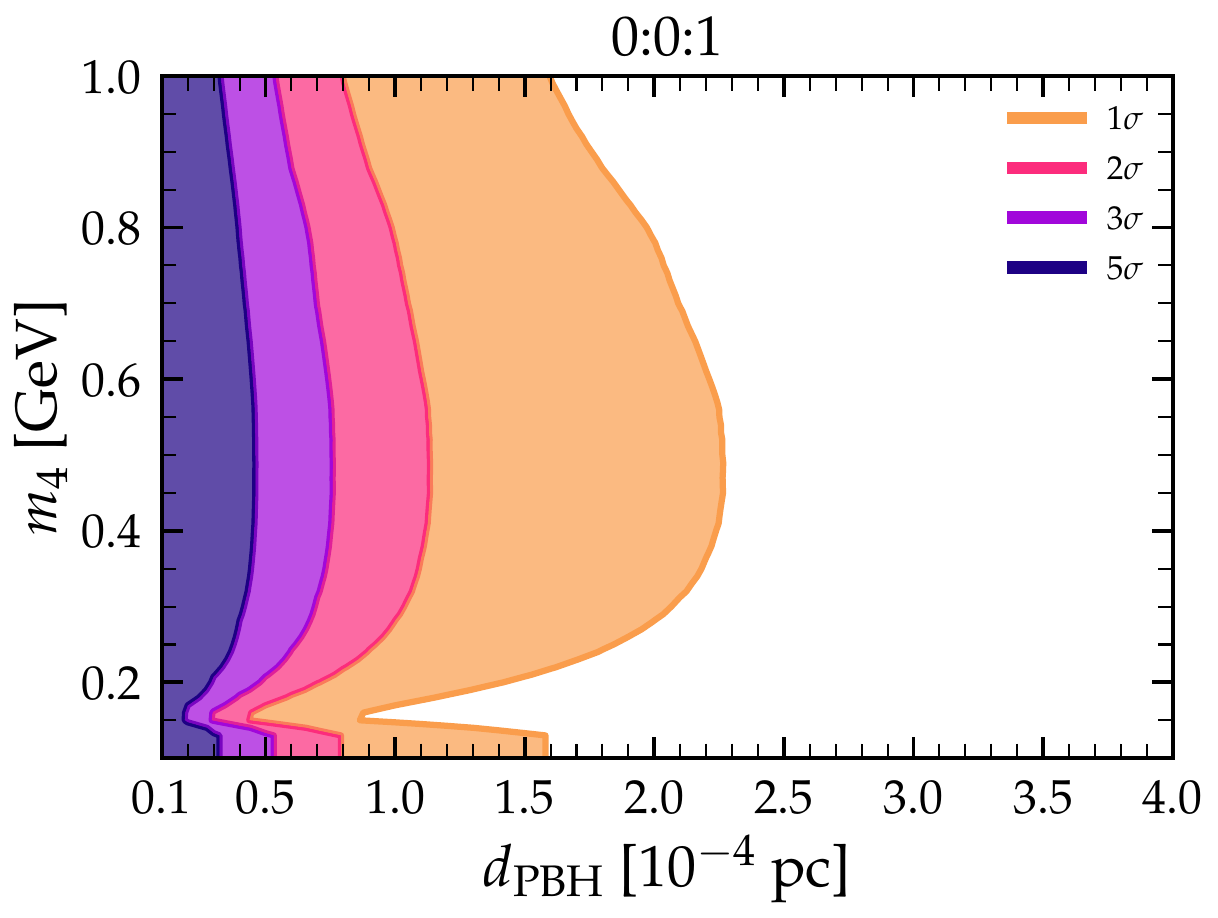}
    \end{subfigure}
    \hfill
    \begin{subfigure}{0.49\textwidth}
        \includegraphics[width=\textwidth]{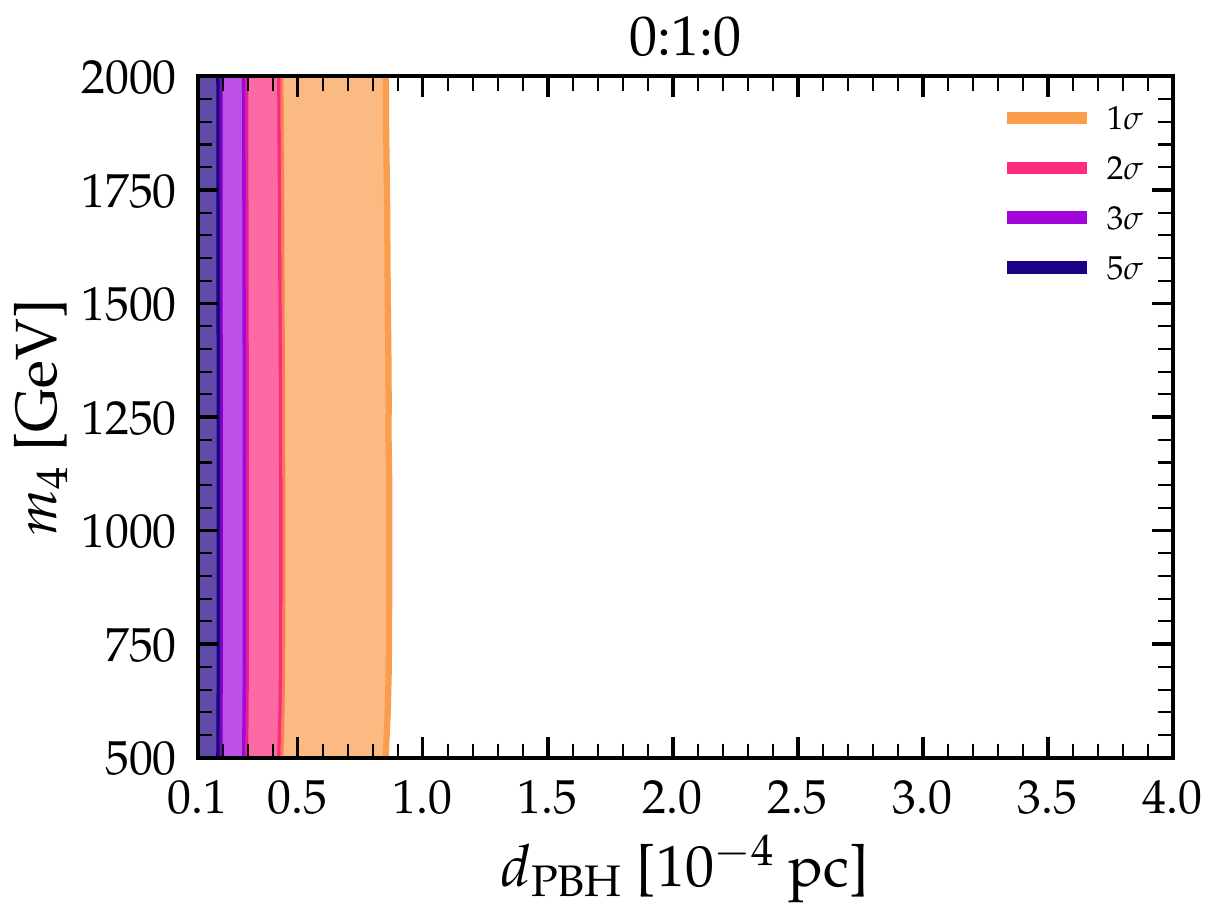}
    \end{subfigure}
\caption{Combined IceCube and HAWC sensitivity to HNLs from a PBH burst lasting 100 s, in terms of the HNL mass and the PBH distance to Earth, for different scenarios for the HNL-active neutrino mixing. See text for more details.}
    \label{fig:sigma_dPBH_mHNL}
\end{figure}

We show in Fig.~\ref{fig:sigma_dPBH_mHNL} the $1,2,3,5 \sigma$ sensitivities (lighter to darker shades) in terms of $m_4$ and $d_{\rm PBH}$. Each panel corresponds to a different mixing scenario, as indicated in its title, and we assume the mixing to be large enough for the HNL decay to occur before reaching the Earth. 
As anticipated, the $\chi^2$ does not depend on $|U_{\alpha 4}|^2$. 
In the case of a HNL with mass in the $[0.1,1]$ GeV range, the mixing scenario 1:0:0 leads to the smallest number of muon neutrino events, as was evident already from Fig.~\ref{fig:Numberofnumu_100s_0.0001pc}. This translates into sensitivities shifted towards smaller distances, as in order to obtain a sizeable signal the PBH must be located closer to Earth. In the other two scenarios, 0:1:0 and 0:0:1, we see that if the PBH explosion occurs at a distance $d_\mathrm{PBH} \sim 10^{-4}$ pc\footnote{For the sake of comparison, this distance is of the same order as the Uranus-Sun distance.}, IceCube and HAWC could basically probe all the mass range $m_4 = [0.1,1]$ GeV at $1\sigma$ CL. Indeed, for the light mass regime, the photon contribution from HNL decay through pion decay is extremely suppressed with respect to the neutrino counterpart, allowing us to pinpoint the PBH distance through the photon signal. \\
HNLs with masses above the electroweak scale (lower-right panel in Fig.~\ref{fig:sigma_dPBH_mHNL})  would instead produce equally large fluxes of muon neutrinos and photons, making the identification of the distance of the explosion more difficult and, as a consequence, translate into milder bounds, requiring the explosion to occur below $d_\mathrm{PBH} \lesssim 10^{-4}$.

\begin{figure}[!tb]
    \centering
    \begin{subfigure}{0.49\textwidth}
        \includegraphics[width=\textwidth]{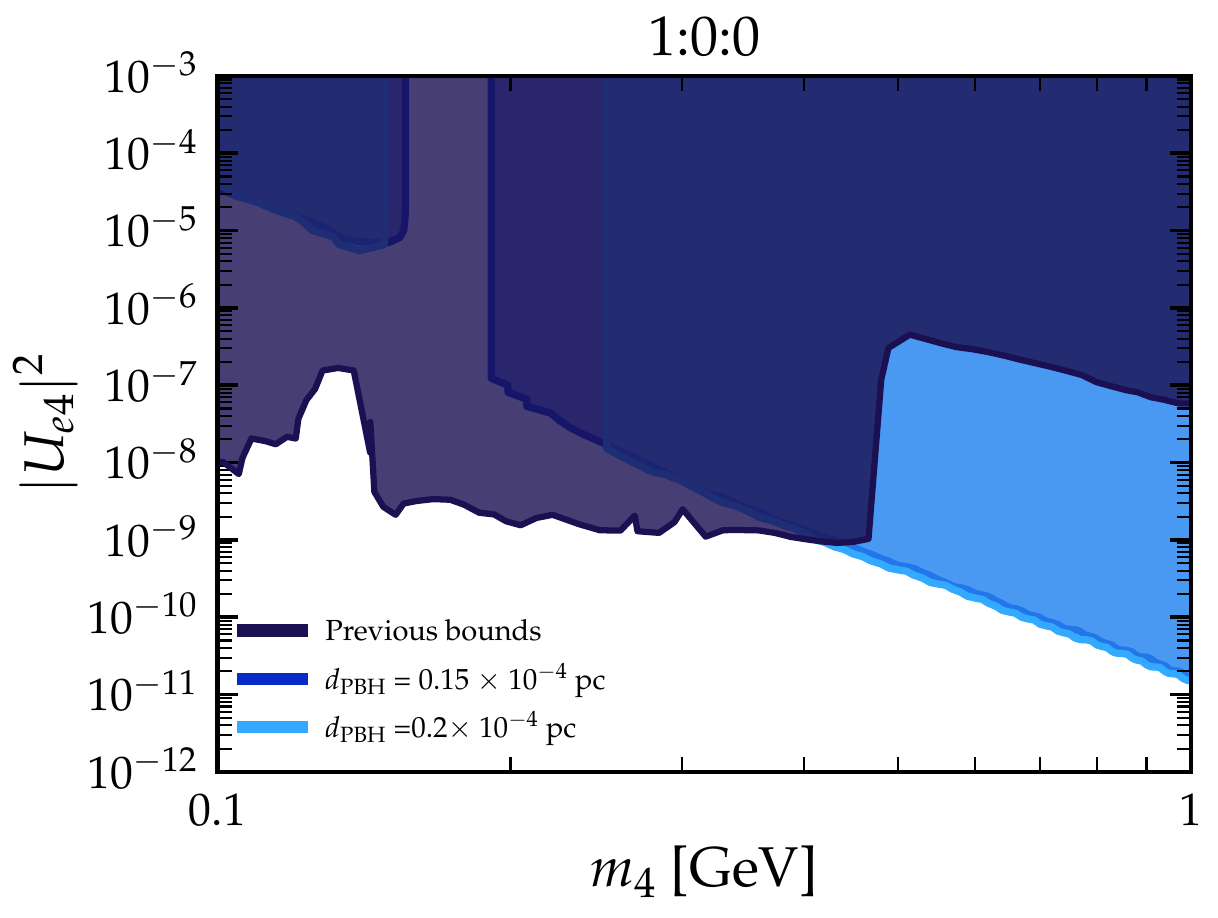}
    \end{subfigure}
    \hfill
    \begin{subfigure}{0.49\textwidth}
        \includegraphics[width=\textwidth]{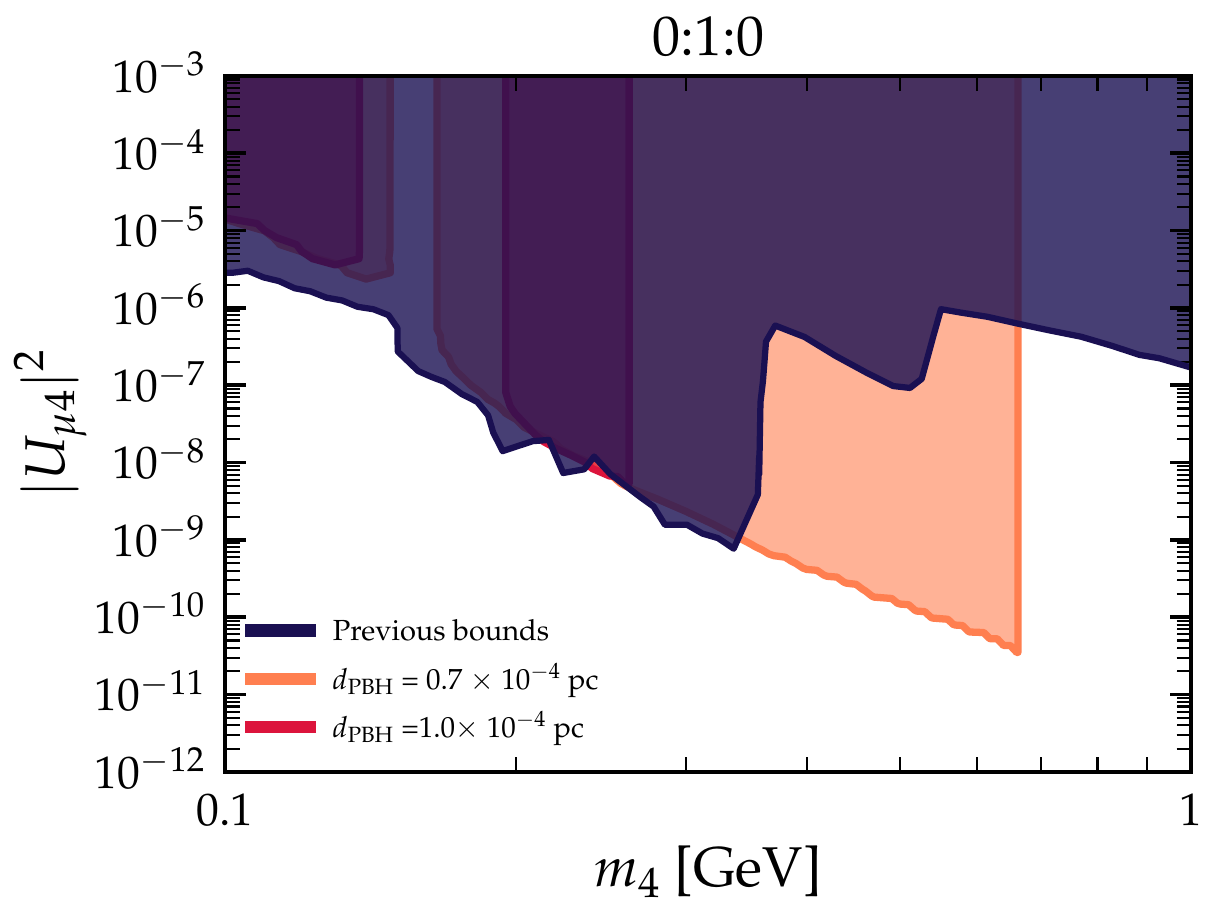}
    \end{subfigure}
    \begin{subfigure}{0.49\textwidth}
        \includegraphics[width=\textwidth]{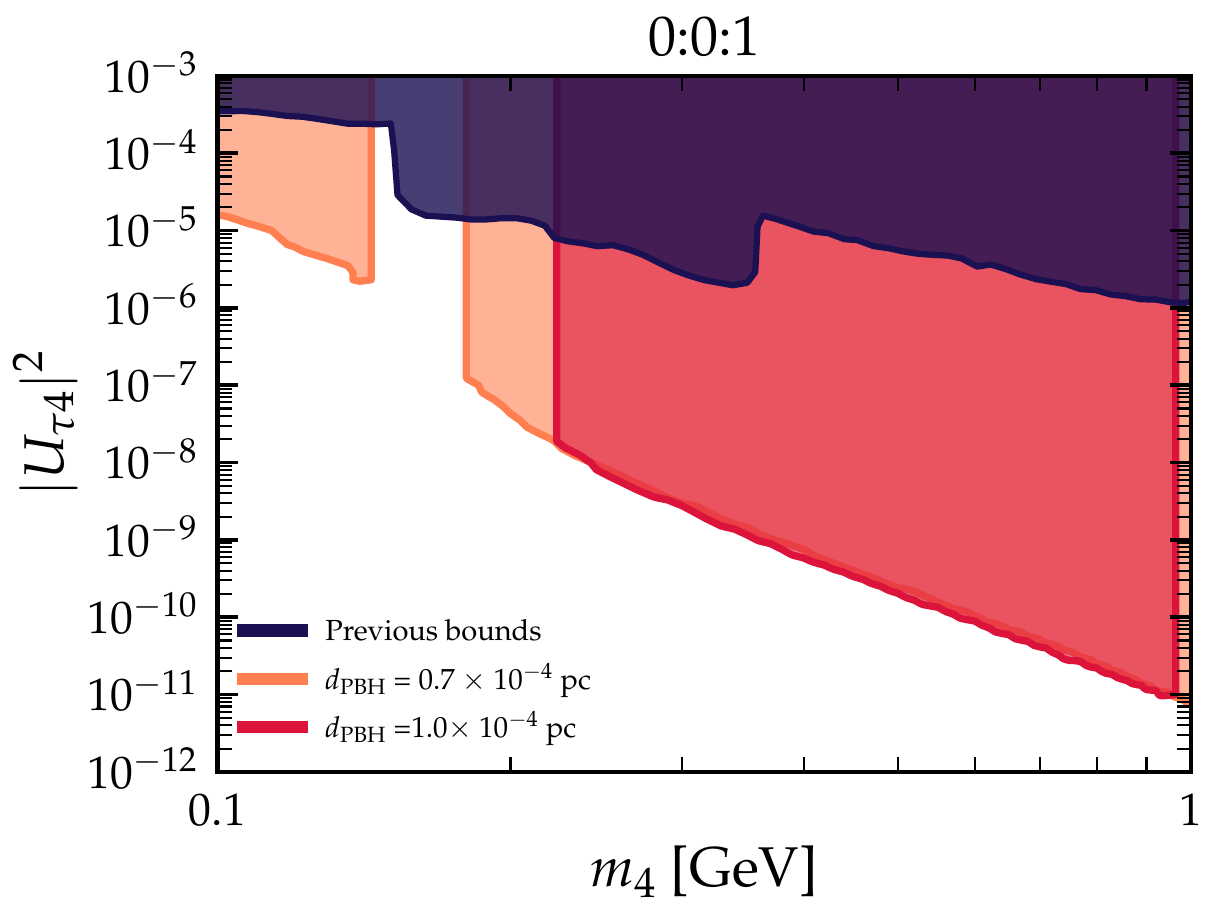}
    \end{subfigure}
    \hfill
    \begin{subfigure}{0.49\textwidth}
        \includegraphics[width=\textwidth]{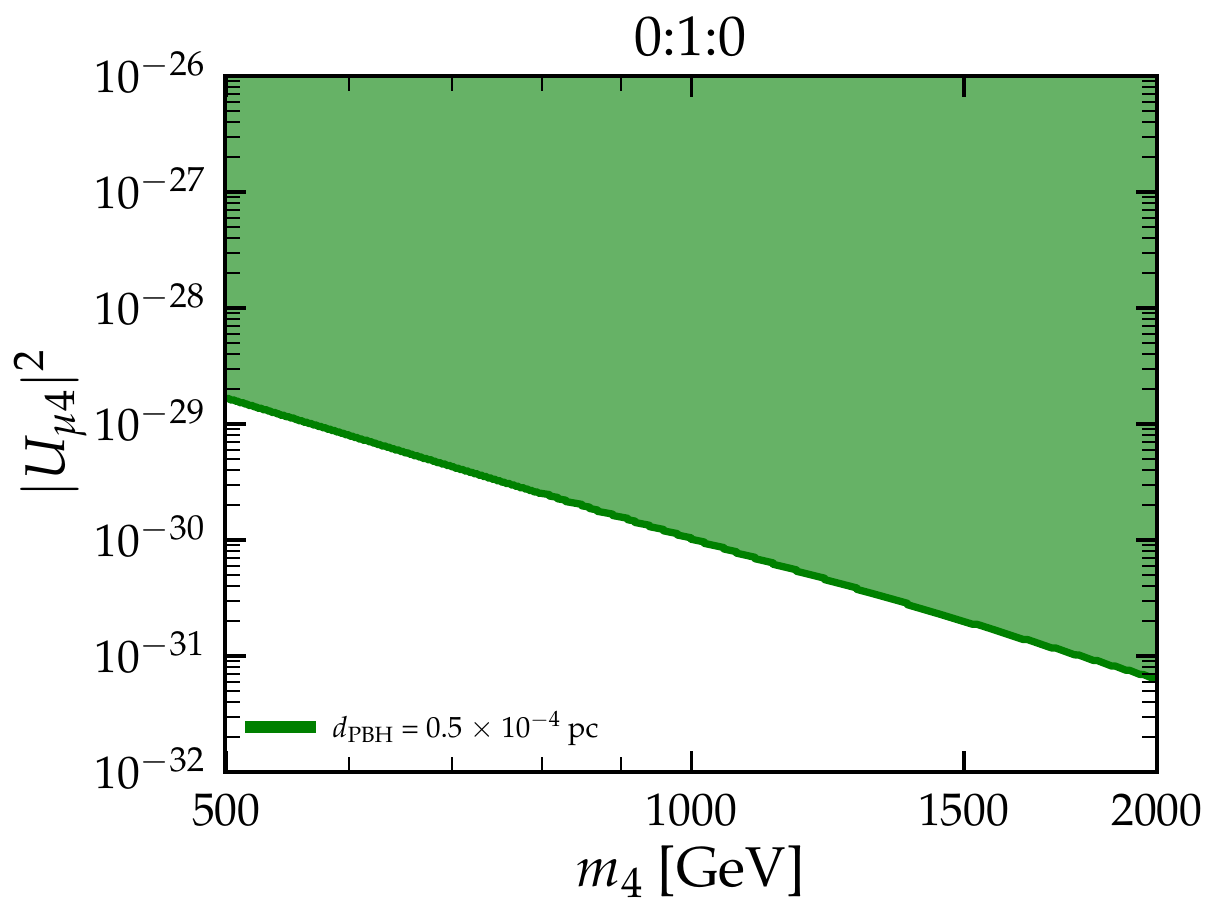}
    \end{subfigure}
\caption{Combined IceCube and HAWC sensitivity (at $90\%$ CL) to the mixing matrix elements $|U_{\alpha 4}|^2$ as a function of the HNL mass, for a PBH burst of $\tau = 100$ s. Each panel correspond to a different mixing scenario, in which the HNL only mixes with one of the active neutrinos with flavor $\alpha$, while the other two mixings are set to zero. Different colors correspond to different PBH distances to Earth. Dark blue shaded areas denote existing constraints~\cite{Bolton:2019pcu,Abdullahi:2022jlv}. 
}
    \label{fig:90_CL_mHNL_mixing}
\end{figure}

In order to compare with previous bounds, we depict in Fig.~\ref{fig:90_CL_mHNL_mixing} our 90\% CL sensitivities in the plane ($|U_{\alpha 4}|^2$, $m_4$) as colored shaded regions. We show a different mixing scenario in each panel, and we fix some benchmark values for the PBH distance $d_\mathrm{PBH}$, as indicated in the legends. Although not directly dependent on $|U_{\alpha 4}|^2$ as already explained, our exclusion regions show a diagonal cut. 
As mentioned before, we impose a lower limit on the HNL lifetime, or, equivalently, an upper limit on the HNL decay length that cannot exceed $d_\mathrm{PBH}$. This constraint excludes mixings below a certain value of $|U_{\alpha 4}|^2$, depending on the HNL mass. Moreover, in each panel we superimpose existing bounds (see e.g. Ref.~\cite{Abdullahi:2022jlv} for a review), indicated as a dark-blue shaded area. In particular, for scenario 1:0:0 bounds from NA62~\cite{NA62:2020mcv}, T2K~\cite{T2K:2019jwa}, PiENU~\cite{PIENU:2017wbj}, BEBC~\cite{Barouki:2022bkt} and PS191~\cite{Bernardi:1987ek} are relevant; for 0:1:0, T2K~\cite{T2K:2019jwa}, MicroBooNE~\cite{MicroBooNE:2019izn}, NuTeV~\cite{NuTeV:1999kej}, E949~\cite{E949:2014gsn}. Finally, for 0:0:1 the most important bounds in the considered mass range are given by T2K~\cite{T2K:2019jwa}, CHARM~\cite{CHARMII:1994jjr}, and constraints from IceCube~\cite{Coloma:2017ppo} looking for low-energy “double-bang”
events. Let us also add that the SN 1987A detection of neutrino events can be used to put constraints on $|U_{\alpha 4}|^2$, for HNLs with masses $\mathcal{O} (100)$ MeV. Although not shown in the plots to not overcrowd the figures, these bounds (see e.g.~\cite{Carenza:2023old}) can cover similar regions of parameter space, thus providing a complementary probe of the same scenario.

Our results show that the combined observation of muon-track events at IceCube and gamma-ray events at HAWC from the explosion of a PBH located at a distance $d_\mathrm{PBH} = 0.7 \times 10^{-4}$ pc would allow to explore the HNL scenario at masses $0.3~(0.1) \lesssim m_4 \lesssim 0.6~(1)$ GeV, in the case where the HNL mixes only with the muon (tau) flavor, testing mixing values below those currently excluded by other searches. In the case of mixing only with the electron flavor, the expected flux of muon neutrinos would be smaller and therefore would require the PBH explosion to occur as close as $d_\mathrm{PBH} = 2 \times 10^{-5}$ pc to Earth to be able to probe part of the parameter space not yet excluded.

Moving to $m_4 \geq 500$ GeV instead, the expected  fluxes of both photons and muon neutrinos are larger and the identification of the HNL signal would require the PBH explosion to occur closer, at a distance of $d_\mathrm{PBH} = 5 \times 10^{-5}$ pc. In this case, the inferred bounds would be much more stringent as they would cover all the parameter space down to very tiny values of $|U_{\alpha4}|^2$. Indeed, such heavy HNLs would be very short-lived and decay right after their production from PBH evaporation. As a consequence, the HNL decay lifetime bound would only apply to mixings $|U_{\alpha4}|^2 \lesssim 10^{-29}$, as visible in the lower-right panel of Fig.~\ref{fig:90_CL_mHNL_mixing}. 

As a final comment, let us recall that all sensitivities have been obtained assuming a burst of $\tau = 100$ s, corresponding to a PBH with mass $M \sim 6.2 \times 10^{9}$ g at the beginning of the explosion. 
Should we consider smaller observation time frames, and hence smaller $M$, our results would change as a $90 \%$ CL detection would require the explosion to occur closer to Earth. To quantify this effect, for an observation time of $\tau = 10$ s, i.e., $M \sim 2.9 \times 10^9$, depending on the mixing scenario, the PBH should be $\sim$ 30-50\% closer to obtain the same $90 \%$ CL sensitivities.

\section{Conclusions}
\label{sec:conc}

Several theories of the early Universe predict the presence of a population of black holes of primordial origin, which in turn may be linked to the DM mystery.
As theorized by Hawking, these objects are expected to evaporate, emitting radiation with thermal spectra. 
The characteristics of such an emission are intimately related to the properties of emitted particles (mass, charge, spin and  number of degrees of freedom) and therefore provide valuable opportunities to probe the existence of new physics.  The final moments of this evaporation phase are particularly relevant, as the emission increases while the PBH loses mass and raises its temperature, leading to the emission of heavier degrees of freedom.

In this work, we investigated a scenario in which one light PBH \textit{explodes}, i.e. ends its evaporation phase close to Earth thus giving rise to a ``burst" of visible particles. We were particularly interested in a BSM scenario in which the SM is extended with one HNL that mixes with active neutrinos. Such a new sterile state will be emitted in the evaporation process and consequently decay into active neutrinos, sizably contributing to the expected flux of observable neutrinos at Earth.

During the rapid explosion and in the subsequent decay chains, large fluxes of gamma rays would also be produced, whose detection would certainly be within the reach of current facilities and would help to identify the PBH distance from Earth. We therefore adopted a multimessenger approach for the statistical analysis.
We estimated the spectra of muon neutrinos expected at a neutrino telescope like IceCube, and of high-energy gamma rays that would be visible at HAWC, taking into account all primary and secondary contributions from the PBH explosion, including the HNL ones. Under the assumption that the PBH explosion occurs sufficiently close to Earth to be visible and lasts 100 s, we inferred sensitivities on the relevant HNL parameter space, in terms of the HNL mass and mixing with the active flavors.
Depending on how close the explosion occurs, a combined observation at IceCube and HAWC could lead to stringent constraints on the HNL parameter space, even improving existing ones from other facilities, in some ranges of the HNL mass and relevant mixing. 

Certainly, our estimates are strongly dependent on the assumptions made for the analysis, namely current IceCube and HAWC's effective areas, their angular resolution, and the capacity of reducing backgrounds. Future IceCube upgrades are anticipated to improve the effective area by a factor $\sim 5$~\cite{IceCube-Gen2:2020qha}, thus implying a substantial enhancement in the detection of high-energy muon tracks. The development of new analysis and detection techniques will further enhance the possibility of observing such events, and eventually improve the sensitivities to BSM physics, translating into observable signals even if the explosion occurs at farther distances.  In the near future, other neutrino telescopes like KM3NeT~\cite{KM3Net:2016zxf}, P-ONE~\cite{P-ONE:2020ljt}, and Baikal-GVD~\cite{BAIKAL:1997iok} will also relevantly contribute to the search for PBH explosions in the proximity of Earth, being sensitive to different directions in the sky and thus allowing to provide a more complete picture.
Regarding photons, future Cherenkov telescope arrays~\cite{Boluna:2023jlo} are also expected to be specially suited to detect light PBHs, like those in the final evaporation stages, and will certainly improve the sensitivities here derived. 

\section*{Acknowledgments}
We are grateful to A. Arbey for support with BlackHawk, and G.~F.~S.~Alves for his input on obtaining the time-integrated spectrum. We also thank the anonymous Referee for useful suggestions that helped improve our analysis.
Y.F.P.G.~would like to thank the warm hospitality of the Astroparticles and High Energy Physics group of IFIC during the visit that marked the inception of this work.
V.D.R. and A.T. acknowledge financial support by the CIDEXG/2022/20 grant (project ``D'AMAGAT'') funded by
Generalitat Valenciana and by the Spanish grants CNS2023-144124 (MCIN/AEI/10.13039/501100011033 and “Next Generation EU”/PRTR) and PID2020-113775GB-I00 (MCIN/AEI/10.13039/501100011033).
Y.F.P.G.~has been funded by the UK Science and Technology Facilities Council (STFC) under grant ST/T001011/1. 
This project has received funding/support from the European Union's Horizon 2020 research and innovation programme under the Marie Sk\l{}odowska-Curie grant agreement No 860881-HIDDeN.

\noindent

\appendix

\bibliographystyle{utphys}
\bibliography{bibliography}  

\providecommand{\href}[2]{#2}\begingroup\raggedright\begin{thebibliography}{100}

\bibitem{Aghanim:2018eyx}
{\bfseries Planck} Collaboration, N.~Aghanim {\em et~al.}
  \href{http://dx.doi.org/10.1051/0004-6361/201833910}{{\em Astron. Astrophys.}
  {\bfseries 641} (2020) A6}, \href{http://arxiv.org/abs/1807.06209}{{\ttfamily
  arXiv:1807.06209 [astro-ph.CO]}}.

\bibitem{Schumann:2019eaa}
M.~Schumann \href{http://dx.doi.org/10.1088/1361-6471/ab2ea5}{{\em J. Phys. G}
  {\bfseries 46} no.~10, (2019) 103003},
  \href{http://arxiv.org/abs/1903.03026}{{\ttfamily arXiv:1903.03026
  [astro-ph.CO]}}.

\bibitem{Billard:2021uyg}
J.~Billard {\em et~al.} \href{http://arxiv.org/abs/2104.07634}{{\ttfamily
  arXiv:2104.07634 [hep-ex]}}.

\bibitem{Carr:2020xqk}
B.~Carr and F.~Kuhnel
  \href{http://dx.doi.org/10.1146/annurev-nucl-050520-125911}{{\em Ann. Rev.
  Nucl. Part. Sci.} {\bfseries 70} (2020) 355--394},
  \href{http://arxiv.org/abs/2006.02838}{{\ttfamily arXiv:2006.02838
  [astro-ph.CO]}}.

\bibitem{Bird:2022wvk}
S.~Bird {\em et~al.} \href{http://dx.doi.org/10.1016/j.dark.2023.101231}{{\em
  Phys. Dark Univ.} {\bfseries 41} (2023) 101231},
  \href{http://arxiv.org/abs/2203.08967}{{\ttfamily arXiv:2203.08967
  [hep-ph]}}.

\bibitem{Green:2024bam}
A.~M. Green
\newblock 2, 2024.
\newblock \href{http://arxiv.org/abs/2402.15211}{{\ttfamily arXiv:2402.15211
  [astro-ph.CO]}}.

\bibitem{Zeldovich:1967lct}
Y.~B. .~N. Zel'dovich, I.~D. {\em Soviet Astron. AJ (Engl. Transl. ),}
  {\bfseries 10} (1967) 602.

\bibitem{Hawking:1971ei}
S.~Hawking {\em Mon. Not. Roy. Astron. Soc.} {\bfseries 152} (1971) 75.

\bibitem{Carr:1974nx}
B.~J. Carr and S.~W. Hawking
  \href{http://dx.doi.org/10.1093/mnras/168.2.399}{{\em Mon. Not. Roy. Astron.
  Soc.} {\bfseries 168} (1974) 399--415}.

\bibitem{Chapline:1975ojl}
G.~F. Chapline \href{http://dx.doi.org/10.1038/253251a0}{{\em Nature}
  {\bfseries 253} no.~5489, (1975) 251--252}.

\bibitem{Abbott:2016blz}
{\bfseries LIGO Scientific, Virgo} Collaboration, B.~P. Abbott {\em et~al.}
  \href{http://dx.doi.org/10.1103/PhysRevLett.116.061102}{{\em Phys. Rev.
  Lett.} {\bfseries 116} no.~6, (2016) 061102},
  \href{http://arxiv.org/abs/1602.03837}{{\ttfamily arXiv:1602.03837 [gr-qc]}}.

\bibitem{Abbott:2020gyp}
{\bfseries LIGO Scientific, Virgo} Collaboration, R.~Abbott {\em et~al.}
  \href{http://dx.doi.org/10.3847/2041-8213/abe949}{{\em Astrophys. J. Lett.}
  {\bfseries 913} no.~1, (2021) L7},
  \href{http://arxiv.org/abs/2010.14533}{{\ttfamily arXiv:2010.14533
  [astro-ph.HE]}}.

\bibitem{Hawking:1974rv}
S.~W. Hawking \href{http://dx.doi.org/10.1038/248030a0}{{\em Nature} {\bfseries
  248} (1974) 30--31}.

\bibitem{Hawking:1974sw}
S.~W. Hawking \href{http://dx.doi.org/10.1007/BF02345020}{{\em Commun. Math.
  Phys.} {\bfseries 43} (1975) 199--220}. [Erratum: Commun.Math.Phys. 46, 206
  (1976)].

\bibitem{1976ApJ...206....8C}
B.~J. {Carr} \href{http://dx.doi.org/10.1086/154351}{{\em \apj} {\bfseries 206}
  (May, 1976) 8--25}.

\bibitem{Lehoucq:2009ge}
R.~Lehoucq, M.~Casse, J.~M. Casandjian, and I.~Grenier
  \href{http://dx.doi.org/10.1051/0004-6361/200911961}{{\em Astron. Astrophys.}
  {\bfseries 502} (2009) 37}, \href{http://arxiv.org/abs/0906.1648}{{\ttfamily
  arXiv:0906.1648 [astro-ph.HE]}}.

\bibitem{Wright:1995bi}
E.~L. Wright \href{http://dx.doi.org/10.1086/176910}{{\em Astrophys. J.}
  {\bfseries 459} (1996) 487},
  \href{http://arxiv.org/abs/astro-ph/9509074}{{\ttfamily
  arXiv:astro-ph/9509074}}.

\bibitem{Arbey:2019mbc}
A.~Arbey and J.~Auffinger
  \href{http://dx.doi.org/10.1140/epjc/s10052-019-7161-1}{{\em Eur. Phys. J. C}
  {\bfseries 79} no.~8, (2019) 693},
  \href{http://arxiv.org/abs/1905.04268}{{\ttfamily arXiv:1905.04268 [gr-qc]}}.

\bibitem{Ballesteros:2019exr}
G.~Ballesteros, J.~Coronado-Bl\'azquez, and D.~Gaggero
  \href{http://dx.doi.org/10.1016/j.physletb.2020.135624}{{\em Phys. Lett. B}
  {\bfseries 808} (2020) 135624},
  \href{http://arxiv.org/abs/1906.10113}{{\ttfamily arXiv:1906.10113
  [astro-ph.CO]}}.

\bibitem{Laha:2020ivk}
R.~Laha, J.~B. Mu\~noz, and T.~R. Slatyer
  \href{http://dx.doi.org/10.1103/PhysRevD.101.123514}{{\em Phys. Rev. D}
  {\bfseries 101} no.~12, (2020) 123514},
  \href{http://arxiv.org/abs/2004.00627}{{\ttfamily arXiv:2004.00627
  [astro-ph.CO]}}.

\bibitem{Boudaud_2019}
M.~Boudaud and M.~Cirelli
  \href{http://dx.doi.org/10.1103/physrevlett.122.041104}{{\em Physical Review
  Letters} {\bfseries 122} no.~4, (Jan, 2019) }.
  \url{http://dx.doi.org/10.1103/PhysRevLett.122.041104}.

\bibitem{Dasgupta:2019cae}
B.~Dasgupta, R.~Laha, and A.~Ray
  \href{http://dx.doi.org/10.1103/PhysRevLett.125.101101}{{\em Phys. Rev.
  Lett.} {\bfseries 125} no.~10, (2020) 101101},
  \href{http://arxiv.org/abs/1912.01014}{{\ttfamily arXiv:1912.01014
  [hep-ph]}}.

\bibitem{DeRocco:2019fjq}
W.~DeRocco and P.~W. Graham
  \href{http://dx.doi.org/10.1103/PhysRevLett.123.251102}{{\em Phys. Rev.
  Lett.} {\bfseries 123} no.~25, (2019) 251102},
  \href{http://arxiv.org/abs/1906.07740}{{\ttfamily arXiv:1906.07740
  [astro-ph.CO]}}.

\bibitem{Laha:2019ssq}
R.~Laha \href{http://dx.doi.org/10.1103/PhysRevLett.123.251101}{{\em Phys. Rev.
  Lett.} {\bfseries 123} no.~25, (2019) 251101},
  \href{http://arxiv.org/abs/1906.09994}{{\ttfamily arXiv:1906.09994
  [astro-ph.HE]}}.

\bibitem{Wang:2020uvi}
S.~Wang, D.-M. Xia, X.~Zhang, S.~Zhou, and Z.~Chang
  \href{http://dx.doi.org/10.1103/PhysRevD.103.043010}{{\em Phys. Rev. D}
  {\bfseries 103} no.~4, (2021) 043010},
  \href{http://arxiv.org/abs/2010.16053}{{\ttfamily arXiv:2010.16053
  [hep-ph]}}.

\bibitem{DeRomeri:2021xgy}
V.~De~Romeri, P.~Mart\'\i{}nez-Mirav\'e, and M.~T\'ortola
  \href{http://dx.doi.org/10.1088/1475-7516/2021/10/051}{{\em JCAP} {\bfseries
  10} (2021) 051}, \href{http://arxiv.org/abs/2106.05013}{{\ttfamily
  arXiv:2106.05013 [hep-ph]}}.

\bibitem{Bernal:2022swt}
N.~Bernal, V.~Mu\~noz Albornoz, S.~Palomares-Ruiz, and P.~Villanueva-Domingo
  \href{http://dx.doi.org/10.1088/1475-7516/2022/10/068}{{\em JCAP} {\bfseries
  10} (2022) 068}, \href{http://arxiv.org/abs/2203.14979}{{\ttfamily
  arXiv:2203.14979 [hep-ph]}}.

\bibitem{Carr:2009jm}
B.~J. Carr, K.~Kohri, Y.~Sendouda, and J.~Yokoyama
  \href{http://dx.doi.org/10.1103/PhysRevD.81.104019}{{\em Phys. Rev. D}
  {\bfseries 81} (2010) 104019},
  \href{http://arxiv.org/abs/0912.5297}{{\ttfamily arXiv:0912.5297
  [astro-ph.CO]}}.

\bibitem{Carr:2020gox}
B.~Carr, K.~Kohri, Y.~Sendouda, and J.~Yokoyama
  \href{http://arxiv.org/abs/2002.12778}{{\ttfamily arXiv:2002.12778
  [astro-ph.CO]}}.

\bibitem{Keith:2020jww}
C.~Keith, D.~Hooper, N.~Blinov, and S.~D. McDermott
  \href{http://arxiv.org/abs/2006.03608}{{\ttfamily arXiv:2006.03608
  [astro-ph.CO]}}.

\bibitem{He:2002vz}
P.~He and L.-Z. Fang \href{http://dx.doi.org/10.1086/340144}{{\em Astrophys. J.
  Lett.} {\bfseries 568} (2002) L1--L4},
  \href{http://arxiv.org/abs/astro-ph/0202218}{{\ttfamily
  arXiv:astro-ph/0202218}}.

\bibitem{Mack:2008nv}
K.~J. Mack and D.~H. Wesley \href{http://arxiv.org/abs/0805.1531}{{\ttfamily
  arXiv:0805.1531 [astro-ph]}}.

\bibitem{Auffinger:2022khh}
J.~Auffinger \href{http://dx.doi.org/10.1016/j.ppnp.2023.104040}{{\em Prog.
  Part. Nucl. Phys.} {\bfseries 131} (2023) 104040},
  \href{http://arxiv.org/abs/2206.02672}{{\ttfamily arXiv:2206.02672
  [astro-ph.CO]}}.

\bibitem{Page:1976df}
D.~N. Page \href{http://dx.doi.org/10.1103/PhysRevD.13.198}{{\em Phys. Rev. D}
  {\bfseries 13} (1976) 198--206}.

\bibitem{Page:1976ki}
D.~N. Page \href{http://dx.doi.org/10.1103/PhysRevD.14.3260}{{\em Phys. Rev. D}
  {\bfseries 14} (1976) 3260--3273}.

\bibitem{MacGibbon:2007yq}
J.~H. MacGibbon, B.~J. Carr, and D.~N. Page
  \href{http://dx.doi.org/10.1103/PhysRevD.78.064043}{{\em Phys. Rev. D}
  {\bfseries 78} (2008) 064043},
  \href{http://arxiv.org/abs/0709.2380}{{\ttfamily arXiv:0709.2380
  [astro-ph]}}.

\bibitem{Boluna:2023jlo}
X.~Boluna, S.~Profumo, J.~Bl\'e, and D.~Hennings
  \href{http://arxiv.org/abs/2307.06467}{{\ttfamily arXiv:2307.06467
  [astro-ph.HE]}}.

\bibitem{Glicenstein:2013vha}
{\bfseries H.E.S.S.} Collaboration, J.-F. Glicenstein, A.~Barnacka, M.~Vivier,
  and T.~Herr in {\em {33rd International Cosmic Ray Conference}}, p.~0930.
\newblock 7, 2013.
\newblock \href{http://arxiv.org/abs/1307.4898}{{\ttfamily arXiv:1307.4898
  [astro-ph.HE]}}.

\bibitem{Tavernier:2019exh}
T.~Tavernier, J.-F. Glicenstein, and F.~Brun
  \href{http://dx.doi.org/10.22323/1.358.0804}{{\em PoS} {\bfseries ICRC2019}
  (2020) 804}, \href{http://arxiv.org/abs/1909.01620}{{\ttfamily
  arXiv:1909.01620 [astro-ph.HE]}}.

\bibitem{Abdo:2014apa}
A.~A. Abdo {\em et~al.}
  \href{http://dx.doi.org/10.1016/j.astropartphys.2014.10.007}{{\em Astropart.
  Phys.} {\bfseries 64} (2015) 4--12},
  \href{http://arxiv.org/abs/1407.1686}{{\ttfamily arXiv:1407.1686
  [astro-ph.HE]}}.

\bibitem{Archambault:2017asc}
{\bfseries VERITAS} Collaboration, S.~Archambault
  \href{http://dx.doi.org/10.22323/1.301.0691}{{\em PoS} {\bfseries ICRC2017}
  (2018) 691}, \href{http://arxiv.org/abs/1709.00307}{{\ttfamily
  arXiv:1709.00307 [astro-ph.HE]}}.

\bibitem{Fermi-LAT:2018pfs}
{\bfseries Fermi-LAT} Collaboration, M.~Ackermann {\em et~al.}
  \href{http://dx.doi.org/10.3847/1538-4357/aaac7b}{{\em Astrophys. J.}
  {\bfseries 857} no.~1, (2018) 49},
  \href{http://arxiv.org/abs/1802.00100}{{\ttfamily arXiv:1802.00100
  [astro-ph.HE]}}.

\bibitem{HAWC:2013kzm}
{\bfseries HAWC} Collaboration, A.~U. Abeysekara {\em et~al.}
  \href{http://arxiv.org/abs/1310.0073}{{\ttfamily arXiv:1310.0073
  [astro-ph.HE]}}.

\bibitem{HAWC:2019wla}
{\bfseries HAWC} Collaboration, A.~Albert {\em et~al.}
  \href{http://dx.doi.org/10.1088/1475-7516/2020/04/026}{{\em JCAP} {\bfseries
  04} (2020) 026}, \href{http://arxiv.org/abs/1911.04356}{{\ttfamily
  arXiv:1911.04356 [astro-ph.HE]}}.

\bibitem{Ukwatta:2010zn}
T.~N. Ukwatta, J.~H. MacGibbon, W.~C. Parke, K.~S. Dhuga, S.~Rhodes,
  A.~Eskandarian, N.~Gehrels, L.~Maximon, and D.~C. Morris in {\em {12th Marcel
  Grossmann Meeting on General Relativity}}, pp.~1588--1590.
\newblock 3, 2010.
\newblock \href{http://arxiv.org/abs/1003.4515}{{\ttfamily arXiv:1003.4515
  [astro-ph.HE]}}.

\bibitem{Perez-Gonzalez:2023uoi}
Y.~F. Perez-Gonzalez \href{http://dx.doi.org/10.1103/PhysRevD.108.083014}{{\em
  Phys. Rev. D} {\bfseries 108} no.~8, (2023) 083014},
  \href{http://arxiv.org/abs/2307.14408}{{\ttfamily arXiv:2307.14408
  [astro-ph.HE]}}.

\bibitem{HESS:2023zzd}
{\bfseries H.E.S.S.} Collaboration, F.~Aharonian {\em et~al.}
  \href{http://dx.doi.org/10.1088/1475-7516/2023/04/040}{{\em JCAP} {\bfseries
  04} (2023) 040}, \href{http://arxiv.org/abs/2303.12855}{{\ttfamily
  arXiv:2303.12855 [astro-ph.HE]}}.

\bibitem{Halzen:1995hu}
F.~Halzen, B.~Keszthelyi, and E.~Zas
  \href{http://dx.doi.org/10.1103/PhysRevD.52.3239}{{\em Phys. Rev. D}
  {\bfseries 52} (1995) 3239--3247},
  \href{http://arxiv.org/abs/hep-ph/9502268}{{\ttfamily arXiv:hep-ph/9502268}}.

\bibitem{Dave:2019epr}
{\bfseries IceCube} Collaboration, P.~Dave and I.~Taboada
  \href{http://dx.doi.org/10.22323/1.358.0863}{{\em PoS} {\bfseries ICRC2019}
  (2021) 863}, \href{http://arxiv.org/abs/1908.05403}{{\ttfamily
  arXiv:1908.05403 [astro-ph.HE]}}.

\bibitem{Capanema:2021hnm}
A.~Capanema, A.~Esmaeili, and A.~Esmaili
  \href{http://dx.doi.org/10.1088/1475-7516/2021/12/051}{{\em JCAP} {\bfseries
  12} no.~12, (2021) 051}, \href{http://arxiv.org/abs/2110.05637}{{\ttfamily
  arXiv:2110.05637 [hep-ph]}}.

\bibitem{Calza:2023iqa}
M.~Calz\`a and J.~a.~G. Rosa \href{http://arxiv.org/abs/2312.09261}{{\ttfamily
  arXiv:2312.09261 [hep-ph]}}.

\bibitem{Ukwatta:2015iba}
T.~N. Ukwatta, D.~R. Stump, J.~T. Linnemann, J.~H. MacGibbon, S.~S. Marinelli,
  T.~Yapici, and K.~Tollefson
  \href{http://dx.doi.org/10.1016/j.astropartphys.2016.03.007}{{\em Astropart.
  Phys.} {\bfseries 80} (2016) 90--114},
  \href{http://arxiv.org/abs/1510.04372}{{\ttfamily arXiv:1510.04372
  [astro-ph.HE]}}.

\bibitem{Baker:2021btk}
M.~J. Baker and A.~Thamm
  \href{http://dx.doi.org/10.21468/SciPostPhys.12.5.150}{{\em SciPost Phys.}
  {\bfseries 12} no.~5, (2022) 150},
  \href{http://arxiv.org/abs/2105.10506}{{\ttfamily arXiv:2105.10506
  [hep-ph]}}.

\bibitem{Calabrese:2021src}
R.~Calabrese, M.~Chianese, D.~F.~G. Fiorillo, and N.~Saviano
  \href{http://dx.doi.org/10.1103/PhysRevD.105.L021302}{{\em Phys. Rev. D}
  {\bfseries 105} no.~2, (2022) L021302},
  \href{http://arxiv.org/abs/2107.13001}{{\ttfamily arXiv:2107.13001
  [hep-ph]}}.

\bibitem{Calabrese:2022rfa}
R.~Calabrese, M.~Chianese, D.~F.~G. Fiorillo, and N.~Saviano
  \href{http://dx.doi.org/10.1103/PhysRevD.105.103024}{{\em Phys. Rev. D}
  {\bfseries 105} no.~10, (2022) 103024},
  \href{http://arxiv.org/abs/2203.17093}{{\ttfamily arXiv:2203.17093
  [hep-ph]}}.

\bibitem{Baker:2022rkn}
M.~J. Baker and A.~Thamm \href{http://dx.doi.org/10.1007/JHEP01(2023)063}{{\em
  JHEP} {\bfseries 01} (2023) 063},
  \href{http://arxiv.org/abs/2210.02805}{{\ttfamily arXiv:2210.02805
  [hep-ph]}}.

\bibitem{Calza:2021czr}
M.~Calz\`a, J.~March-Russell, and J.~a.~G. Rosa
  \href{http://arxiv.org/abs/2110.13602}{{\ttfamily arXiv:2110.13602
  [astro-ph.CO]}}.

\bibitem{Calza:2022ljw}
M.~Calz\`a and J.~a.~G. Rosa
  \href{http://dx.doi.org/10.1007/JHEP12(2022)090}{{\em JHEP} {\bfseries 12}
  (2022) 090}, \href{http://arxiv.org/abs/2210.06500}{{\ttfamily
  arXiv:2210.06500 [gr-qc]}}.

\bibitem{Calza:2023gws}
M.~Calz\`a and J.~a.~G. Rosa \href{http://arxiv.org/abs/2311.12930}{{\ttfamily
  arXiv:2311.12930 [gr-qc]}}.

\bibitem{Atre:2009rg}
A.~Atre, T.~Han, S.~Pascoli, and B.~Zhang
  \href{http://dx.doi.org/10.1088/1126-6708/2009/05/030}{{\em JHEP} {\bfseries
  05} (2009) 030}, \href{http://arxiv.org/abs/0901.3589}{{\ttfamily
  arXiv:0901.3589 [hep-ph]}}.

\bibitem{Drewes:2015iva}
M.~Drewes and B.~Garbrecht
  \href{http://dx.doi.org/10.1016/j.nuclphysb.2017.05.001}{{\em Nucl. Phys. B}
  {\bfseries 921} (2017) 250--315},
  \href{http://arxiv.org/abs/1502.00477}{{\ttfamily arXiv:1502.00477
  [hep-ph]}}.

\bibitem{Abdullahi:2022jlv}
A.~M. Abdullahi {\em et~al.}
  \href{http://dx.doi.org/10.1088/1361-6471/ac98f9}{{\em J. Phys. G} {\bfseries
  50} no.~2, (2023) 020501}, \href{http://arxiv.org/abs/2203.08039}{{\ttfamily
  arXiv:2203.08039 [hep-ph]}}.

\bibitem{Bugaev:2001xr}
E.~V. Bugaev, M.~G. Elbakidze, and K.~V. Konishchev
  \href{http://dx.doi.org/10.1134/1.1563709}{{\em Phys. Atom. Nucl.} {\bfseries
  66} (2003) 476--480}, \href{http://arxiv.org/abs/astro-ph/0110660}{{\ttfamily
  arXiv:astro-ph/0110660}}.

\bibitem{Fujita:2014hha}
T.~Fujita, M.~Kawasaki, K.~Harigaya, and R.~Matsuda
  \href{http://dx.doi.org/10.1103/PhysRevD.89.103501}{{\em Phys. Rev. D}
  {\bfseries 89} no.~10, (2014) 103501},
  \href{http://arxiv.org/abs/1401.1909}{{\ttfamily arXiv:1401.1909
  [astro-ph.CO]}}.

\bibitem{Morrison:2018xla}
L.~Morrison, S.~Profumo, and Y.~Yu
  \href{http://dx.doi.org/10.1088/1475-7516/2019/05/005}{{\em JCAP} {\bfseries
  05} (2019) 005}, \href{http://arxiv.org/abs/1812.10606}{{\ttfamily
  arXiv:1812.10606 [astro-ph.CO]}}.

\bibitem{Ambrosone:2021lsx}
A.~Ambrosone, R.~Calabrese, D.~F.~G. Fiorillo, G.~Miele, and S.~Morisi
  \href{http://dx.doi.org/10.1103/PhysRevD.105.045001}{{\em Phys. Rev. D}
  {\bfseries 105} no.~4, (2022) 045001},
  \href{http://arxiv.org/abs/2106.11980}{{\ttfamily arXiv:2106.11980
  [hep-ph]}}.

\bibitem{Hooper:2020otu}
D.~Hooper and G.~Krnjaic
  \href{http://dx.doi.org/10.1103/PhysRevD.103.043504}{{\em Phys. Rev. D}
  {\bfseries 103} no.~4, (2021) 043504},
  \href{http://arxiv.org/abs/2010.01134}{{\ttfamily arXiv:2010.01134
  [hep-ph]}}.

\bibitem{Perez-Gonzalez:2020vnz}
Y.~F. Perez-Gonzalez and J.~Turner
  \href{http://dx.doi.org/10.1103/PhysRevD.104.103021}{{\em Phys. Rev. D}
  {\bfseries 104} no.~10, (2021) 103021},
  \href{http://arxiv.org/abs/2010.03565}{{\ttfamily arXiv:2010.03565
  [hep-ph]}}.

\bibitem{Bernal:2022pue}
N.~Bernal, C.~S. Fong, Y.~F. Perez-Gonzalez, and J.~Turner
  \href{http://dx.doi.org/10.1103/PhysRevD.106.035019}{{\em Phys. Rev. D}
  {\bfseries 106} no.~3, (2022) 035019},
  \href{http://arxiv.org/abs/2203.08823}{{\ttfamily arXiv:2203.08823
  [hep-ph]}}.

\bibitem{Calabrese:2023bxz}
R.~Calabrese, M.~Chianese, J.~Gunn, G.~Miele, S.~Morisi, and N.~Saviano
  \href{http://arxiv.org/abs/2311.13276}{{\ttfamily arXiv:2311.13276
  [hep-ph]}}.

\bibitem{Calabrese:2023key}
R.~Calabrese, M.~Chianese, J.~Gunn, G.~Miele, S.~Morisi, and N.~Saviano
  \href{http://dx.doi.org/10.1103/PhysRevD.107.123537}{{\em Phys. Rev. D}
  {\bfseries 107} no.~12, (2023) 123537},
  \href{http://arxiv.org/abs/2305.13369}{{\ttfamily arXiv:2305.13369
  [hep-ph]}}.

\bibitem{Ghoshal:2023fno}
A.~Ghoshal, Y.~F. Perez-Gonzalez, and J.~Turner
  \href{http://arxiv.org/abs/2312.06768}{{\ttfamily arXiv:2312.06768
  [hep-ph]}}.

\bibitem{IceCube:2018pgc}
{\bfseries IceCube} Collaboration, M.~G. Aartsen {\em et~al.}
  \href{http://dx.doi.org/10.1103/PhysRevD.99.032004}{{\em Phys. Rev. D}
  {\bfseries 99} no.~3, (2019) 032004},
  \href{http://arxiv.org/abs/1808.07629}{{\ttfamily arXiv:1808.07629
  [hep-ex]}}.

\bibitem{IceCube:2019cia}
{\bfseries IceCube} Collaboration, M.~G. Aartsen {\em et~al.}
  \href{http://dx.doi.org/10.1103/PhysRevLett.124.051103}{{\em Phys. Rev.
  Lett.} {\bfseries 124} no.~5, (2020) 051103},
  \href{http://arxiv.org/abs/1910.08488}{{\ttfamily arXiv:1910.08488
  [astro-ph.HE]}}.

\bibitem{Hawking:1976ra}
S.~W. Hawking \href{http://dx.doi.org/10.1103/PhysRevD.14.2460}{{\em Phys. Rev.
  D} {\bfseries 14} (1976) 2460--2473}.

\bibitem{Almheiri:2020cfm}
A.~Almheiri, T.~Hartman, J.~Maldacena, E.~Shaghoulian, and A.~Tajdini
  \href{http://dx.doi.org/10.1103/RevModPhys.93.035002}{{\em Rev. Mod. Phys.}
  {\bfseries 93} no.~3, (2021) 035002},
  \href{http://arxiv.org/abs/2006.06872}{{\ttfamily arXiv:2006.06872
  [hep-th]}}.

\bibitem{Buoninfante:2021ijy}
L.~Buoninfante, F.~Di~Filippo, and S.~Mukohyama
  \href{http://dx.doi.org/10.1007/JHEP10(2021)081}{{\em JHEP} {\bfseries 10}
  (2021) 081}, \href{http://arxiv.org/abs/2107.05662}{{\ttfamily
  arXiv:2107.05662 [hep-th]}}.

\bibitem{Page:1993wv}
D.~N. Page \href{http://dx.doi.org/10.1103/PhysRevLett.71.3743}{{\em Phys. Rev.
  Lett.} {\bfseries 71} (1993) 3743--3746},
  \href{http://arxiv.org/abs/hep-th/9306083}{{\ttfamily arXiv:hep-th/9306083}}.

\bibitem{Page:2013dx}
D.~N. Page \href{http://dx.doi.org/10.1088/1475-7516/2013/09/028}{{\em JCAP}
  {\bfseries 09} (2013) 028}, \href{http://arxiv.org/abs/1301.4995}{{\ttfamily
  arXiv:1301.4995 [hep-th]}}.

\bibitem{Alexandre:2024nuo}
A.~Alexandre, G.~Dvali, and E.~Koutsangelas
  \href{http://arxiv.org/abs/2402.14069}{{\ttfamily arXiv:2402.14069
  [hep-ph]}}.

\bibitem{Thoss:2024hsr}
V.~Thoss, A.~Burkert, and K.~Kohri
  \href{http://arxiv.org/abs/2402.17823}{{\ttfamily arXiv:2402.17823
  [astro-ph.CO]}}.

\bibitem{Balaji:2024hpu}
S.~Balaji, G.~Dom\`enech, G.~Franciolini, A.~Ganz, and J.~Tr\"ankle
  \href{http://arxiv.org/abs/2403.14309}{{\ttfamily arXiv:2403.14309 [gr-qc]}}.

\bibitem{Haque:2024eyh}
M.~R. Haque, S.~Maity, D.~Maity, and Y.~Mambrini
  \href{http://arxiv.org/abs/2404.16815}{{\ttfamily arXiv:2404.16815
  [hep-ph]}}.

\bibitem{Green:2016xgy}
A.~M. Green \href{http://dx.doi.org/10.1103/PhysRevD.94.063530}{{\em Phys. Rev.
  D} {\bfseries 94} no.~6, (2016) 063530},
  \href{http://arxiv.org/abs/1609.01143}{{\ttfamily arXiv:1609.01143
  [astro-ph.CO]}}.

\bibitem{Dolgov:2008wu}
A.~D. Dolgov, M.~Kawasaki, and N.~Kevlishvili
  \href{http://dx.doi.org/10.1016/j.nuclphysb.2008.08.029}{{\em Nucl. Phys. B}
  {\bfseries 807} (2009) 229--250},
  \href{http://arxiv.org/abs/0806.2986}{{\ttfamily arXiv:0806.2986 [hep-ph]}}.

\bibitem{Cheek:2022mmy}
A.~Cheek, L.~Heurtier, Y.~F. Perez-Gonzalez, and J.~Turner
  \href{http://dx.doi.org/10.1103/PhysRevD.108.015005}{{\em Phys. Rev. D}
  {\bfseries 108} no.~1, (2023) 015005},
  \href{http://arxiv.org/abs/2212.03878}{{\ttfamily arXiv:2212.03878
  [hep-ph]}}.

\bibitem{MacGibbon:1990zk}
J.~H. MacGibbon and B.~R. Webber
  \href{http://dx.doi.org/10.1103/PhysRevD.41.3052}{{\em Phys. Rev. D}
  {\bfseries 41} (1990) 3052--3079}.

\bibitem{MacGibbon:1991tj}
J.~H. MacGibbon \href{http://dx.doi.org/10.1103/PhysRevD.44.376}{{\em Phys.
  Rev. D} {\bfseries 44} (1991) 376--392}.

\bibitem{Cheek:2021odj}
A.~Cheek, L.~Heurtier, Y.~F. Perez-Gonzalez, and J.~Turner
  \href{http://dx.doi.org/10.1103/PhysRevD.105.015022}{{\em Phys. Rev. D}
  {\bfseries 105} no.~1, (2022) 015022},
  \href{http://arxiv.org/abs/2107.00013}{{\ttfamily arXiv:2107.00013
  [hep-ph]}}.

\bibitem{Page:1977um}
D.~N. Page \href{http://dx.doi.org/10.1103/PhysRevD.16.2402}{{\em Phys. Rev. D}
  {\bfseries 16} (1977) 2402--2411}.

\bibitem{Arbey:2021mbl}
A.~Arbey and J.~Auffinger
  \href{http://dx.doi.org/10.1140/epjc/s10052-021-09702-8}{{\em Eur. Phys. J.
  C} {\bfseries 81} (2021) 910},
  \href{http://arxiv.org/abs/2108.02737}{{\ttfamily arXiv:2108.02737 [gr-qc]}}.

\bibitem{Doran:2005vm}
C.~Doran, A.~Lasenby, S.~Dolan, and I.~Hinder
  \href{http://dx.doi.org/10.1103/PhysRevD.71.124020}{{\em Phys. Rev. D}
  {\bfseries 71} (2005) 124020},
  \href{http://arxiv.org/abs/gr-qc/0503019}{{\ttfamily arXiv:gr-qc/0503019}}.

\bibitem{Dolan:2006vj}
S.~Dolan, C.~Doran, and A.~Lasenby
  \href{http://dx.doi.org/10.1103/PhysRevD.74.064005}{{\em Phys. Rev. D}
  {\bfseries 74} (2006) 064005},
  \href{http://arxiv.org/abs/gr-qc/0605031}{{\ttfamily arXiv:gr-qc/0605031}}.

\bibitem{Lunardini:2019zob}
C.~Lunardini and Y.~F. Perez-Gonzalez
  \href{http://dx.doi.org/10.1088/1475-7516/2020/08/014}{{\em JCAP} {\bfseries
  08} (2020) 014}, \href{http://arxiv.org/abs/1910.07864}{{\ttfamily
  arXiv:1910.07864 [hep-ph]}}.

\bibitem{Bauer:2020jay}
C.~W. Bauer, N.~L. Rodd, and B.~R. Webber
  \href{http://dx.doi.org/10.1007/JHEP06(2021)121}{{\em JHEP} {\bfseries 06}
  (2021) 121}, \href{http://arxiv.org/abs/2007.15001}{{\ttfamily
  arXiv:2007.15001 [hep-ph]}}.

\bibitem{Minkowski:1977sc}
P.~Minkowski
\href{http://dx.doi.org/10.1016/0370-2693(77)90435-X}{{\em Phys. Lett.}
  {\bfseries 67B} (1977) 421--428}.

\bibitem{Yanagida:1979as}
T.~Yanagida {\em Conf. Proc. C} {\bfseries 7902131} (1979) 95--99.

\bibitem{GellMann:1980vs}
M.~Gell-Mann, P.~Ramond, and R.~Slansky {\em Conf. Proc.} {\bfseries C790927}
  (1979) 315--321, \href{http://arxiv.org/abs/1306.4669}{{\ttfamily
  arXiv:1306.4669 [hep-th]}}.

\bibitem{Mohapatra:1979ia}
R.~N. Mohapatra and G.~Senjanovic
  \href{http://dx.doi.org/10.1103/PhysRevLett.44.912}{{\em Phys. Rev. Lett.}
  {\bfseries 44} (1980) 912}.

\bibitem{Schechter:1980gr}
J.~Schechter and J.~W.~F. Valle
  \href{http://dx.doi.org/10.1103/PhysRevD.22.2227}{{\em Phys.Rev.D} {\bfseries
  22} (1980) 2227}.

\bibitem{deSalas:2020pgw}
P.~F. de~Salas, D.~V. Forero, S.~Gariazzo, P.~Mart\'\i{}nez-Mirav\'e, O.~Mena,
  C.~A. Ternes, M.~T\'ortola, and J.~W.~F. Valle
  \href{http://dx.doi.org/10.1007/JHEP02(2021)071}{{\em JHEP} {\bfseries 02}
  (2021) 071}, \href{http://arxiv.org/abs/2006.11237}{{\ttfamily
  arXiv:2006.11237 [hep-ph]}}.

\bibitem{Capozzi:2021fjo}
F.~Capozzi, E.~Di~Valentino, E.~Lisi, A.~Marrone, A.~Melchiorri, and A.~Palazzo
  \href{http://dx.doi.org/10.1103/PhysRevD.104.083031}{{\em Phys. Rev. D}
  {\bfseries 104} no.~8, (2021) 083031},
  \href{http://arxiv.org/abs/2107.00532}{{\ttfamily arXiv:2107.00532
  [hep-ph]}}.

\bibitem{Esteban:2020cvm}
I.~Esteban, M.~C. Gonzalez-Garcia, M.~Maltoni, T.~Schwetz, and A.~Zhou
  \href{http://dx.doi.org/10.1007/JHEP09(2020)178}{{\em JHEP} {\bfseries 09}
  (2020) 178}, \href{http://arxiv.org/abs/2007.14792}{{\ttfamily
  arXiv:2007.14792 [hep-ph]}}.

\bibitem{Mohapatra:1986bd}
R.~N. Mohapatra and J.~W.~F. Valle
  \href{http://dx.doi.org/10.1103/PhysRevD.34.1642}{{\em Phys. Rev. D}
  {\bfseries 34} (1986) 1642}.

\bibitem{Akhmedov:1995ip}
E.~K. Akhmedov, M.~Lindner, E.~Schnapka, and J.~W.~F. Valle
  \href{http://dx.doi.org/10.1016/0370-2693(95)01504-3}{{\em Phys. Lett. B}
  {\bfseries 368} (1996) 270--280},
  \href{http://arxiv.org/abs/hep-ph/9507275}{{\ttfamily arXiv:hep-ph/9507275}}.

\bibitem{Akhmedov:1995vm}
E.~K. Akhmedov, M.~Lindner, E.~Schnapka, and J.~W.~F. Valle
  \href{http://dx.doi.org/10.1103/PhysRevD.53.2752}{{\em Phys. Rev. D}
  {\bfseries 53} (1996) 2752--2780},
  \href{http://arxiv.org/abs/hep-ph/9509255}{{\ttfamily arXiv:hep-ph/9509255}}.

\bibitem{Malinsky:2005bi}
M.~Malinsky, J.~C. Romao, and J.~W.~F. Valle
  \href{http://dx.doi.org/10.1103/PhysRevLett.95.161801}{{\em Phys. Rev. Lett.}
  {\bfseries 95} (2005) 161801},
  \href{http://arxiv.org/abs/hep-ph/0506296}{{\ttfamily arXiv:hep-ph/0506296}}.

\bibitem{Shrock:1980ct}
R.~E. Shrock \href{http://dx.doi.org/10.1103/PhysRevD.24.1232}{{\em Phys. Rev.
  D} {\bfseries 24} (1981) 1232}.

\bibitem{Helo:2010cw}
J.~C. Helo, S.~Kovalenko, and I.~Schmidt
  \href{http://dx.doi.org/10.1016/j.nuclphysb.2011.07.020}{{\em Nucl. Phys. B}
  {\bfseries 853} (2011) 80--104},
  \href{http://arxiv.org/abs/1005.1607}{{\ttfamily arXiv:1005.1607 [hep-ph]}}.

\bibitem{Bondarenko:2018ptm}
K.~Bondarenko, A.~Boyarsky, D.~Gorbunov, and O.~Ruchayskiy
  \href{http://dx.doi.org/10.1007/JHEP11(2018)032}{{\em JHEP} {\bfseries 11}
  (2018) 032}, \href{http://arxiv.org/abs/1805.08567}{{\ttfamily
  arXiv:1805.08567 [hep-ph]}}.

\bibitem{Bryman:2019ssi}
D.~A. Bryman and R.~Shrock
  \href{http://dx.doi.org/10.1103/PhysRevD.100.053006}{{\em Phys. Rev. D}
  {\bfseries 100} no.~5, (2019) 053006},
  \href{http://arxiv.org/abs/1904.06787}{{\ttfamily arXiv:1904.06787
  [hep-ph]}}.

\bibitem{Bryman:2019bjg}
D.~A. Bryman and R.~Shrock
  \href{http://dx.doi.org/10.1103/PhysRevD.100.073011}{{\em Phys. Rev. D}
  {\bfseries 100} (2019) 073011},
  \href{http://arxiv.org/abs/1909.11198}{{\ttfamily arXiv:1909.11198
  [hep-ph]}}.

\bibitem{Mastrototaro:2019vug}
L.~Mastrototaro, A.~Mirizzi, P.~D. Serpico, and A.~Esmaili
  \href{http://dx.doi.org/10.1088/1475-7516/2020/01/010}{{\em JCAP} {\bfseries
  01} (2020) 010}, \href{http://arxiv.org/abs/1910.10249}{{\ttfamily
  arXiv:1910.10249 [hep-ph]}}.

\bibitem{Coloma:2020lgy}
P.~Coloma, E.~Fern\'andez-Mart\'\i{}nez, M.~Gonz\'alez-L\'opez,
  J.~Hern\'andez-Garc\'\i{}a, and Z.~Pavlovic
  \href{http://dx.doi.org/10.1140/epjc/s10052-021-08861-y}{{\em Eur. Phys. J.
  C} {\bfseries 81} no.~1, (2021) 78},
  \href{http://arxiv.org/abs/2007.03701}{{\ttfamily arXiv:2007.03701
  [hep-ph]}}.

\bibitem{Akita:2023iwq}
K.~Akita, S.~H. Im, M.~Masud, and S.~Yun
  \href{http://arxiv.org/abs/2312.13627}{{\ttfamily arXiv:2312.13627
  [hep-ph]}}.

\bibitem{Oberauer:1993yr}
L.~Oberauer, C.~Hagner, G.~Raffelt, and E.~Rieger
  \href{http://dx.doi.org/10.1016/0927-6505(93)90004-W}{{\em Astropart. Phys.}
  {\bfseries 1} (1993) 377--386}.

\bibitem{Syvolap:2023trc}
V.~Syvolap \href{http://arxiv.org/abs/2301.07052}{{\ttfamily arXiv:2301.07052
  [hep-ph]}}.

\bibitem{Mastrototaro:2021wzl}
L.~Mastrototaro, P.~D. Serpico, A.~Mirizzi, and N.~Saviano
  \href{http://dx.doi.org/10.1103/PhysRevD.104.016026}{{\em Phys. Rev. D}
  {\bfseries 104} no.~1, (2021) 016026},
  \href{http://arxiv.org/abs/2104.11752}{{\ttfamily arXiv:2104.11752
  [hep-ph]}}.

\bibitem{Fischer:2022zwu}
{\bfseries IceCube} Collaboration, L.~Fischer
  \href{http://dx.doi.org/10.22323/1.414.0190}{{\em PoS} {\bfseries ICHEP2022}
  (11, 2022) 190}.

\bibitem{ParticleDataGroup:2022pth}
{\bfseries Particle Data Group} Collaboration, R.~L. Workman {\em et~al.}
  \href{http://dx.doi.org/10.1093/ptep/ptac097}{{\em PTEP} {\bfseries 2022}
  (2022) 083C01}.

\bibitem{Cirelli:2010xx}
M.~Cirelli, G.~Corcella, A.~Hektor, G.~Hutsi, M.~Kadastik, P.~Panci, M.~Raidal,
  F.~Sala, and A.~Strumia
  \href{http://dx.doi.org/10.1088/1475-7516/2012/10/E01}{{\em JCAP} {\bfseries
  03} (2011) 051}, \href{http://arxiv.org/abs/1012.4515}{{\ttfamily
  arXiv:1012.4515 [hep-ph]}}. [Erratum: JCAP 10, E01 (2012)].

\bibitem{Sjostrand:2014zea}
T.~Sj\"ostrand, S.~Ask, J.~R. Christiansen, R.~Corke, N.~Desai, P.~Ilten,
  S.~Mrenna, S.~Prestel, C.~O. Rasmussen, and P.~Z. Skands
  \href{http://dx.doi.org/10.1016/j.cpc.2015.01.024}{{\em Comput. Phys.
  Commun.} {\bfseries 191} (2015) 159--177},
  \href{http://arxiv.org/abs/1410.3012}{{\ttfamily arXiv:1410.3012 [hep-ph]}}.

\bibitem{HAWC:2011gts}
{\bfseries HAWC} Collaboration, A.~U. Abeysekara {\em et~al.}
  \href{http://dx.doi.org/10.1016/j.astropartphys.2012.02.001}{{\em Astropart.
  Phys.} {\bfseries 35} (2012) 641--650},
  \href{http://arxiv.org/abs/1108.6034}{{\ttfamily arXiv:1108.6034
  [astro-ph.HE]}}.

\bibitem{Esmaili:2012us}
A.~Esmaili, A.~Ibarra, and O.~L.~G. Peres
  \href{http://dx.doi.org/10.1088/1475-7516/2012/11/034}{{\em JCAP} {\bfseries
  11} (2012) 034}, \href{http://arxiv.org/abs/1205.5281}{{\ttfamily
  arXiv:1205.5281 [hep-ph]}}.

\bibitem{Honda:2015fha}
M.~Honda, M.~Sajjad~Athar, T.~Kajita, K.~Kasahara, and S.~Midorikawa
  \href{http://dx.doi.org/10.1103/PhysRevD.92.023004}{{\em Phys. Rev. D}
  {\bfseries 92} no.~2, (2015) 023004},
  \href{http://arxiv.org/abs/1502.03916}{{\ttfamily arXiv:1502.03916
  [astro-ph.HE]}}.

\bibitem{NA62:2020mcv}
{\bfseries NA62} Collaboration, E.~Cortina~Gil {\em et~al.}
  \href{http://dx.doi.org/10.1016/j.physletb.2020.135599}{{\em Phys. Lett. B}
  {\bfseries 807} (2020) 135599},
  \href{http://arxiv.org/abs/2005.09575}{{\ttfamily arXiv:2005.09575
  [hep-ex]}}.

\bibitem{T2K:2019jwa}
{\bfseries T2K} Collaboration, K.~Abe {\em et~al.}
  \href{http://dx.doi.org/10.1103/PhysRevD.100.052006}{{\em Phys. Rev. D}
  {\bfseries 100} no.~5, (2019) 052006},
  \href{http://arxiv.org/abs/1902.07598}{{\ttfamily arXiv:1902.07598
  [hep-ex]}}.

\bibitem{PIENU:2017wbj}
{\bfseries PIENU} Collaboration, A.~Aguilar-Arevalo {\em et~al.}
  \href{http://dx.doi.org/10.1103/PhysRevD.97.072012}{{\em Phys. Rev. D}
  {\bfseries 97} no.~7, (2018) 072012},
  \href{http://arxiv.org/abs/1712.03275}{{\ttfamily arXiv:1712.03275
  [hep-ex]}}.

\bibitem{Barouki:2022bkt}
R.~Barouki, G.~Marocco, and S.~Sarkar
  \href{http://dx.doi.org/10.21468/SciPostPhys.13.5.118}{{\em SciPost Phys.}
  {\bfseries 13} (2022) 118}, \href{http://arxiv.org/abs/2208.00416}{{\ttfamily
  arXiv:2208.00416 [hep-ph]}}.

\bibitem{Bernardi:1987ek}
G.~Bernardi {\em et~al.}
  \href{http://dx.doi.org/10.1016/0370-2693(88)90563-1}{{\em Phys. Lett. B}
  {\bfseries 203} (1988) 332--334}.

\bibitem{MicroBooNE:2019izn}
{\bfseries MicroBooNE} Collaboration, P.~Abratenko {\em et~al.}
  \href{http://dx.doi.org/10.1103/PhysRevD.101.052001}{{\em Phys. Rev. D}
  {\bfseries 101} no.~5, (2020) 052001},
  \href{http://arxiv.org/abs/1911.10545}{{\ttfamily arXiv:1911.10545
  [hep-ex]}}.

\bibitem{NuTeV:1999kej}
{\bfseries NuTeV, E815} Collaboration, A.~Vaitaitis {\em et~al.}
  \href{http://dx.doi.org/10.1103/PhysRevLett.83.4943}{{\em Phys. Rev. Lett.}
  {\bfseries 83} (1999) 4943--4946},
  \href{http://arxiv.org/abs/hep-ex/9908011}{{\ttfamily arXiv:hep-ex/9908011}}.

\bibitem{E949:2014gsn}
{\bfseries E949} Collaboration, A.~V. Artamonov {\em et~al.}
  \href{http://dx.doi.org/10.1103/PhysRevD.91.052001}{{\em Phys. Rev. D}
  {\bfseries 91} no.~5, (2015) 052001},
  \href{http://arxiv.org/abs/1411.3963}{{\ttfamily arXiv:1411.3963 [hep-ex]}}.
  [Erratum: Phys.Rev.D 91, 059903 (2015)].

\bibitem{CHARMII:1994jjr}
{\bfseries CHARM II} Collaboration, P.~Vilain {\em et~al.}
  \href{http://dx.doi.org/10.1016/0370-2693(94)01422-9}{{\em Phys. Lett. B}
  {\bfseries 343} (1995) 453--458}.

\bibitem{Coloma:2017ppo}
P.~Coloma, P.~A.~N. Machado, I.~Martinez-Soler, and I.~M. Shoemaker
  \href{http://dx.doi.org/10.1103/PhysRevLett.119.201804}{{\em Phys. Rev.
  Lett.} {\bfseries 119} no.~20, (2017) 201804},
  \href{http://arxiv.org/abs/1707.08573}{{\ttfamily arXiv:1707.08573
  [hep-ph]}}.

\bibitem{Carenza:2023old}
P.~Carenza, G.~Lucente, L.~Mastrototaro, A.~Mirizzi, and P.~D. Serpico
  \href{http://dx.doi.org/10.1103/PhysRevD.109.063010}{{\em Phys. Rev. D}
  {\bfseries 109} no.~6, (2024) 063010},
  \href{http://arxiv.org/abs/2311.00033}{{\ttfamily arXiv:2311.00033
  [hep-ph]}}.

\bibitem{Bolton:2019pcu}
P.~D. Bolton, F.~F. Deppisch, and P.~S. Bhupal~Dev
  \href{http://dx.doi.org/10.1007/JHEP03(2020)170}{{\em JHEP} {\bfseries 03}
  (2020) 170}, \href{http://arxiv.org/abs/1912.03058}{{\ttfamily
  arXiv:1912.03058 [hep-ph]}}.

\bibitem{IceCube-Gen2:2020qha}
{\bfseries IceCube-Gen2} Collaboration, M.~G. Aartsen {\em et~al.}
  \href{http://dx.doi.org/10.1088/1361-6471/abbd48}{{\em J. Phys. G} {\bfseries
  48} no.~6, (2021) 060501}, \href{http://arxiv.org/abs/2008.04323}{{\ttfamily
  arXiv:2008.04323 [astro-ph.HE]}}.

\bibitem{KM3Net:2016zxf}
{\bfseries KM3Net} Collaboration, S.~Adrian-Martinez {\em et~al.}
  \href{http://dx.doi.org/10.1088/0954-3899/43/8/084001}{{\em J. Phys. G}
  {\bfseries 43} no.~8, (2016) 084001},
  \href{http://arxiv.org/abs/1601.07459}{{\ttfamily arXiv:1601.07459
  [astro-ph.IM]}}.

\bibitem{P-ONE:2020ljt}
{\bfseries P-ONE} Collaboration, M.~Agostini {\em et~al.}
  \href{http://dx.doi.org/10.1038/s41550-020-1182-4}{{\em Nature Astron.}
  {\bfseries 4} no.~10, (2020) 913--915},
  \href{http://arxiv.org/abs/2005.09493}{{\ttfamily arXiv:2005.09493
  [astro-ph.HE]}}.

\bibitem{BAIKAL:1997iok}
{\bfseries BAIKAL} Collaboration, I.~A. Belolaptikov {\em et~al.}
  \href{http://dx.doi.org/10.1016/S0927-6505(97)00022-4}{{\em Astropart. Phys.}
  {\bfseries 7} (1997) 263--282}.

\end{thebibliography}\endgroup

\end{document}